# Machine-Learned Interatomic Potential for the $MoS_2$/Sapphire System

Emir Bilgili[1*], Nicholas Taormina[1], Simon R. Phillpot[2], Youping Chen[1]

[1] Department of Mechanical and Aerospace Engineering, University of Florida, Gainesville, FL, 32611
[2] Department of Materials Science and Engineering, University of Florida, Gainesville, FL, 32611

[*] *Corresponding author: emir.bilgili@ufl.edu*

## Abstract

We develop a machine-learned neuroevolution potential (NEP) for the $MoS_2$/α-$Al_2O_3$(0001) system. The potential is trained on 63,611 PBE+D3 structures distilled from ~1.1 million curated density-functional-theory (DFT) calculations, using an energy-difference loss function that directly targets interfacial and orientation-dependent energies. On 2,820 held-out test structures from the DFT pool, the energy and force errors are 29 meV/atom and 0.23 eV/Å. Validated against DFT on a suite of physical properties, the NEP reproduces the lattice constants within 0.3 %, the elastic constants within 17 %, the phonon dispersions with mean deviations below 2.5 %, and the surface energies within 1.3 %. At the $MoS_2$/sapphire interface, it reproduces the adhesion energy minimum within 0.8 %, the interfacial energy within 2.2 % and the most stable orientation of the monolayer. It preserves the DFT-predicted energetic ordering of point defects (Spearman $\rho$ = 0.995) and mirror-twin boundaries ($\rho$ = 1.0), and reproduces tilt grain-boundary energies within 8 % ($\rho$ = 0.98) and stacking-fault energy barriers within 14 %.

## I. Introduction

As silicon field-effect transistors approach their fundamental scaling limits[1,2], two-dimensional (2D) transition metal dichalcogenides (TMDs), especially $MoS_2$, have emerged as leading channel materials for post-silicon nanoelectronics[3,4]. Heteroepitaxy, the growth of crystalline thin films on oriented substrates, is the only commercially viable manufacturing method for atomically sharp interfaces required for solid-state devices[5,6]; c-plane sapphire, α-$Al_2O_3$(0001), is the state-of-the-art substrate for $MoS_2$[7–10]. Despite many laboratory demonstrations of highly crystalline $MoS_2$ overlayers on sapphire[8–11], no commercial TMD devices yet exist. The critical bottleneck remains the formation of defects, particularly grain boundaries, which form during domain coalescence: low-angle tilt boundaries between slightly misoriented domains and, most commonly, mirror-twin boundaries between antiparallel domains rotated by 60°[12–14].

Atomistic simulations are natural and powerful tools for studying such processes, but their success depends on the fidelity of the available interatomic potentials. Classical potentials for $MoS_2$ (REBO, Stillinger–Weber, ReaxFF)[15–18] were parameterized for Mo–S only and do not describe the $MoS_2$/sapphire interface. Machine-learned interatomic potentials (MLIPs) fit energies and forces from DFT as functions of local atomic environments and offer near-DFT accuracy at a fraction of its cost[19,20]. Representative frameworks include neural-network[21,22] and Gaussian approximation potentials[23], linear[24–26] and equivariant models[27–29], which differ in their balance of accuracy, data efficiency and computational cost, and hence in the problems for which they are best suited[30]. Universal MLIPs[31–33] cover most of the periodic table but systematically soften the potential energy surface and underestimate surface and defect energies[34]. Significant system- and application-specific effort therefore remains necessary for the development of reliable potentials[33,34], and we have previously developed such MLIPs for AlN[35] and $MoS_2$[36].

The $MoS_2$/sapphire system, though, is considerably more challenging, as it is a four-component system with covalent, ionic and van der Waals (vdW) bonding. An appropriate potential must accurately describe the sapphire substrate, single- and multilayer $MoS_2$ and the interface between them, and be efficient enough to reach the length and time scales required to realistically model extended defects and interfaces. MLIPs have been used to study Mo deposition and sulfurization on monolayer $WS_2$ in small cells without a sapphire substrate[37] and to pre-screen 2D Mo–S phases on sapphire for DFT[38]. To our knowledge, however, no general-purpose MLIP for the $MoS_2/Al_2O_3$ heterosystem has yet been developed and comprehensively validated against DFT across bulk, interfacial and defect properties. Low energy and force errors on training or test data do not guarantee accurate defect energies or other derived properties[39], which must therefore be validated directly[40].

The neuroevolution potential (NEP) framework[41–44] combines a compact radial/angular descriptor with a single-hidden-layer neural network trained by a separable natural evolution strategy, runs natively on GPUs and maintains comparable accuracy to other state-of-the-art MLIPs at order(s)-of-magnitude lower computational cost[41,43,44]. The potential developed here runs molecular dynamics simulations of a 55,404-atom $MoS_2/Al_2O_3$(0001) superlattice at 300 K at 7.8 ns/day on a single RTX PRO 6000 Blackwell GPU.

In this work, we develop a neuroevolution potential (NEP) for the $MoS_2$/c-plane sapphire, α-$Al_2O_3$(0001), system. The training set covers bulk phases, surfaces, $MoS_2/Al_2O_3$ interfaces, flakes, defects and high-temperature configurations. We validate the NEP against DFT for (i) lattice parameters and equations of state; (ii) elastic constants and phonon dispersions; (iii) surface, adhesion and interfacial energies; (iv) the rotation energetics of $MoS_2$ monolayers and flakes on sapphire; and (v) the energies of edges, point defects, grain boundaries and stacking faults, both freestanding and on sapphire. The resulting potential provides a DFT-validated atomistic description of the $MoS_2$/sapphire system. Combining NEP with the concurrent atomistic-continuum (CAC) method[45–47], we open the way to mechanistic studies of $MoS_2$/sapphire interfaces, extended defects, and growth at device-relevant length scales.

## II. Methodology

The NEP model was developed for α-$Al_2O_3$, single- and multilayer $MoS_2$, and the $MoS_2/Al_2O_3(0001)$ interface. DFT data were generated, curated and subsampled into a training set (Secs. 2.1–2.2); NEP models were fit (Sec. 2.3); and each candidate was screened against a suite of DFT validation metrics (Sec. 2.4). Failures were addressed by targeted DFT data, by active-learning data or by changes to the training-set composition and weights or to the loss.

### 2.1 DFT reference calculations

All reference data were computed with the projector augmented-wave (PAW) method[48,49] as implemented in VASP 6.4.1[50,51], using the PBE exchange–correlation functional[52] and the PAW potentials Al, O, Mo_sv and S. Dispersion was included through the DFT-D3 correction with zero damping[53]. A plane-wave cutoff of 520 eV, PREC = Accurate, an electronic convergence criterion of $10^{-6}$ eV, and Gaussian smearing of 0.05 eV were used throughout, without spin polarization. Brillouin zones were sampled with Γ-centered meshes with k-point spacings of about 0.13–0.38 $Å^{-1}$ (mean 0.23 $Å^{-1}$), depending on the configuration; large supercells used the Γ point only (Note S1). The NEP was trained on DFT energies referenced to spin-polarized isolated atoms. All energies reported here, and those of the released model, are instead referenced to non-spin-polarized isolated atoms, consistent with the non-spin-polarized DFT data. The two references differ by a constant per element (Note S1), which leaves forces, stresses, and dynamics, as well as all energy differences at fixed composition, unchanged. Most training ab initio molecular dynamics (aiMD) trajectories used PREC = Normal and a $10^{-5}$ eV criterion. Equations of state, elastic constants, and phonons were computed with EDIFF = $10^{-8}$ eV and projection operators evaluated in reciprocal space. We refer the reader to Note S1 and Table S1 of the Supplemental Material for more information.

### 2.2 Training data

Before filtering, all ionic steps of the DFT calculations were collected into a single database; calculations performed with settings other than those of Sec. 2.1 (e.g., with a lower plane-wave cutoff) were excluded. Spurious or corrupted frames were then discarded. Frames with unconverged electronic steps, unphysically short interatomic contacts, atomic forces above 50 eV/Å, a net force above 1 eV/Å along any axis, or energies more than 0.3 eV/atom below the convex hull of reference structures, or energies more than 0.5 eV/atom above the spin-polarized isolated-atom reference used in training (except isolated atom pairs) were discarded, as were steps of structure-search relaxations whose volume had changed by more than 20 %, for which the fixed plane-wave basis no longer matches the cell. For the same reason, the stresses of later steps of variable-cell relaxations, which carry Pulay (incomplete-basis) errors, were removed from the virial loss (2,812 training structures); for those whose volume had changed by more than 5 %, the energy was also removed from the loss (1,581 structures), so that only their forces are fitted.

The resulting pool of 1,090,322 frames was distilled into the training set by stratified subsampling. Frames were grouped into 11 strata by origin and physical type, each with a frame quota and a per-structure weight. Within a stratum, the quota was divided among its subsets in proportion to the square root of their size. Within each subset, frames were selected by farthest-point sampling (FPS) in a per-structure feature space built from the per-species mean and standard deviation of the 49-dimensional NEP descriptor vectors and from the composition, standardized within the stratum. The test set consists of 2,820 structures drawn from entire trajectories held out from the training set.

The curated dataset was designed to cover both near-equilibrium and highly non-equilibrium environments. It comprises: (1) distorted crystals of α-$Al_2O_3$ and of the $MoS_2$ polytypes (1H, 1T, 1T′, 2H, 3R), including strained, equation-of-state, elastic and phonon-displacement cells and targeted interlayer compression, shear and sliding sets; (2) stable and metastable structures from grand-canonical genetic-algorithm searches (GASP)[54,55] over all Mo–S–Al–O binaries, ternaries and the quaternary, together with Materials Project compounds[56]; (3) static and heated $Al_2O_3$ and $MoS_2$ slabs; (4) isolated and sapphire-supported $MoS_2$ flakes, including rotation scans and point defects; (5) $MoS_2$/$Al_2O_3$ heterostructures, random epilayers, solid solutions and aiMD trajectories of the interface at 300–1600 K; (6) rigid layer–substrate separation scans of $MoS_2$ monolayers, flakes and bilayers on sapphire, which sample the adhesion as a function of distance; (7) adatom potential-energy-surface scans, Mo–S nanoparticles, deposition snapshots, and targeted interfacial cells; (8) active-learning cells from MD simulations (see below); and (9) the $MoS_2$ dataset of our previous UF3 potential[36]. The composition of the training set is summarized in Table I.

**TABLE I.** Composition of the training set of the final model: 62,182 distinct structures (7.09 million atoms) and 1,429 copies, i.e. 63,611 entries (7.18 million atoms); the test set contains 2,820 structures from whole held-out runs. Per-structure weights enter the loss squared. The entries with weight 0.3 within the categories are the 102 reference structures of the energy-difference combinations (Sec. 2.3).

| Category | Contents | Structures | Weight |
|---|---|---|---|
| **Bulk crystals** | α-$Al_2O_3$ and $MoS_2$ polytypes (1H, 1T, 1T′, 2H, 3R) under strain, compression, shear and sliding; elastic and phonon displacements; α-S ($S_8$) under pressure | 13,958 | 0.3–8 |
| **Structure search and compounds** | GASP structures over all Mo–S–Al–O binaries, ternaries and the quaternary; Materials Project compounds; elemental phases | 3,030 | 1–2 |
| **Bulk aiMD** | aiMD of α-$Al_2O_3$, $MoS_2$ and other Mo–S–Al–O compounds | 4,500 | 1 |
| **Surfaces** | static and heated α-$Al_2O_3$(0001) and $MoS_2$(0001) slabs, stepped and passivated surfaces | 13,512 | 0.3–2 |
| **$MoS_2$ flakes** | isolated and sapphire-supported 1H flakes, rotation scans, edges, grain boundaries and point defects | 6,111 | 0.3–1.5 |
| **$MoS_2$/$Al_2O_3$ interfaces** | periodic heterostructures, random epilayers, solid solutions, interface aiMD (300–1600 K) | 13,905 | 0.3–2 |
| **Layer–substrate separation scans** | rigid adhesion scans of $MoS_2$ monolayers, flakes and bilayers on sapphire | 426 | 0.3–3 |
| **Growth-relevant configurations** | adatom energy-surface scans, Mo–S clusters and nanoparticles, deposition snapshots, targeted interface and wall-scan cells | 4,790 | 2–3 |
| **Active learning** | extrapolative growth environments cut into periodic cells and labeled by DFT | 885 | 2 |
| **$MoS_2$ dataset of Ref.[36]** | the $MoS_2$ training data of our previous UF3 potential | 1,065 | 1 |
| **Copies for energy-difference batching** | renamed duplicates of structures shared by energy-difference combinations in different mini-batches (Sec. 2.3) | 1,429 | 0.3 |
| **Total** | | 63,611 | |

The dataset was iteratively extended via active learning on finite-temperature MD simulations including deposition trajectories. The most extrapolative, non-redundant environments were embedded into periodic cells of a few hundred atoms with a *periodic substrate-window cutter*, labeled by DFT single-point calculations, and added to the training set, and the cycle was repeated. The final training set contains 885 active-learning structures (Table I). The methodological details of the active-learning procedure and of the periodic substrate-window cutter will be presented in a forthcoming paper and are not discussed further here.

### 2.3 NEP model and training

In the NEP framework[41–43], the total energy is a sum of site energies $U_i$, each given by a feed-forward neural network with one hidden layer of $N_{neu}$ neurons that acts on a descriptor vector $q^i$ of atom i:

$$E = \sum_i U_i(q^i),\ U_i = \sum_{\mu=1}^{N_{\mathrm{neu}}} w_\mu^{(1)} \tanh\left(\sum_\nu w_{\mu\nu}^{(0)} q_\nu^i - b_\mu^{(0)}\right) - b^{(1)} \tag{1}$$

The radial descriptor components are built from radial functions of the neighbor distances, each a linear combination of Chebyshev-polynomial basis functions with trainable coefficients that depend on the pair of elements; the angular components combine the same type of radial functions with spherical harmonics of the bond directions into three-body and four-body invariants[42]. In NEP4[43], the network parameters $w^{(0)}$, $b^{(0)}$ and $w^{(1)}$ are specific to the element of atom i, whereas the output bias $b^{(1)}$ is shared by all elements.

We use NEP4 with element-specific radial and angular cutoffs. Several cutoff schemes were tested. Per-element cutoffs are efficient because a long interaction range is unnecessary inside the stiff oxide substrate but is essential across the interface and, even more so, within and between $MoS_2$ layers. The final cutoffs are 9.5 Å (radial) and 5.0 Å (angular) for Mo and S, and 5.5 Å and 3.5 Å for Al and O; a pair of different elements uses the mean of the two cutoffs, i.e. 7.5 Å and 4.25 Å across the $MoS_2/Al_2O_3$ interface. The descriptor uses $n_{max} = 8$ (7) and basis sizes of 12 (7) for the radial (angular) part, three-body angular terms up to $l_{max} = 4$ and the default four-body term, which gives 9 radial and 40 angular, i.e. 49, components. The network has 46 hidden neurons, and the model has 12,281 trainable parameters, of which 2,896 are in the descriptor. The universal Ziegler–Biersack–Littmark (ZBL) potential[57] is added to the NEP energy at short interatomic distances[58]: it acts in full below 0.75 Å and is smoothly switched off at 1.5 Å. The hyperparameters were chosen heuristically over the development period; the complete input file is given in Table S2 (Note S2) of the Supplemental Material.

The parameters are optimized with the separable natural evolution strategy (SNES)[41,59] on the loss

$$L = \lambda_{\mathrm{e}}\mathrm{RMSE}_E + \lambda_{\mathrm{f}}\mathrm{RMSE}_F + \lambda_{\mathrm{v}}\mathrm{RMSE}_W + \lambda_1 \frac{1}{N_{\mathrm{par}}}\sum_n |w_n| + \lambda_2 \sqrt{\frac{1}{N_{\mathrm{par}}}\sum_n w_n{}^2} + \lambda_{\mathrm{d}}\mathrm{RMSE}_{\Delta E} \tag{2}$$

where $RMSE_E$ and $RMSE_W$ are the root-mean-square errors of the energies and virials per atom, and $RMSE_F$ that of the force components. The two regularization terms are the mean absolute value and the root-mean-square value of the $N_{par}$ parameters $w_n$[41], with the *nep 5.9.1* default weights $\lambda_1 = \lambda_2 = \sqrt{(N_{par}/N_{types})}/1000 = 0.055$ ($N_{par}$ = 12,281, $N_{types}$ = 4). Per-structure weights (Table I) enter the energy, force and virial terms squared. The shear components of the virial are weighted by $\lambda_s = 5$, so that they enter with an effective weight $\lambda_v\ \lambda_s$[60]. The force term is mildly biased toward small forces[60]: the squared force error of atom i is multiplied by $\delta/(\delta + |F_i^{DFT}|)$ with $\delta = 60$ eV/Å, which reduces the weight of an atom with a 10 eV/Å reference force by 14 %. In NEP4, SNES evaluates Eq. (2) separately for each element, over the structures containing it and that element's parameters, and updates each element's parameters by its own loss[43].

The last term of Eq. (2) is the energy-difference loss, available since GPUMD 5.9[61]. Most interfacial properties are defined by energy differences between related structures, including adhesion, rotation, surface, stacking-fault and defect energies. The accuracy with which a potential reproduces them therefore depends on how well the errors of the individual structures cancel. The energy-difference term instead fits user-defined linear combinations of total energies directly[60]:

$$\mathrm{RMSE}_{\Delta E} = \sqrt{\frac{1}{N_\mathrm{c}}\sum_k w_k\left(\sum_i c_{ki}\left(E_i^{\mathrm{NEP}} - E_i^{\mathrm{DFT}}\right)\right)^2}, \quad \sum_i c_{ki} N_i^{\alpha} = 0 \quad \forall\alpha \tag{3}$$

where $c_{ki}$ is the coefficient of structure i in combination k, $w_k$ the weight of the combination, $N_i^\alpha$ the number of atoms of element $\alpha$ in structure i, and $N_c$ the number of combinations in the current mini-batch. The second condition, which the code enforces, makes each combination balanced in every element, so that isolated-atom references cancel. GPUMD keeps all structures of a combination in the same mini-batch, and the per-structure weights do not enter this term. We scale the $c_{ki}$ by the inverse number of atoms of the target structure, so that each residual is an energy per target atom, and set $w_k = (1\ \mathrm{meV/atom}/\tau)^2$, where $\tau$ is a tolerance of 0.15–1 meV/atom for most families, set per family, and up to 8 and 35 meV/atom for individual surface and edge combinations. The combinations were built automatically: separation, registry, rotation and sliding series whose references are rigid copies of subsystems of the target cell; equation-of-state, small-strain and stacking-fault series relative to a structure of the same composition; and surface, edge and defect energies relative to bulk or chemical-potential references. The final model uses $\lambda_d = 2.5$ with 1,302 training combinations in 32 families; a further 116 combinations are held out and evaluated with the test set. The 102 reference structures that were not already in the training set were added with a per-structure weight of 0.3. Because combinations that share a structure must be in the same mini-batch, a structure shared by combinations in different mini-batches was duplicated under another name, also with a weight of 0.3 (1,429 copies).

Training used the nep executable of GPUMD 5.9.1[42] with $\lambda_e = 1$, $\lambda_f = 0.9$, $\lambda_v = 1$, $\lambda_s = 5$, a population of 50 and mini-batches of about 1,000 structures (64 per pass through the training set). The final model was obtained in three consecutive runs of $3\times10^5$ generations each ($9\times10^5$ in total): (1) a model trained from scratch, without the energy-difference term, on an earlier version of the training set (67,118 structures); (2) a continuation of this model on the same structures after the stresses of later steps of variable-cell relaxations had been removed (Sec. 2.2); and (3) a continuation with the energy-difference term on the final training set. Each run took 12–13 h on a single NVIDIA RTX PRO 6000 Blackwell GPU. The final training set contains 63,611 entries (62,182 distinct structures with 7.09 million atoms, plus the 1,429 copies), and the test set contains 2,820 structures.

### 2.4 Validation protocol

Every property was computed with DFT and NEP on identical structures or with identical protocols. NEP statics and relaxations used calorine[62] and ASE[63]; phonons used phonopy[64]; molecular dynamics used LAMMPS[65]. In property validation, NEP was evaluated as a single point on the DFT-relaxed geometry and, where applicable, by relaxing that geometry with NEP.

Quantities that change the Mo:S ratio, such as the energies of off-stoichiometric edges and grain boundaries, depend on the sulfur chemical potential, $\mu_S$, and are reported at its two limits: the Mo-rich limit, in which $MoS_2$ is in equilibrium with bcc-Mo, and the S-rich limit, in which it is in equilibrium with α-sulfur. With $\mu_{MoS_2}$ the energy per formula unit of the 1H monolayer, these limits are

$$\mu_S^{Mo-rich} = \frac{1}{2}\left(\mu_{MoS_2} - \mu_{Mo}^{bcc}\right), \qquad \mu_S^{S-rich} = \mu_S^{\alpha-S}$$

In DFT, the formation enthalpy, $\Delta H_f$, of 1H-$MoS_2$ relative to bcc-Mo and α-S is −2.399 eV per formula unit, which defines the window $\Delta\mu_S = \mu_S - \mu_S^{\alpha-S} \in [-1.20, 0]$ eV. Relaxing the reference structures with NEP gives $\Delta H_f$ = −2.543 eV per formula unit and hence the window [−1.27, 0] eV. In validation, each method is evaluated at its own two limits. Comparisons are therefore between the same equilibrium conditions ($MoS_2$ coexisting with bcc Mo or with α-S), as predicted by each model, i.e., DFT references do not enter NEP predictions for such calculations.

Unless stated otherwise, the substrate is an Al-terminated α-$Al_2O_3$(0001) slab one hexagonal unit cell thick (six $O_3$ planes, i.e. 18 atomic layers, ≈12 Å). In finite-sized interfacial configurations, the bottom two $O_3$ planes, together with the Al atoms below and between them, are held fixed at bulk positions. The DFT settings of the validation calculations are given in Note S1, and the cells and computational details of each property in Notes S4–S9 of the Supplemental Material.

## III. Results and Discussion

### 3.1 Energy, force and virial accuracy

We compare the NEP with DFT on the 63,611 training entries (Sec. 2.2) and on a test set of 2,820 held-out structures. The test set consists of entire runs from the ~1.1-million DFT pool held out from the training data, so that test structures are not correlated with training structures through a shared trajectory. As shown in Fig. 1a–c, the NEP reaches root-mean-square errors (RMSEs) on the test set of 29.2 meV/atom in energy, 0.230 eV/Å in force and 43.0 meV/atom in virial; on the training set, the values are 33.4 meV/atom, 0.276 eV/Å and 75.1 meV/atom. Only two training structures, Mo–Al dimers on the repulsive wall at 1.6 and 1.7 Å, lie above the isolated-atom reference (+1.9 and +0.4 eV/atom).

The error distributions of the two sets are centered at zero and nearly coincide as shown in Fig. 1d–f. Table II provides the test errors by category. The global energy RMSE is dominated by the structure-search category, which contains strongly distorted, high-energy structures and has the largest errors (93.9 meV/atom and 0.494 eV/Å): with 5 % of the test structures, it carries 55 % of the squared energy error, but only 3 % and 9 % of the squared force and virial errors. For most other categories, including those associated with heterostructures, the energy errors are much smaller: 7.6 meV/atom for bulk crystals, 9.5 meV/atom for $MoS_2$ flakes, 14.7 meV/atom for $MoS_2/Al_2O_3$ interfaces, 8.6 meV/atom for growth-relevant configurations and 1.3 meV/atom for the scans of the layer–substrate separation (Sec. 3.3.1). The surface category, which is dominated by heated slabs, reaches 33.9 meV/atom. The NEP also reproduces 300 high-temperature (273–1729 K) interface configurations from aiMD with errors of 3.5 meV/atom and 0.136 eV/Å (see Note S3, Figs. S1 and S2, and Table S3 of the Supplemental Material).

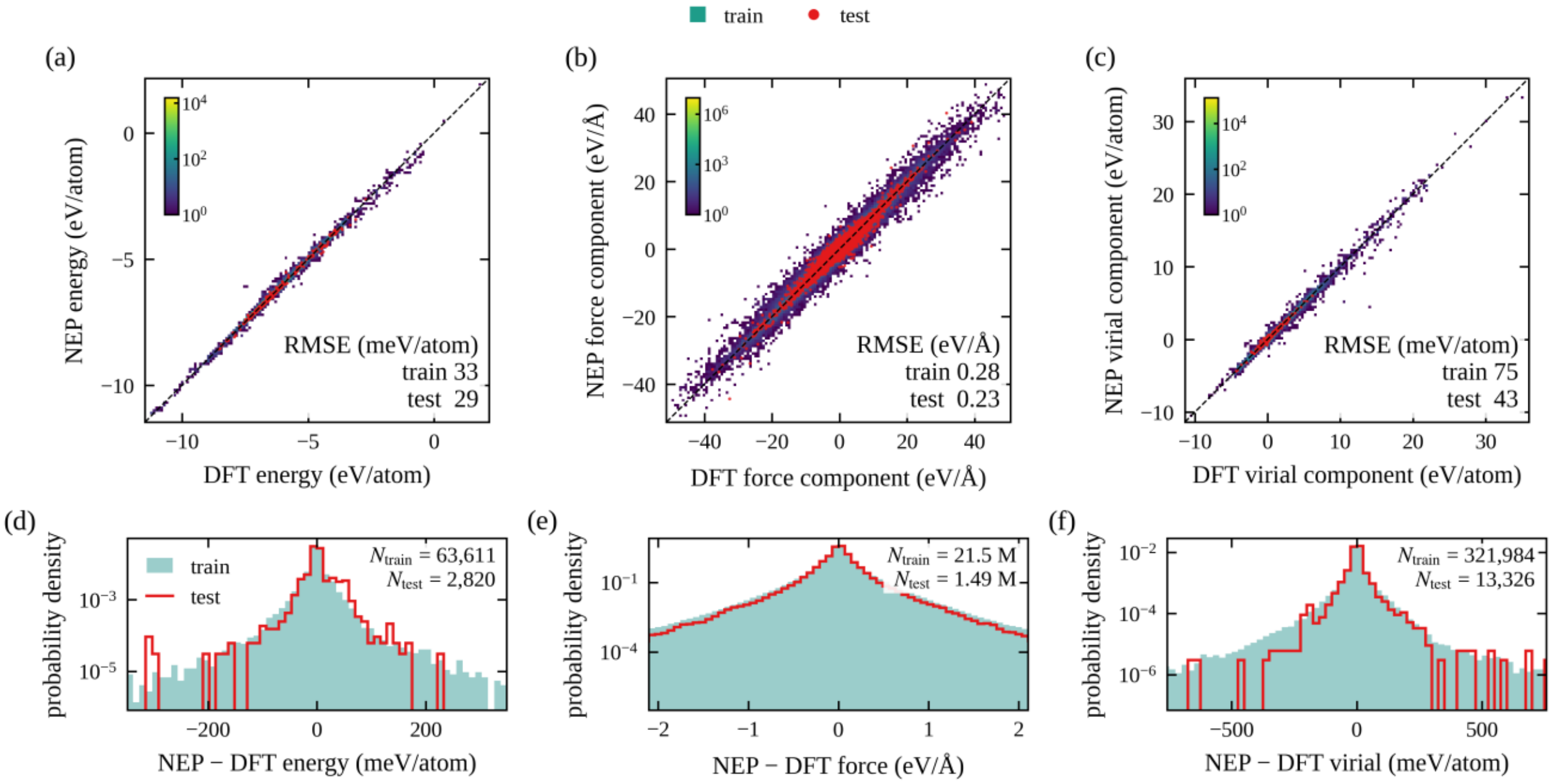

**FIG. 1. Accuracy of the NEP against DFT**. (a–c) Parity of energies, force components, and virial components for the training set, shown as the number of points per bin on a logarithmic color scale, with every test-set point overlaid in red. Dashed lines mark perfect agreement; RMSEs are given in each panel. (d–f) Distributions of the NEP − DFT errors for the training (filled) and test (outline) sets on a logarithmic scale, over all 63,611 training and 2,820 test structures (21.5 M and 1.49 M force components; 321,984 and 13,326 virial components).

**TABLE II.** Test-set root-mean-square errors of the NEP by data category (categories as in Table I; whole held-out runs). Energies and virials are per atom. Virial errors are computed over the structures that carry a DFT stress tensor (last column); stresses of heated slabs and of later steps of variable-cell relaxations are not used.

| Category | Structures | *E* (meV/atom) | *F* (eV/Å) | *W* (meV/atom) | Structures with *W* |
|---|---|---|---|---|---|
| **Bulk crystals** | 420 | 7.6 | 0.061 | 51.1 | 420 |
| **Structure search and compounds** | 150 | 93.9 | 0.494 | 146.7 | 17 |
| **Bulk aiMD** | 225 | 13.5 | 0.453 | 59.7 | 225 |
| **Surfaces** | 675 | 33.9 | 0.402 | 44.2 | 250 |
| **$MoS_2$ flakes** | 290 | 9.5 | 0.141 | 15.7 | 290 |
| **$MoS_2/Al_2O_3$ interfaces** | 662 | 14.7 | 0.170 | 30.7 | 660 |
| **Layer–substrate separation scans** | 10 | 1.3 | 0.084 | 8.7 | 10 |
| **Growth-relevant configurations** | 285 | 8.6 | 0.241 | 12.8 | 246 |
| **Active learning** | 15 | 30.4 | 0.251 | 22.4 | 15 |
| **$MoS_2$ dataset of Ref.[36]** | 88 | 19.1 | 0.361 | 82.8 | 88 |
| **All test data** | 2,820 | 29.2 | 0.230 | 43.0 | 2,221 |

### 3.2 Bulk properties of $MoS_2$ and $Al_2O_3$

3.2.1 Lattice Constants and Equations of State

The lattice constants of $MoS_2$ and $Al_2O_3$ determine the lattice mismatch and strain in the $MoS_2$ film on the sapphire substrate. Lattice constants for α-$Al_2O_3$ and the 1H (monolayer), 2H (bulk) and 3R (bulk) polytypes of $MoS_2$ were obtained by relaxing the cell and atomic positions of each structure. The van der Waals gap, $d_{\mathrm{vdW}}$, was measured as the distance between the S planes of adjacent layers in the relaxed 2H and 3R crystals. Equation of state (EOS) curves were obtained by computing the energy at volumes $V/V_0$ = 0.94–1.06 with the cell shape and the atoms relaxed at each volume (Fig. 2a-c). The curves were fit to the third-order Birch–Murnaghan form[66] to obtain the equilibrium volume $V_0$ and the bulk modulus $B_0$ (see Note S4 and Fig. S3 of the Supplemental Material). Because the binding of $MoS_2$ multilayers is dominated by van der Waals forces, we also compute the energy of 2H- and 3R-$MoS_2$ as a function of the interlayer lattice parameter $c$ at fixed in-plane lattice parameter $a$ (Fig. 2e-f). For the 1H monolayer, we also consider its energy as a function of biaxial strain (Fig. 2d).

Table III lists the structural parameters and EOS constants predicted by both NEP and DFT. The NEP reproduces the DFT lattice constants of all four systems within 0.3 % (Table III). The van der Waals gaps of 2H- and 3R-$MoS_2$ are 0.018 and 0.014 Å (0.6 % and 0.5 %) larger than in DFT, and this small excess accounts for the slightly larger $c$ of the NEP. As shown in Fig. 2, the NEP energy curves follow DFT closely in all six EOS curves. The equilibrium volumes agree within 0.1 %, and the bulk moduli of α-$Al_2O_3$ and 2H-$MoS_2$ within 0.7 %. The largest deviation is the bulk modulus of 3R-$MoS_2$, which the NEP overestimates by 8 %. Consistently, the NEP energy of 3R-$MoS_2$ rises slightly too steeply when the layers are pushed together, lying about 2.5 meV/atom above DFT at $c/c_0$ = 0.92 as shown in Fig. 2f.

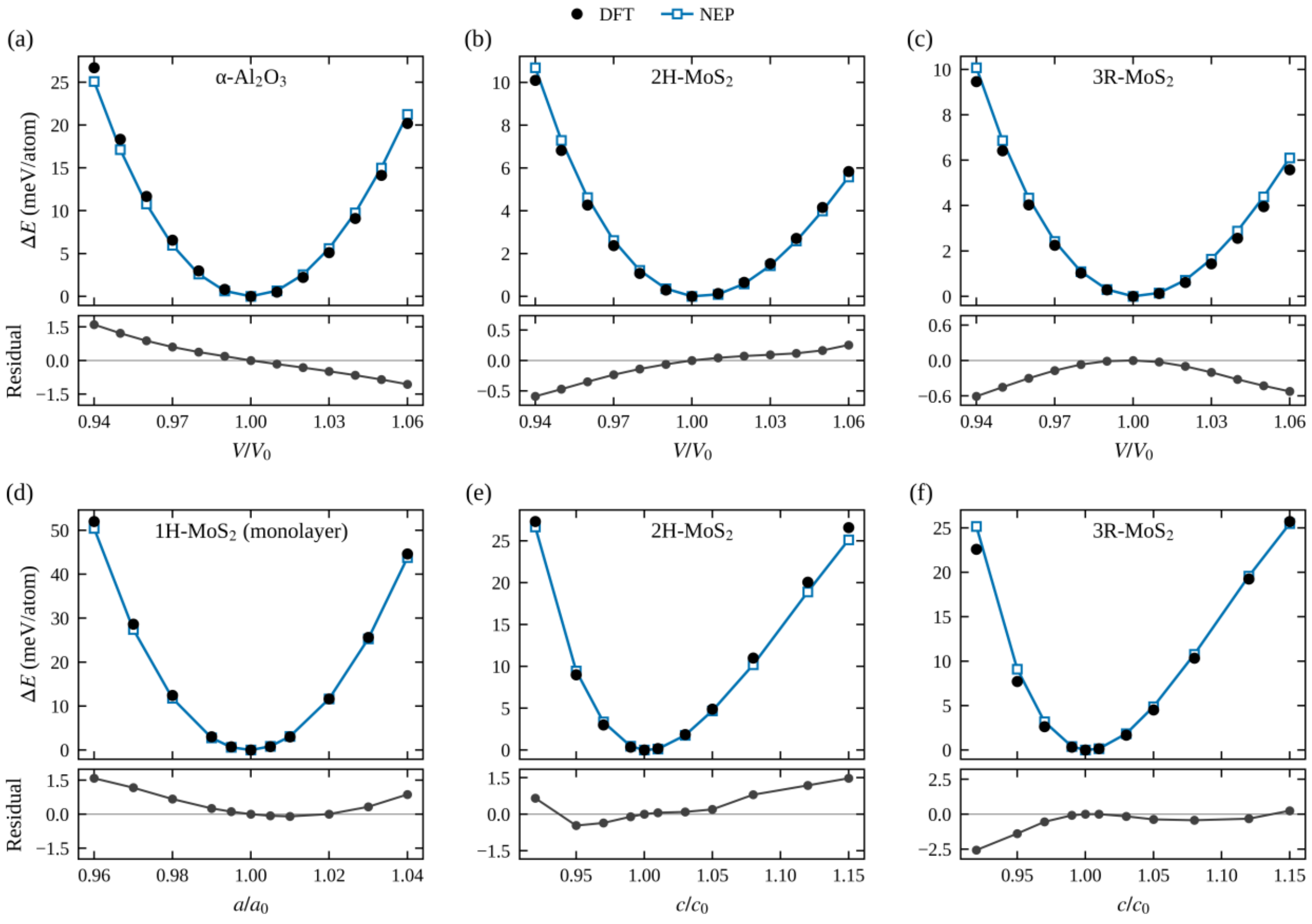

**FIG. 2**. **Equations of state computed with identical protocols for DFT (circles) and the NEP (squares; lines are guides to the eye).** (a–c) E(V) of α-$Al_2O_3$, 2H-$MoS_2$ and 3R-$MoS_2$ for $V/V_0$ = 0.94–1.06, with cell shape and ions relaxed at each volume. (d) Biaxial strain of the 1H-$MoS_2$ monolayer, $a/a_0$ = 0.96–1.04. (e, f) Interlayer compression and expansion of 2H- and 3R-$MoS_2$, $c/c_0$ = 0.92–1.15 at fixed a with ions relaxed. All axes are normalized to the DFT equilibrium cell, and each curve is referenced to its own minimum, $\Delta E = E - E_{min}$. The strip below each panel shows the residual $\Delta E^{DFT} - \Delta E^{NEP}$ (meV/atom), the difference between the two curves above.

**TABLE III.** Lattice parameters from full relaxation, van der Waals gap, $d_{vdW}$, between the S planes of adjacent layers in the relaxed 2H and 3R crystals, and equilibrium volume $V_0$ and bulk modulus $B_0$ from third-order Birch–Murnaghan fits.

| System | Quantity | Unit | DFT | NEP | Error (%) |
|---|---|---|---|---|---|
| α-$Al_2O_3$ | *a* | Å | 4.786 | 4.789 | 0.0 |
| | *c* | Å | 13.06 | 13.04 | −0.1 |
| | *V*/atom | Å$^3$ | 8.637 | 8.635 | 0.0 |
| | $V_0$/atom (EOS) | Å$^3$ | 8.647 | 8.635 | −0.1 |
| | $B_0$ | GPa | 238.4 | 237.4 | −0.4 |
| 2H-$MoS_2$ | *a* | Å | 3.163 | 3.163 | 0.0 |
| | *c* | Å | 12.33 | 12.37 | +0.3 |
| | $d_{vdW}$ (S–S) | Å | 3.040 | 3.058 | +0.6 |
| | *V*/atom | Å$^3$ | 17.81 | 17.86 | +0.3 |
| | $V_0$/atom (EOS) | Å$^3$ | 17.84 | 17.86 | +0.1 |
| | $B_0$ | GPa | 38.20 | 38.47 | +0.7 |
| 3R-$MoS_2$ | *a* | Å | 3.165 | 3.164 | 0.0 |
| | *c* | Å | 18.42 | 18.47 | +0.2 |
| | $d_{vdW}$ (S–S) | Å | 3.017 | 3.031 | +0.5 |
| | *V*/atom | Å$^3$ | 17.76 | 17.79 | +0.2 |
| | $V_0$/atom (EOS) | Å$^3$ | 17.79 | 17.79 | 0.0 |
| | $B_0$ | GPa | 36.24 | 39.14 | +8.0 |
| 1H-$MoS_2$ (monolayer) | *a* | Å | 3.167 | 3.165 | 0.0 |

## 3.2.2 Elastic Constants

Elastic constants describe the anisotropic behavior of $MoS_2$ and sapphire to applied stress; accurate elastic constants are essential for the reliable modeling of mechanical properties and the simulation of deformation processes. The relaxed-ion elastic tensors were computed with the same finite-strain protocol for both NEP and DFT, each on its own relaxed cell. Strains of ±0.5 % and ±1 % were applied to one Voigt component at a time, the ions were relaxed, and the elastic constants were obtained from linear fits of the stress–strain relation $\sigma_i = C_{ij}\varepsilon_j$.

As shown in Table IV, all 23 elastic constants agree with DFT within 17 %. The largest deviations are $C_{13}$ and $C_{33}$ of α-$Al_2O_3$ (+16.9 % and −14.8 %), $C_{12}$ of the monolayer (−14.6 %) and $C_{33}$ of 3R-$MoS_2$ (+14.3 %), consistent with the stiffer interlayer response of 3R-$MoS_2$ seen in its bulk modulus (Sec. 3.2.1). The constants that describe the interlayer response of 2H-$MoS_2$, $C_{33}$, $C_{13}$ and $C_{44}$, agree with DFT within 13.2 %.

**TABLE IV.** Predicted elastic constants of α-$Al_2O_3$, 2H-$MoS_2$, 3R-$MoS_2$, and 1H-$MoS_2$. $C_{66}$ is computed directly and satisfies $C_{66} = (C_{11} - C_{12})/2$ within 0.2 %. Errors are (NEP − DFT)/|DFT|.

| System | Constant | Unit | DFT | NEP | Error (%) |
|---|---|---|---|---|---|
| α-$Al_2O_3$ | $C_{11}$ | GPa | 466.8 | 465.7 | −0.2 |
| | $C_{12}$ | GPa | 153.1 | 153.7 | +0.4 |
| | $C_{13}$ | GPa | 112.5 | 131.6 | +16.9 |
| | $C_{33}$ | GPa | 466.8 | 397.9 | −14.8 |
| | $C_{44}$ | GPa | 138.8 | 131.3 | −5.4 |
| | $C_{66} = (C_{11} - C_{12})/2$ | GPa | 156.8 | 155.9 | −0.5 |
| | $C_{14}$ | GPa | −20.06 | −20.72 | −3.3 |
| 2H-$MoS_2$ | $C_{11}$ | GPa | 221.8 | 217.9 | −1.7 |
| | $C_{12}$ | GPa | 52.85 | 48.47 | −8.3 |
| | $C_{13}$ | GPa | 9.148 | 7.937 | −13.2 |
| | $C_{33}$ | GPa | 47.95 | 49.82 | +3.9 |
| | $C_{44}$ | GPa | 15.10 | 13.13 | −13.0 |
| | $C_{66} = (C_{11} - C_{12})/2$ | GPa | 84.36 | 84.80 | +0.5 |
| 3R-$MoS_2$ | $C_{11}$ | GPa | 222.7 | 218.3 | −2.0 |
| | $C_{12}$ | GPa | 54.17 | 47.95 | −11.5 |
| | $C_{13}$ | GPa | 12.03 | 11.62 | −3.4 |
| | $C_{33}$ | GPa | 42.80 | 48.93 | +14.3 |
| | $C_{44}$ | GPa | 15.46 | 14.32 | −7.4 |
| | $C_{66} = (C_{11} - C_{12})/2$ | GPa | 84.18 | 85.20 | +1.2 |
| | $C_{14}$ | GPa | 3.527 | 3.806 | +7.9 |
| 1H-$MoS_2$ (monolayer) | $C_{11}$ | N/m | 135.3 | 135.8 | +0.4 |
| | $C_{12}$ | N/m | 31.08 | 26.54 | −14.6 |
| | $C_{66} = (C_{11} - C_{12})/2$ | N/m | 52.06 | 54.65 | +5.0 |

### 3.2.3 Phonon Dispersion

Phonon dispersions measure the curvature of the potential-energy surface for the vibrational modes of a crystal, governing the atomic vibrations and lattice dynamics in molecular dynamics simulations. Accurate phonon spectra are therefore essential for the reliable modeling and simulation of thermal transport processes.

Phonons were computed by finite displacements of 0.01 Å with phonopy[64]. The NEP dispersions were computed both on the NEP-relaxed cell and on the DFT-relaxed cell, to separate errors in the forces from errors in the lattice constants. We quantify the deviation at each point of the band path by $\delta_p = \frac{|f^{\mathrm{NEP}} - f^{\mathrm{DFT}}|}{\max(f^{\mathrm{DFT}}, 1\ \mathrm{THz})}$, where $f$ is the phonon frequency; the 1 THz floor prevents the near-zero acoustic frequencies close to Γ from dominating the error. As shown in Fig. 3, the NEP reproduces the DFT dispersions of all four systems, with mean deviations of 1.7–2.4 % and no imaginary modes. The acoustic and low-frequency branches agree closely. The largest absolute errors occur in the highest optical branches, which the NEP places too low, at about 13.1–13.2 versus 13.8–14.0 THz at Γ for $MoS_2$ and about 23 versus 26 THz near Z for α-$Al_2O_3$, and in the longitudinal acoustic branch at the zone boundary (M), which is about 0.7–0.9 THz too low in all $MoS_2$ polytypes. The supercells used in the phonon calculations are shown in Fig. S3(e–h) of the Supplemental Material.

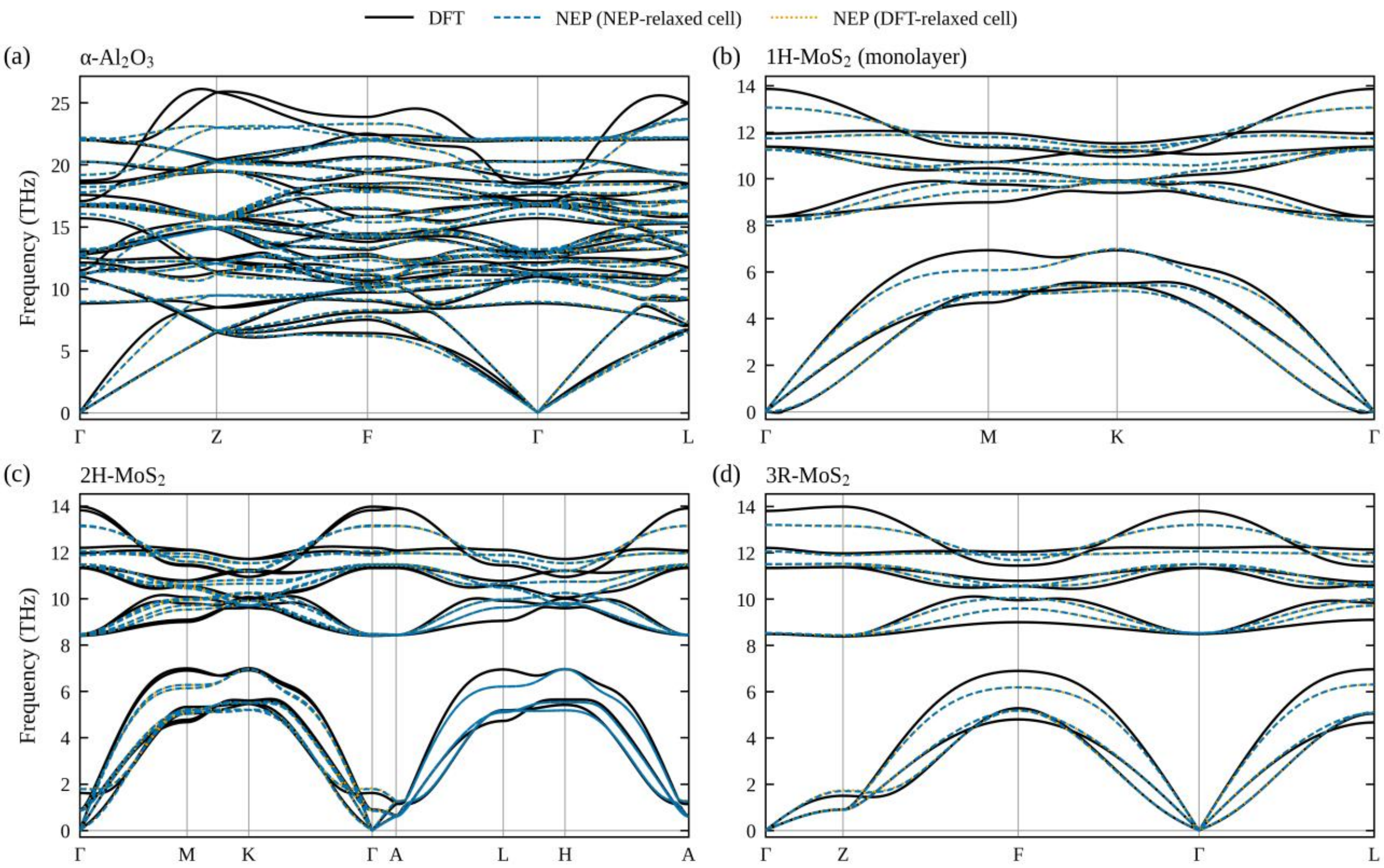

**FIG. 3**. **Phonon dispersions of (a) α-$Al_2O_3$, (b) the 1H-$MoS_2$ monolayer, (c) 2H-$MoS_2$ and (d) 3R-$MoS_2$** from DFT (black, solid) and the NEP, evaluated on the NEP-relaxed cell (blue, dashed) and on the DFT-relaxed cell (orange, dotted). The two NEP curves differ by less than 0.05 THz and largely overlap.

### 3.3 Interfacial Properties

3.3.1 Surface, Adhesion, and Interfacial Energies

Surface energy quantifies the work required to create a surface per unit area and is an important physical property, governing processes such as fracture, epitaxial growth, and wetting. Surface energies, $\gamma_s$, were computed for slabs of increasing thickness for $Al_2O_3$(0001) and 2H-$MoS_2$(0001), as

$$\gamma_s = \frac{E_{slab} - E_{bulk}}{2A},$$

where $E_{slab}$ is the energy of the slab, $E_{bulk}$ is the energy of an equivalent amount of bulk crystal, and $A$ is the surface area (see Note S5.1 and Fig. S4). Both surface energies converge within a few layers, and the NEP reproduces the relaxed values of the thickest slabs within 1.3 % as shown in Fig. 4c-d (1.927 versus 1.941 J/m$^2$ for $Al_2O_3$ and 0.239 versus 0.242 J/m$^2$ for $MoS_2$).

The binding strength of a $MoS_2$ monolayer on sapphire was quantified by computing the adhesion energy, $E_{ad}$, per $MoS_2$ unit:

$$E_{\mathrm{ad}} = \frac{E_{\mathrm{het}} - E_{\mathrm{slab}} - E_{\mathrm{layer}}}{N_{\mathrm{MoS_2}}},$$

where $E_{\text{het}}$, $E_{\text{slab}}$, and $E_{\text{layer}}$ are the energies of the heterostructure, the bare slab, and the free monolayer, respectively, all computed in the same cell, and $N_{\text{MoS}_2}$ is the number of $MoS_2$ formula units. The potential energy surface of the binding was probed by rigidly displacing a relaxed monolayer along the surface normal, plotting $E_{ad}$ as a function of the gap, *d*, between the lower S plane and the topmost O atom (see Note S5.2 and Fig. S5 for details). As shown in Fig. 4a, the NEP reasonably reproduces the binding curve of the monolayer. The depth of the adhesion minimum agrees with DFT within 0.8 % at θ = 0° (−375.5 versus −378.4 meV/$MoS_2$) and within 6.5 % at θ = 10.9°, and the gap at the minimum within 0.01 Å. The only systematic discrepancy occurs at large interlayer gaps, where DFT-D3 retains an attraction of approximately −17 meV/$MoS_2$ beyond the NEP cutoff. For the fully relaxed monolayer at θ = 0°, the NEP gives $E_{\text{ad}}$ = −312.6 meV/$MoS_2$, compared with −313.1 meV/$MoS_2$ in DFT.

The interfacial energy, $\gamma_{\text{int}}$, was obtained from a periodic superlattice of four 2H-$MoS_2$ layers and the $Al_2O_3$(0001) slab, which contains two interfaces per cell,

$$\gamma_{\text{int}} = \frac{E_{\text{SL}} - n_1 e_{\text{Al}_2\text{O}_3} - n_2 e_{\text{MoS}_2}}{2A},$$

where $E_{\text{SL}}$ is the energy of the superlattice, $n_1$ and $n_2$ the numbers of $Al_2O_3$ and $MoS_2$ formula units, and $e_{\text{Al}_2\text{O}_3}$ and $e_{\text{MoS}_2}$ the corresponding bulk energies per formula unit. Three lateral positions (registries) of the $MoS_2$ block relative to the substrate were considered (see Note S5.3 and Fig. S6 for details). As shown in Fig. 4b, the DFT-predicted interfacial energy varies by less than 1 % between the three registries (1.613–1.627 J/m$^2$), and the NEP reproduces it within 2.2 %, both on the DFT-relaxed superlattices (1.595–1.601 J/m$^2$) and after its own relaxation (1.584–1.591 J/m$^2$).

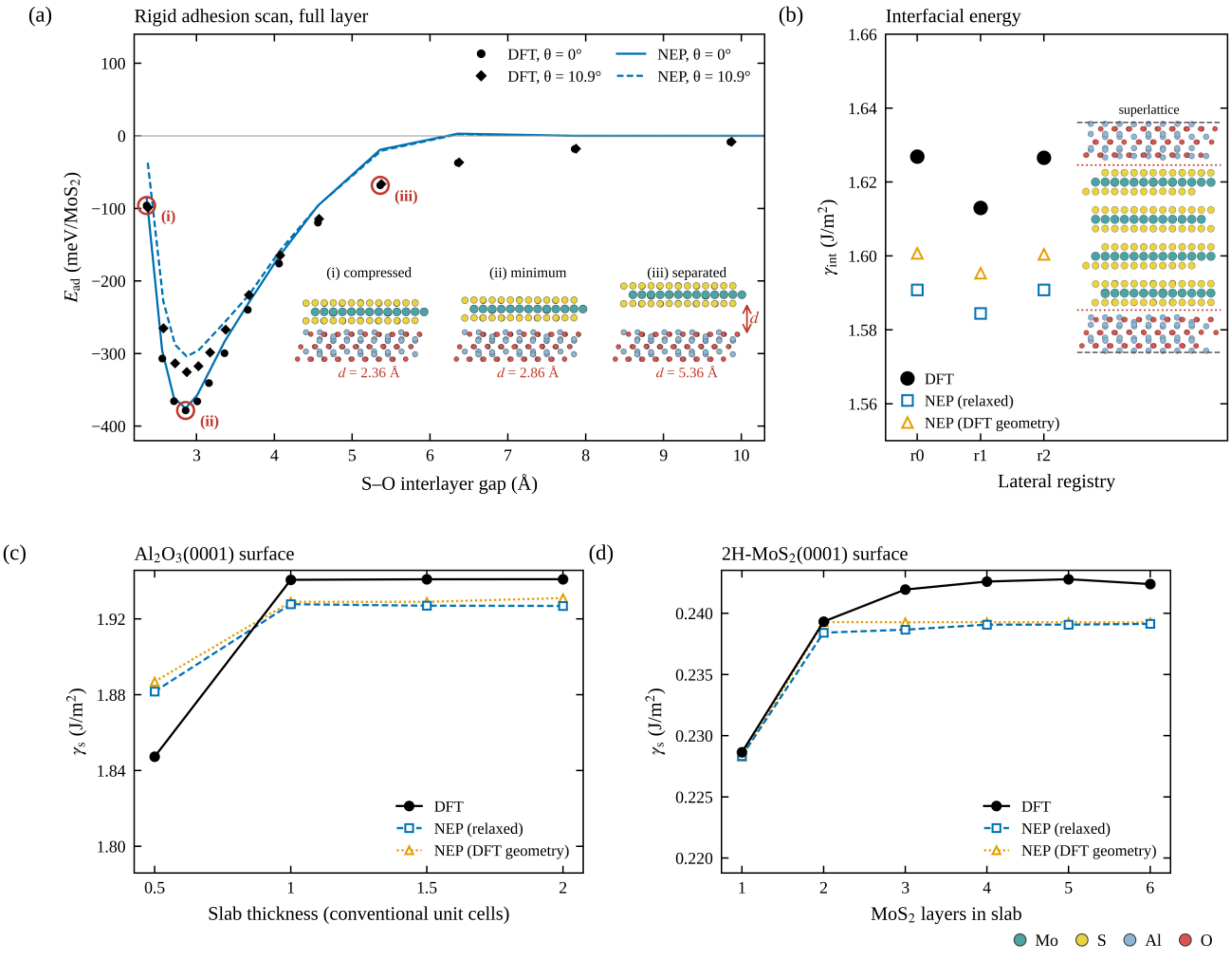

**FIG. 4**. **Surface, adhesion, and interfacial energetics of $MoS_2$ on $Al_2O_3(0001)$.** (a) Rigid adhesion scan of a full $MoS_2$ monolayer on the Al-terminated $Al_2O_3(0001)$ slab: adhesion energy versus the S–O interlayer gap $d$, for twist angles θ = 0° (DFT circles, NEP solid line) and 10.9° (DFT diamonds, NEP dashed line). Insets: side views of the 0° cell at the circled points (i) d = 2.36 Å (compressed), (ii) 2.86 Å (minimum) and (iii) 5.36 Å (separated). (b) Interfacial energy of a periodic superlattice of four 2H-$MoS_2$ layers and the $Al_2O_3(0001)$ slab (228 atoms), for three lateral registries r0–r2. Inset: side view of the superlattice. (c, d) Relaxed surface energies of (c) $Al_2O_3(0001)$ and (d) 2H-$MoS_2(0001)$ as a function of slab thickness.

### 3.3.2 Rotation Energetics

MoS₂ grown on c-plane sapphire is aligned, i.e., epitaxial. Depending on growth conditions, its domains adopt either the 0°/60° orientation family[12,13,67,68] or the 30°/90° family[8–10,69,70]. Which family forms has been linked to the S/Mo precursor ratio and the resulting interfacial structure[67,71]. DFT calculations attribute the alignment to the orientation dependence of the $MoS_2$–sapphire interaction. This dependence differs between continuous layers and finite flakes and is sensitive to the surface termination[11,69,72,73]. A potential for this interface must therefore reasonably reproduce how the interfacial energy changes with the relative in-plane orientation of the two lattices. We define the twist angle θ such that θ = 0° denotes $MoS_2[11\bar{2}0] \parallel Al_2O_3[11\bar{2}0]$.

Full monolayers were computed in commensurate cells at eleven twist angles between 0° and 120°, with 111–1002 atoms and biaxial strains of −1.66 % to +0.94 % in the layer. The adhesion energy $E_{\mathrm{ad}}$ (Sec. 3.3.1) is referenced to the free layer at the same strain, so the elastic energy of the strained layer cancels. For each cell, 16 lateral registries were screened (4 for the 1002-atom 30° cell) and the two lowest relaxed (only the lowest for 30°). In DFT, θ = 0° is the most stable orientation of the full monolayer. The antiparallel orientation, θ = 60°, lies 2.4 meV/$MoS_2$ higher, 79.1° and 100.9° lie 3.6 and 4.2 meV/$MoS_2$ higher, and all other angles lie 11–16 meV/$MoS_2$ above the minimum (Fig. 5a). Relaxations of the same cell started from different registries differ by up to 1.3 meV/$MoS_2$ (6.5 meV/$MoS_2$ for −10.9°), so differences of a few meV/$MoS_2$ are at the resolution of these calculations. After its own relaxation, the NEP keeps θ = 0° as the most stable orientation and reproduces the four lowest orientations (0°, 60°, 79.1° and 100.9°) in the DFT order. For eight of the ten non-zero angles, it agrees with DFT within 5 meV/$MoS_2$. At +10.9° and 30°, it overestimates the energy penalty about twofold and threefold (30.4 versus 15.8 and 37.1 versus 13.9 meV/$MoS_2$). Over all ten angles, the mean absolute error is 5.9 meV/$MoS_2$.

Furthermore, a triangular $Mo_{21}S_{66}$ flake with S-dimer-terminated zigzag Mo edges was placed on a 6×6 $Al_2O_3$(0001) slab (1167 atoms) and rotated in steps of 30°. Its rotation energy is quantified as ΔE(φ) = E(φ) − E(φ = 0), where the rotation angle φ follows the same convention as θ, so that at φ = 0° the flake lattice has the orientation of the full layer at θ = 0°. As shown in Fig. 5b, the rotation energetics of this finite flake differ from those of a full layer. In DFT, φ = 0°, the most stable orientation of the full layer, is the least stable of the four flake orientations: rotating the flake to φ = 30°, 90° and 60° lowers its energy by 0.42, 0.27 and 0.07 eV, respectively. These differences amount to at most 5 meV per flake atom. After relaxation, the NEP also favors φ = 30° and 90° and reproduces the energy gain at 30° (−0.40 versus −0.42 eV), but it overestimates the gain at 90° (−0.48 versus −0.27 eV), so that 90° rather than 30° becomes the lowest orientation, and it gives the wrong sign at φ = 60° (+0.16 versus −0.07 eV). Thus, although the NEP does not reproduce the energy differences quantitatively, it favors the same 30°/90° family as DFT for this flake on sapphire.

Notes S5.4 and S5.5 and Figs. S7–S9 of the Supplemental Material provide the structures, the computational details, and a comparison with previous DFT studies of the rotation energetics of both the commensurate layers and the finite flakes.

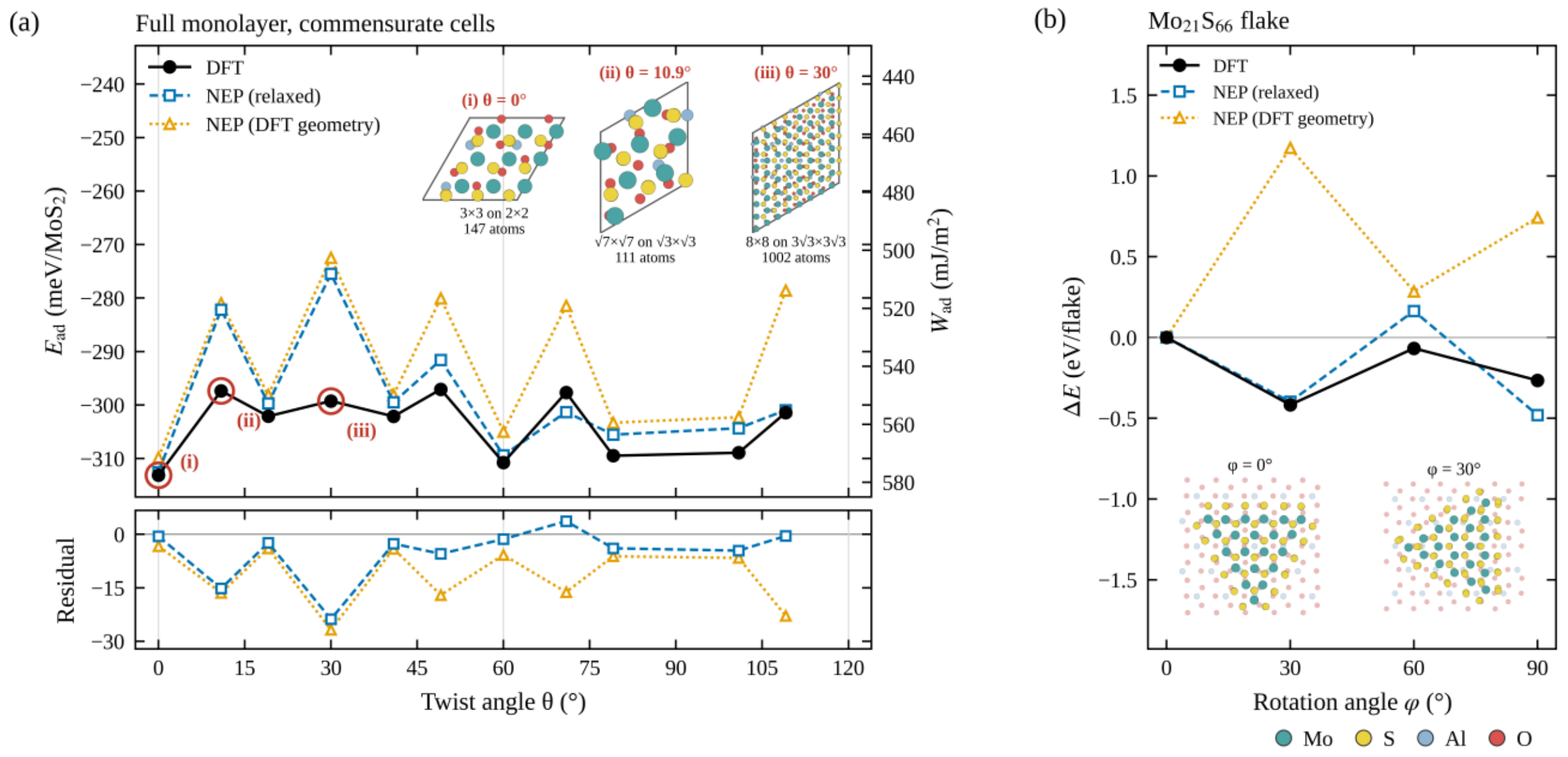

**FIG. 5. Rotational energetics of $MoS_2$ on $Al_2O_3(0001)$.** (a) Adhesion energy $E_{ad} = [E_{het} - E_{slab} - E_{layer}]/N_{MoS_2}$ of a full $MoS_2$ monolayer versus the twist angle $\theta$ between the $MoS_2$ and sapphire lattices, computed in commensurate cells at eleven angles between 0° and 120° (109.1° ≡ −10.9°); for each cell the lowest-energy relaxed registry is used. DFT is shown as black circles, and the NEP after its own relaxation (blue squares) and as a single point on the DFT-relaxed structures (orange triangles). The right axis gives the corresponding work of adhesion, $W_{ad} = -E_{ad}/A$ (mJ/m$^2$), with $A = 8.69$ Å$^2$ the area per formula unit of the unstrained DFT monolayer. The strip below shows the residual $E_{ad}^{DFT} - E_{ad}^{NEP}$ (meV/$MoS_2$); negative values mean that the NEP underbinds. Insets: top views of the $MoS_2$ layer and the topmost substrate layer of the cells circled in the plot (supercell outlined, each drawn at its own scale): (i) $\theta = 0°$, 3×3 $MoS_2$ on 2×2 $Al_2O_3$ (147 atoms, +0.94 % strain on $MoS_2$); (ii) $\theta = 10.9°$, √7×√7 on √3×√3 (111 atoms, −0.88 %); (iii) $\theta = 30°$, 8×8 on 3√3×3√3 (1002 atoms, −1.66 %). (b) Rotation energy $\Delta E(\varphi) = E(\varphi) - E(\varphi = 0)$ of a triangular $Mo_{21}S_{66}$ flake with S-dimer-terminated zigzag Mo edges on the 6×6 $Al_2O_3(0001)$ slab (1167 atoms), versus the angle $\varphi$, defined as in (a). Insets: top views of the relaxed flake at $\varphi = 0°$ and 30° (substrate atoms faded).

The NEP MLIP also reproduces the $MoS_2$–$Al_2O_3$ gap within 0.035 Å for the full monolayer at all eleven twist angles and for both interfaces of the superlattices, and places the interior of the supported $Mo_{21}S_{66}$ flake up to 0.1 Å closer to the substrate than DFT (Note S5.6 and Fig. S10). The data shown in Figs. 4 and 5 are listed in Table S4 (Note S5.7) of the Supplemental Material.

### 3.4 Edges, point defects, grain boundaries and stacking faults

#### 3.4.1 Edge Energies

The equilibrium shapes of $MoS_2$ flakes are determined by the energies of their exposed edges, which govern anisotropic crystal growth and morphological evolution. $MoS_2$ has two zigzag edges, terminated by Mo or by S atoms, which can carry different amounts of sulfur, and the armchair edge, which is stoichiometric. We label the zigzag edges by their S coverage, from the bare Mo edge (0 %) to one $S_2$ dimer per edge Mo atom (100 %). Edges that add or remove sulfur depend on the sulfur chemical potential $\mu_S$, which we vary between the Mo-rich and the S-rich limit (Sec. 2.4). Because a zigzag ribbon always has one Mo edge and one S edge, the individual zigzag edge energies were obtained from triangular flakes, whose three edges are identical. For triangles with $n$ = 4–8 Mo atoms along each edge we assume,

$$E_n - N_{\mathrm{Mo}}\,\mu_{\mathrm{MoS_2}} - \Delta N_{\mathrm{S}}\,\mu_{\mathrm{S}} = 3\gamma n a + 3E_{\mathrm{c}},$$

where $E_n$ is the total energy of the flake, $N_{\mathrm{Mo}}$ its number of Mo atoms, $\mu_{\mathrm{MoS_2}}$ the energy per formula unit of the monolayer, $\Delta N_{\mathrm{S}} = N_{\mathrm{S}} - 2N_{\mathrm{Mo}}$ the excess of S atoms, $a$ the lattice constant and $E_{\mathrm{c}}$ the energy of one corner. The edge energy, $\gamma$, is thus obtained from the slope of a linear fit over $n$. The armchair edge was computed from stoichiometric ribbons with two identical edges, $\gamma_{\mathrm{AC}} = (E - N_{\mathrm{Mo}}\,\mu_{\mathrm{MoS_2}})/2L$, where $L$ is the length of the cell along the edge. The same calculations were repeated for flakes and ribbons on the $Al_2O_3(0001)$ slab.

As shown in Fig. 6, the NEP reproduces the edge energies of the freestanding monolayer and of the monolayer on sapphire. Over all 18 values, nine edges at the two limits of $\mu_{\mathrm{S}}$, it deviates from DFT by 9.4 % on average and by at most 25 %. The largest relative error occurs for the smallest edge energy, that of the Mo edge with 100 % S coverage on sapphire under S-rich conditions (0.149 versus 0.119 eV/Å). The NEP also reproduces which edge is most stable. In both methods, the Mo edge with 50 % S coverage has the lowest energy under Mo-rich conditions and the Mo edge with 100 % S coverage under S-rich conditions. On sapphire, the Mo edge with 100 % S coverage is again the most stable edge under S-rich conditions. The atomic structures of the relaxed edges are shown in Fig. 6c,d. We refer the reader to Note S6, Figs. S11–S16 and Tables S5–S7 of the Supplemental Material for further details regarding the modeling and computation of edge energies.

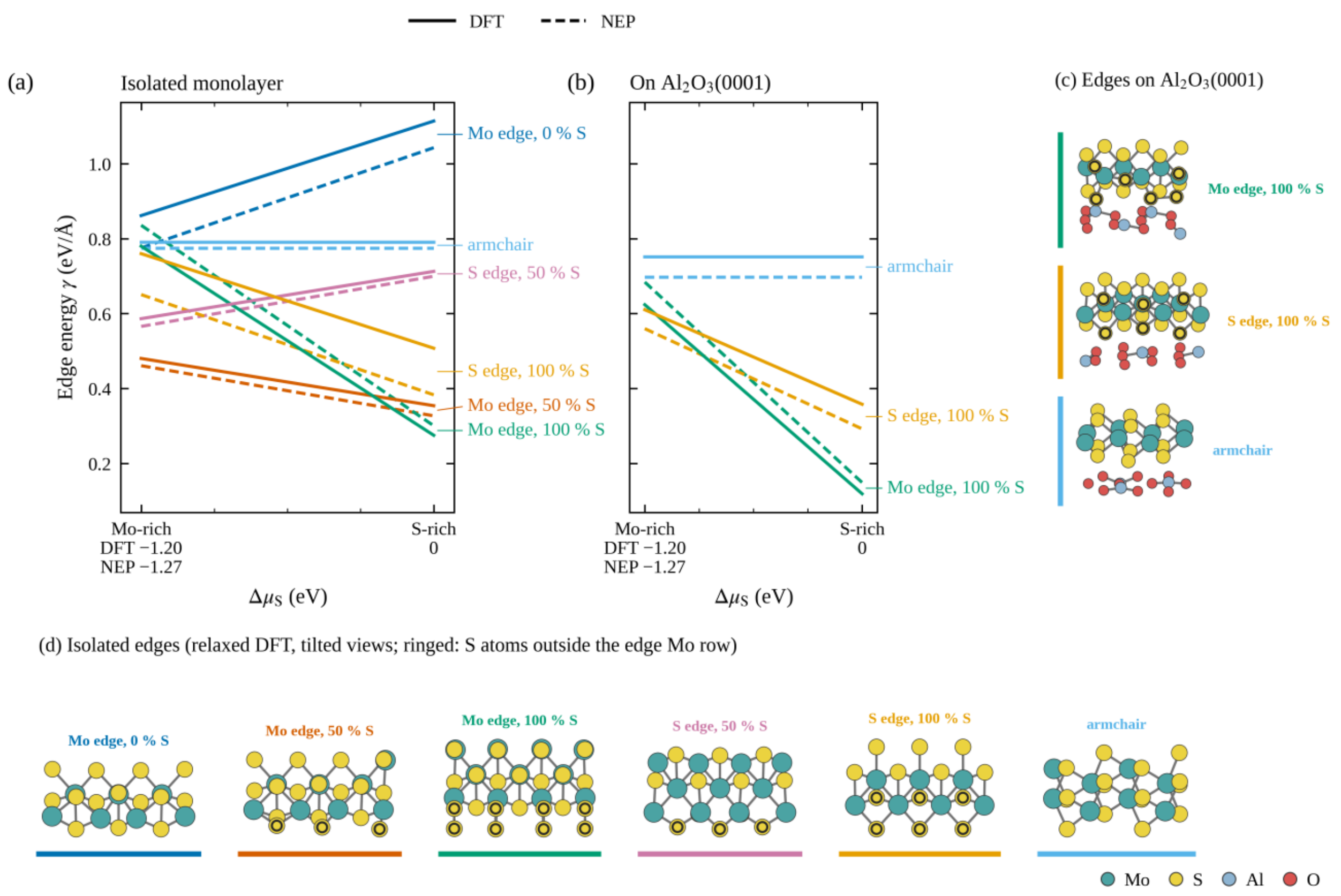

**FIG. 6. Edge energies of 1H-$MoS_2$ versus the sulfur chemical potential, $\Delta\mu_S = \mu_S - \mu_S^{\alpha\text{-}S}$, from DFT (solid lines) and the NEP (dashed lines):** (a) for the freestanding monolayer and (b) on $Al_2O_3$(0001). Each method is evaluated between its own limits, from Mo-rich ($\Delta\mu_S = -1.20$ eV in DFT and $-1.27$ eV in the NEP) to S-rich ($\Delta\mu_S$ = 0). Because the computed edge energy depends on $\mu_S$ only through its excess or deficit of S atoms, it is exactly linear in $\Delta\mu_S$, and each line connects the values at the two limits. Zigzag edges are labelled by their S coverage. Individual zigzag edge energies are obtained from triangular flakes of increasing size, and armchair edge energies from ribbons (Note S6). (c, d) Relaxed DFT structures of the edges, in tilted views and color-matched to the lines: (c) on $Al_2O_3$(0001), with the topmost substrate layer, and (d) in the freestanding monolayer. Ringed atoms are S atoms outside the edge Mo row.

### 3.4.2 Point Defects

Point defects, such as vacancies and antisites, are commonly observed in $MoS_2$ films and have a strong influence on their electronic and chemical properties[74–76]. Thirteen point defects, comprising vacancies, vacancy complexes and antisites, were placed in a 6×6 1H-$MoS_2$ monolayer on a 4×4 $Al_2O_3$(0001) slab (Note S7, Fig. S17). Ten of them have been observed experimentally[74,75]. We compute the defect energy as $\Delta E = E(\text{defect}) - E(\text{pristine})$, where E(defect) and E(pristine) are the total energies of the relaxed defective and pristine cells, with all energies on the DFT total-energy scale (NEP energies, which are trained relative to isolated atoms, are shifted back by the isolated-atom energies).

As shown in Fig. 7a-b, the NEP ranks the defects in nearly the same order as DFT: the Spearman rank correlation is 0.995 for both DFT-relaxed and NEP-relaxed defective heterostructures. The mean absolute error in $\Delta E$ is 0.57 eV on the DFT-relaxed structures and 0.60 eV after NEP relaxation. The largest errors are +1.5 eV for a Mo atom on an $S_2$ site at the DFT geometry and −2.6 eV for the Mo vacancy after relaxation. In the latter case, the NEP-relaxed $MoS_2$ layer slides by about 1.6 Å over the substrate, which accounts for much of the difference. Figure 7c–f shows the DFT-relaxed structures of two experimentally observed defects, the S vacancy and a Mo atom on an S site of the top layer. For the S vacancies, the NEP-relaxed structures agree with DFT within 0.11 Å. The defect energies are listed in Table S8 of the Supplemental Material, and the formation energies ($E_f = \Delta E + \Sigma_i n_i \mu_i$, where $n_i$ is the number of atoms of species removed and $\mu_i$ its chemical potential) of both methods are compared in Fig. S18.

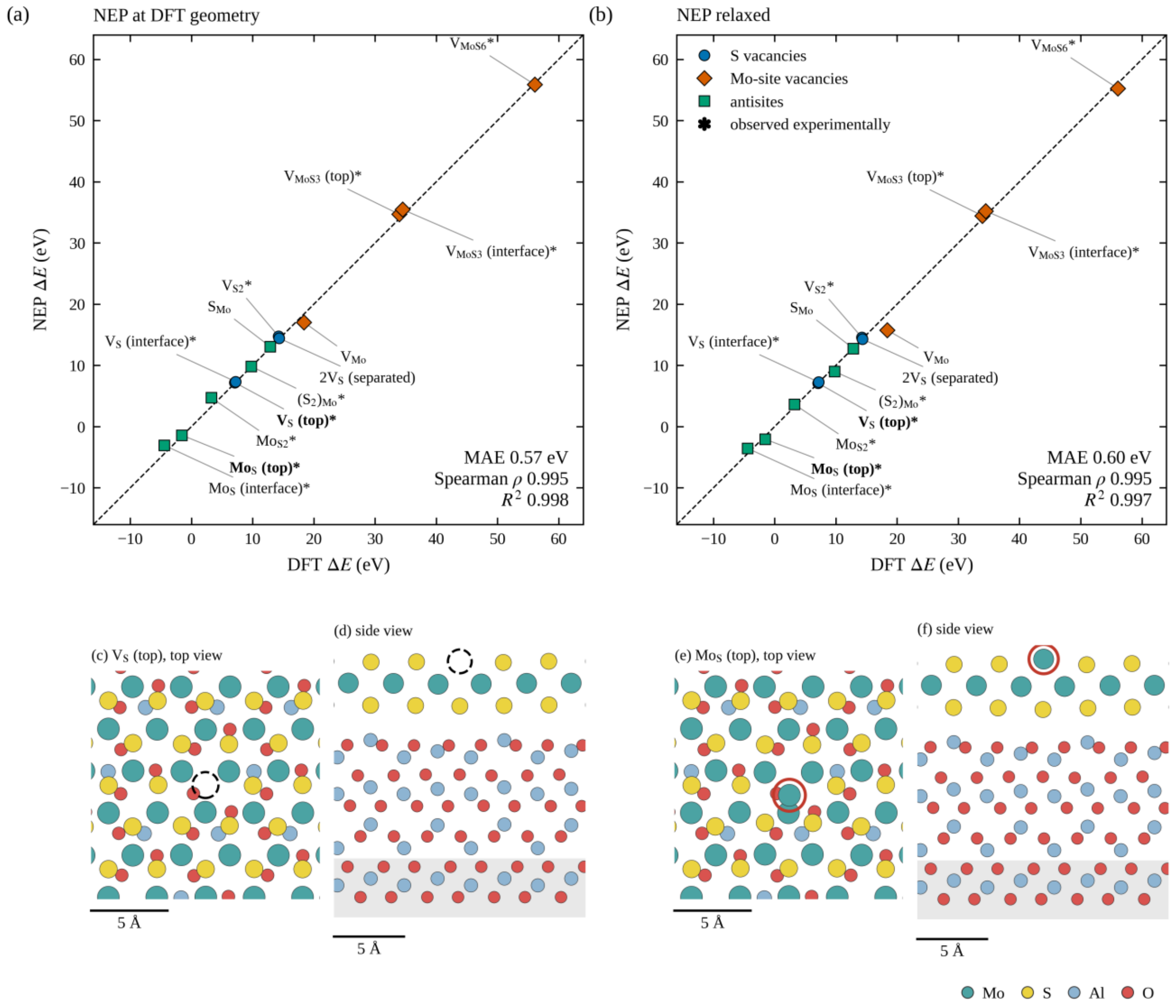

**FIG. 7.** **NEP versus DFT point-defect energies, ΔE = E(defect) − E(pristine)**, (a) with the NEP evaluated on the DFT-relaxed structures and (b) after NEP relaxation from the same starting structures. The defects are placed in a 6×6 1H-$MoS_2$ monolayer on a 4×4 Al-terminated $Al_2O_3$(0001) slab (588 atoms in the pristine cell). ‘Top’ and ‘interface’ denote the S layer facing the vacuum or the sapphire, and an asterisk marks defects observed experimentally. (c–f) DFT-relaxed structures of the two experimentally observed defects shown in bold in (a, b). (c) S vacancy in the top S layer: top view of the Mo layer, the upper S layer and the topmost substrate layer around the defect (dashed circle: removed S atom). (d) Side view of a thin slice through the defect. (e, f) The same for a Mo atom on an S site of the top layer (ring: antisite Mo). Shading in the side views marks the fixed substrate atoms.

3.4.3 Mirror-Twin and Tilt Boundaries

Grain boundaries form where $MoS_2$ domains of different orientation meet, and mirror-twin boundaries (MTBs), between domains rotated by 60°, are common in $MoS_2$ films on sapphire[12,70]. We computed the 4|4P and 4|4E MTBs[77] and symmetric tilt boundaries with misorientations of 21.8° (Σ7), 13.2° (Σ19) and 9.4° (Σ37), freestanding and, for the 4|4P and 4|4E MTBs and a Σ7 boundary between ±10.9° grains, on the $Al_2O_3(0001)$ slab. The MTBs were computed in ribbons that contain one boundary and two S-passivated Mo edges, and the energy per lattice constant is

$$E_{\mathrm{MTB}} = \left[E - N_{\mathrm{Mo}}\,\mu_{\mathrm{MoS_2}} - \Delta N_{\mathrm{S}}\,\mu_{\mathrm{S}} - 2\gamma_{\mathrm{Mo}}L\right]\frac{a}{L},$$

where $E$ is the total energy of the ribbon, $\gamma_{\mathrm{Mo}}$ the energy of its outer Mo edges (Sec. 3.4.1), $L$ the length of the cell along the boundary, and the other symbols are as in Sec. 3.4.1. The tilt boundaries were computed in periodic cells that contain two boundaries,

$$\gamma_{\mathrm{GB}} = \frac{E - N_{\mathrm{Mo}}\,\mu_{\mathrm{MoS_2}} - \Delta N_{\mathrm{S}}\,\mu_{\mathrm{S}}}{2L}.$$

As shown in Fig. 8a, the NEP evaluated on DFT-relaxed structures reproduces the 4|4P MTB within 4–12 %, freestanding and on sapphire, but overestimates the 4|4E MTB by 9–27 %. It keeps the DFT order in every case: the 4|4P boundary is the lower-energy MTB, both freestanding and on sapphire and at both limits of $\mu_S$, and the Spearman rank correlation over all eight MTB energies is 1.0. As shown in Fig. 8b, the NEP reproduces the tilt-boundary energies within 8 %, with a mean absolute error of 0.021 eV/Å and a Spearman rank correlation of 0.98 over all boundaries, terminations and $\mu_S$ limits. The stoichiometric boundaries agree within 3.6 %. The NEP reproduces the decrease of $\gamma_{GB}$ with decreasing misorientation, from Σ7 (21.8°) through Σ19 (13.2°) to Σ37 (9.4°) (0.620, 0.512 and 0.413 versus 0.624, 0.496 and 0.410 eV/Å). The S-deficient cores are slightly less accurate (up to +7.4 %). Even so, the NEP selects the same lowest-energy core as DFT in five of the six cases. The exception is Σ19 at the Mo-rich limit, where DFT separates the two cores by only 0.014 eV/Å. On sapphire, DFT lowers the energy of the Σ7 boundary between ±10.9° grains by 0.10 eV/Å relative to the same boundary freestanding at equal strain. The NEP reproduces this stabilization but overestimates it by about one third (0.13 eV/Å). For the S-rich core at the S-rich limit, it overestimates it by about 50 % (0.14 versus 0.09 eV/Å). The computational details and all numerical values are given in Note S8, Figs. S19–S21, and Table S9 of the Supplemental Material.

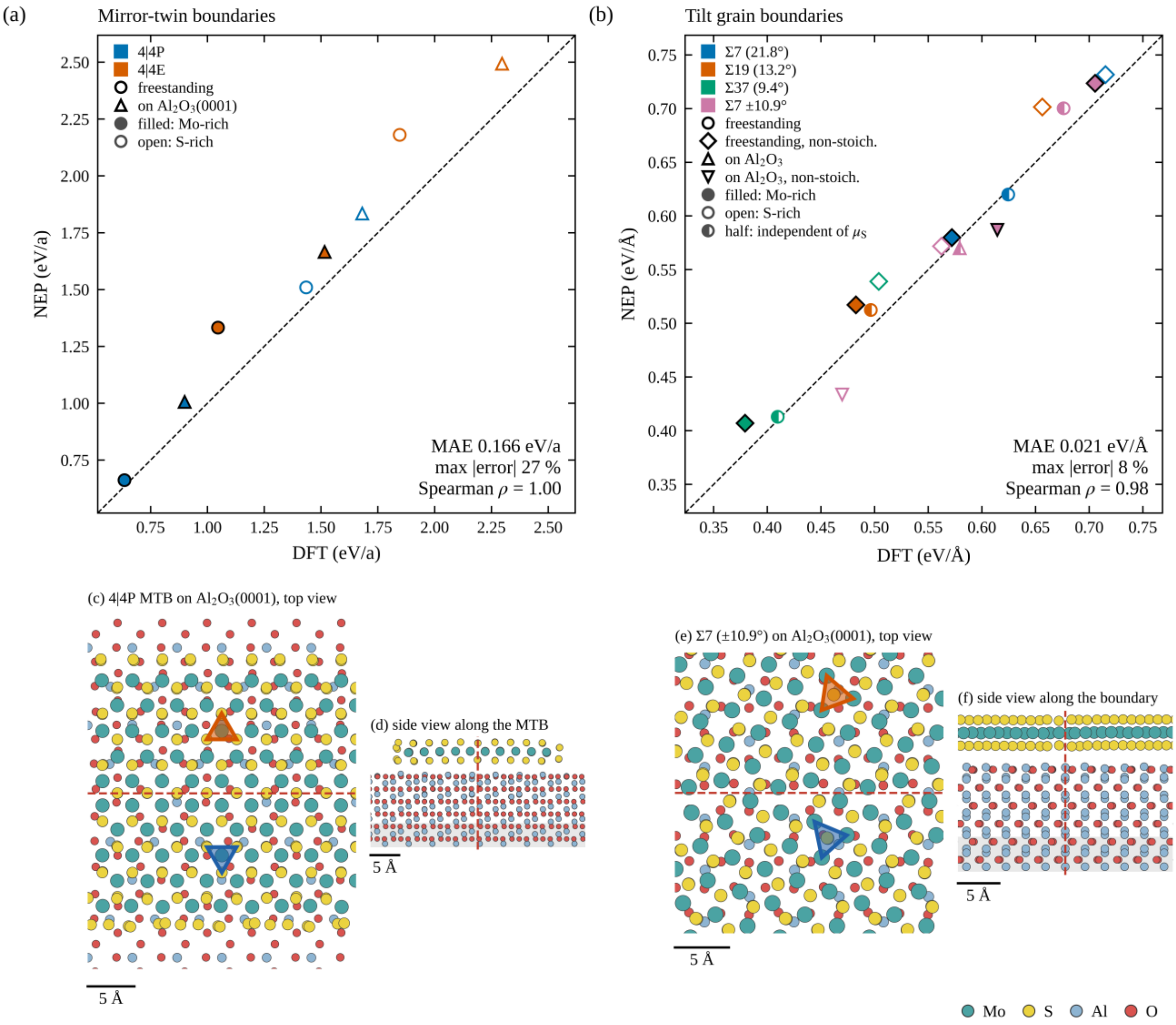

**FIG. 8. Grain-boundary energies of 1H-$MoS_2$, NEP versus DFT, with the NEP evaluated on the DFT-relaxed structures.** (a) Mirror-twin boundaries (4|4P, 4|4E), $E_{MTB}$ per lattice constant a, freestanding (circles) and on $Al_2O_3$(0001) (triangles). (b) Symmetric tilt boundaries, $\gamma_{GB}$: Σ7 (21.8°), Σ19 (13.2°) and Σ37 (9.4°) freestanding, and a Σ7 boundary between ±10.9° grains freestanding and on $Al_2O_3$(0001) at equal strain, for the stoichiometric and the lowest non-stoichiometric translations (S-deficient for Σ7 (21.8°), Σ19 and Σ37; S-rich core for the ±10.9° boundary). Filled symbols denote the Mo-rich limit and open symbols the S-rich limit (α-S); stoichiometric boundaries do not depend on $\mu_S$ and are drawn once, half-filled. (c–f) Relaxed DFT structures of the supported boundaries. (c) 4|4P MTB in a $MoS_2$ ribbon with S-passivated Mo edges on the $Al_2O_3$(0001) slab (816 atoms), top view showing three periods along the boundary, and (d) side view along the boundary. (e) Σ7 boundary between ±10.9° grains on the slab (882 atoms), top view, and (f) side view along the boundary. Dashed red lines mark the boundaries. Shaded triangles mark the orientation of each mirror domain in (c), drawn through the three upper-layer S atoms around one Mo atom, and of each grain in (e), drawn as a unit triangle of Mo lattice sites. The top views show the $MoS_2$ layer and the topmost substrate layer, and shading in the side views marks the fixed substrate atoms.

### 3.4.4 Stacking Faults

The energy needed to slide one $MoS_2$ layer over the next controls the stacking of multilayer films, the formation of stacking faults, and the resistance of the layers to shear. We describe it by the generalized stacking-fault energy

$$\gamma(t) = \frac{E(t) - E(0)}{A},$$

where $E(t)$ is the total energy when the upper part of the cell is displaced by a fraction $t$ of one lattice period along the sliding direction, and $A$ is the area of the faulted interface. Paths along the zigzag and armchair directions, starting from 2H or 3R stacking, were computed for a freestanding bilayer, a bulk crystal with one faulted interface and a bilayer on sapphire. Each path was computed rigidly, with all atoms at the spacing of the unfaulted cell, and with the atoms relaxed only perpendicular to the layers.

Across all paths, the NEP reproduces the highest energy along each path, $\gamma_{\text{max}}$, with a mean deviation of 3.5 % for the rigid and 6.0 % for the relaxed curves, and by at most 14 %. For the bilayer on sapphire (Fig. 9a–c), $\gamma_{\text{max}}$ along the 2H armchair path is 17.84 meV/Å$^2$ in the NEP and 17.85 meV/Å$^2$ in DFT, and along the 2H zigzag path 7.77 versus 8.11 meV/Å$^2$ (rigid). Figure 9d–h shows how the stacking changes along the 2H armchair path. The interface passes from 2H stacking through a stacking with the Mo atoms of the two layers on top of each other, which is a local energy minimum at $t$ = 1/3, to a stacking with the S atoms on top of each other, which is the maximum at $t$ = 2/3. The NEP reproduces this local minimum at the same position in all armchair curves, but places it lower relative to 2H than DFT: on sapphire, it lies 1.50 meV/Å$^2$ above 2H in DFT and 0.11 meV/Å$^2$ below 2H in the NEP. Note S9 describes the sliding convention and all sliding paths used; Figs. S22 and S23 and Table S10 of the Supplemental Material give the curves and the values of $\gamma_{\text{max}}$.

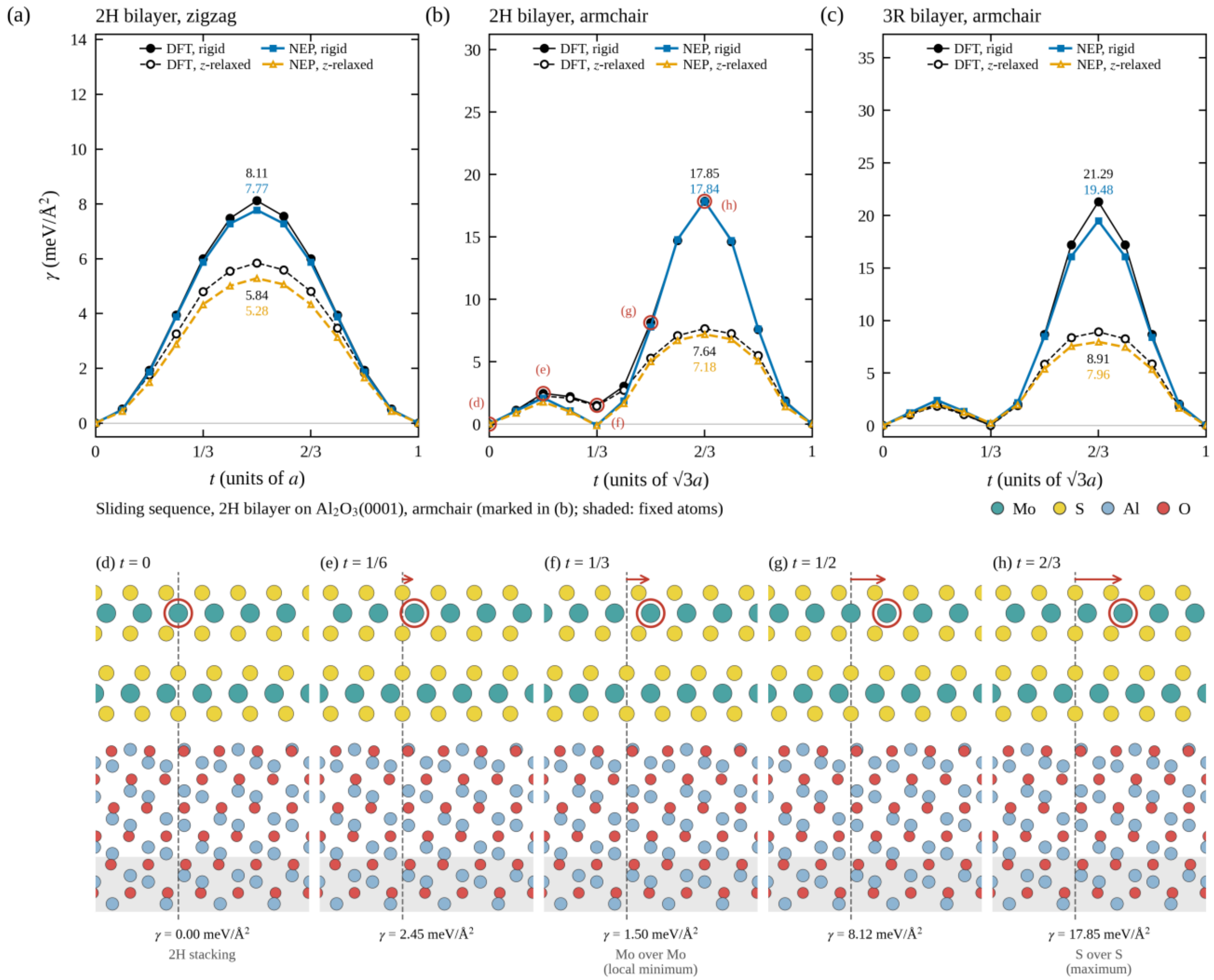

**FIG. 9. Generalized stacking-fault energy of a $MoS_2$ bilayer on $Al_2O_3$(0001)** (0° registry, 174 atoms) for sliding the upper layer (a) along the zigzag direction from 2H stacking (t in units of a), and along the armchair direction (t in units of √3a) from (b) 2H and (c) 3R stacking. (d–h) Sliding sequence of the 2H bilayer along the armchair path at t = 0, 1/6, 1/3, 1/2 and 2/3 (red circles in (b)). The rigid DFT cells are viewed along the zigzag direction, with the upper layer sliding to the right. The red ring follows one Mo atom of the upper layer, the dashed line marks its position at t = 0, and the shaded region marks the fixed substrate atoms. The stacking passes from 2H (d) through a stacking with the Mo atoms of the two layers on top of each other, the local minimum at t = 1/3 (f), to a stacking with the S atoms on top of each other, the maximum at t = 2/3 (h). The γ values below the snapshots are the rigid DFT energies.

## IV. Conclusion and Outlook

We have developed a neuroevolution potential (NEP) for the $MoS_2$/α-$Al_2O_3$(0001) system. It is trained on 63,611 PBE+D3 configurations distilled from about 1.1 million DFT-labeled configurations. These cover bulk phases, surfaces, $MoS_2$/$Al_2O_3$ interfaces, flakes, defects and high-temperature configurations. An energy-difference loss which fits interfacial energies directly was used. We validated the developed model against DFT, from the bulk crystals through the $MoS_2$/sapphire interface to extended defects.

On the held-out test set, its errors are 29 meV/atom in energy and 0.23 eV/Å in force. The NEP model reproduces the lattice constants within 0.3 %, all 23 elastic constants within 17 %, the phonon dispersions with mean deviations of 1.7–2.4 %, and surface energies within 1.3 %. At the heterointerface, the NEP predicts the depth of the monolayer adhesion minimum within 0.8 % and the interfacial energy within 2.2 %. It maintains θ = 0° as the most stable orientation of the full monolayer, with the four lowest orientations in the DFT order. For extended defects, the NEP preserves the energetic ordering: it identifies the most stable edge at both limits of the sulfur chemical potential and ranks thirteen point defects (Spearman ρ = 0.995), the mirror-twin boundaries (ρ = 1.0) and the tilt boundaries (ρ = 0.98) in close agreement with DFT. It reproduces the tilt-boundary energies within 8 % and the maximum energy of each stacking-fault path within 14 %. Nonetheless, for a triangular $Mo_{21}S_{66}$ flake on sapphire the NEP reproduces only the preferred 30°/90° orientation family; it does not resolve the energy differences between orientations, which amount to a few meV per atom. Furthermore, the NEP model lacks the weak dispersion attraction beyond its cutoff and overestimates the energy of the 4|4E mirror-twin boundary.

Two directions follow. First, the potential serves as an extensively validated base model for the $MoS_2/Al_2O_3(0001)$ system; in forthcoming work, the model will be fine-tuned by active learning on configurations sampled during growth, enabling molecular dynamics simulations of $MoS_2$ epitaxy on sapphire in which crystallization of nuclei, growth and coalescence of domains, and the formation of grain boundaries can be followed atom by atom. Second, NEP has been implemented in our GPU-accelerated concurrent atomistic-continuum (CAC) code[47]. CAC enables accurate modeling of mesoscopic substrates using finite elements, while the interface and epilayer are modeled with atoms; this allows the epitaxial growth process, including defect formation, to be simulated with atomic resolution[78,79]. Like MD, the CAC method does not rely on additional assumptions or empirical rules (e.g., defect nucleation criteria or diffusion constants) beyond an interatomic potential; both atomic and finite-element regions are governed by Newton's second law. Combining NEP with CAC opens the route to atomically resolved mechanistic studies of $MoS_2$ heteroepitaxial growth at device-relevant scales.

**Acknowledgements**

This work is based on research supported by the US National Science Foundation under Award No. CMMI-2349160. The work of S. R. Phillpot is supported by NSF Award No. DMR-2121895, and the work of Y. Chen is also partially supported by the same award. Computational resources were provided by the Advanced Cyberinfrastructure Coordination Ecosystem: Services & Support (ACCESS) under allocation DMR190008, with further support under allocation MAT260001 for E. Bilgili.

## AI Disclosure

Claude Opus 5.5 (Anthropic) was used to assist with Python scripting for data processing, property evaluation, and plotting, as well as manuscript editing. All outputs were reviewed and verified by the authors, who take full responsibility for the content of this work.

## Data Availability

The NEP model as well as its training and test data will be publicly available on Zenodo (https://doi.org/10.5281/zenodo.24233698) upon journal publication. Two equivalent versions of the model are provided: *nep.txt* gives energies relative to non-spin-polarized isolated atoms, as reported here, and *nep_as_trained.txt* gives energies relative to spin-polarized isolated atoms, as used in training. The two differ only by a constant energy per element and produce identical dynamics.

## Supplemental Material

# Machine-Learned Interatomic Potential for the $MoS_2$/Sapphire System

Emir Bilgili[1*], Nicholas Taormina[1], Simon R. Phillpot[2], Youping Chen[1]

[1] Department of Mechanical and Aerospace Engineering, University of Florida, Gainesville, FL, 32611
[2] Department of Materials Science and Engineering, University of Florida, Gainesville, FL, 32611

### Note S1. DFT Calculations and Energy Reference

The PAW potentials were the VASP PAW_PBE datasets Al (04Jan2001), O (08Apr2002), Mo_sv (02Feb2006) and S (06Sep2000), and dispersion was included with the D3 correction in its zero-damping form (IVDW = 11). Reported DFT energies are the free energies of the final ionic step, including the D3 contribution. Ions were relaxed with the conjugate-gradient algorithm, except for a few edge and grain-boundary cells that were finished with the quasi-Newton algorithm.

The validation calculations fall into two convergence classes, which differ only in the settings listed in Table S1. The tight class is used for the bulk and monolayer reference cells, whose lattice constants, equations of state, elastic constants and phonons require well-converged stresses and forces. The data-collection class reproduces the settings of the training data and is used for all surface, interface, rotation, edge, defect, grain-boundary and stacking-fault cells. Within this class, the force criterion is tightened to 0.01 eV/Å for some edge and grain-boundary cells, to 0.005 eV/Å for the z-relaxed stacking-fault cells and to $10^{-3}$ eV/Å for the chemical-potential references.

*k-point meshes*

All meshes were Γ-centered. Bulk and monolayer cells used dense meshes (e.g., 6×6×2 for α-$Al_2O_3$, 15×15×4 for 2H-$MoS_2$ and 18×18×1 for the monolayer), and slab, interface and rotation cells used 2×2×1 to 4×4×1 in the plane (6×6×1 and 15×15×1 for the bare $Al_2O_3$ and $MoS_2$ slabs). Cells with all periodic lengths above about 15 Å, i.e., isolated flakes, point-defect cells and large supercells, used the Γ point only. The k-point spacing along each sampled direction, $2\pi|b_i|/N_i$, is 0.13–0.38 $Å^{-1}$ for 90% of the calculations (mean 0.23 $Å^{-1}$) and at most 0.47 $Å^{-1}$.

*Isolated-atom energy reference*

The NEP was trained on DFT energies minus the energies of spin-polarized isolated atoms, $E_{o,sp}$ (Mo −4.6484, S −1.0804, O −1.9079, Al −0.3144 eV). All energies in this work are instead given relative to non-spin-polarized isolated atoms, $E_{o,nsp}$, computed with the reference DFT settings (main text, Sec. 2.1) in a 15 Å cubic cell (Mo −0.2622, S −0.0930, O −0.0905, Al −0.0582 eV), consistent with the non-spin-polarized reference data. The released model nep.txt is the trained model with the constant $[E_{o,sp}(X) - E_{o,nsp}(X)]$ added to the energy of each atom of element X, implemented as one additional hidden neuron per element with zero input weights; its forces, stresses and dynamics are identical to those of the as-trained model, which is also released (nep_as_trained.txt).

**TABLE S1.** DFT settings used for "tight" and "data-collection" classes.

| | Tight class | Data-collection class |
|---|---|---|
| Properties | lattice constants, equations of state, elastic constants, phonons | surfaces, adhesion, interface, rotation, edges, point defects, grain boundaries, stacking faults |
| Calculations | 188 | 899 |
| Electronic convergence (eV) | $10^{-8}$ | $10^{-6}$ |
| Real-space projection (LREAL) | .FALSE. | Auto |
| Ionic force criterion (eV/Å) | $10^{-3}$ | ≤0.02 |

## Note S2. NEP Input File

Table S2 lists the complete input file of the final model described in Sec. 2.3 of the main text.

**TABLE S2.** Complete NEP input file (nep.in) of the final model. The cutoff line lists radial and angular cutoffs (Å) per element in the order of the type line; zbl gives the outer ZBL cutoff (Å); save_potential and seed only control output and the random-number seed. The model was continued from a model trained for $6\times10^5$ generations without the energy-difference term (Sec. 2.3).

| Keyword | Value |
|---|---|
| type | 4 Al O Mo S |
| cutoff | 5.5 3.5 5.5 3.5 9.5 5.0 9.5 5.0 |
| n_max | 8 7 |
| basis_size | 12 7 |
| neuron | 46 |
| zbl | 1.5 |
| lambda_e | 1.0 |
| lambda_f | 0.9 |
| lambda_v | 1.0 |
| lambda_shear | 5 |
| force_delta | 60 |
| batch | 1000 |
| population | 50 |
| generation | 300000 |
| save_potential | 10000 1 1 |
| seed | 1 |
| lambda_d | 2.5 |

## Note S3. High-Temperature Configurations

To test the potential on strongly distorted interface configurations, we used 300 frames from 20 DFT ab initio MD heating runs of fully periodic $MoS_2/Al_2O_3(0001)$ cells between 273 and 1729 K (Table S3, Fig. S2). The runs comprise one or two $MoS_2$ layers on the $Al_2O_3(0001)$ slab at several twist angles, and periodic superlattices of four 2H- or 3R-$MoS_2$ layers and the slab that differ in twist angle, lateral registry and in how the lattice mismatch is shared between the two materials. Energies in Fig. S1 are given relative to non-spin-polarized isolated atoms (Note S1). Over all frames, the NEP errors are 3.5 meV/atom in energy, 0.136 eV/Å in force and 16.0 meV/atom in virial.

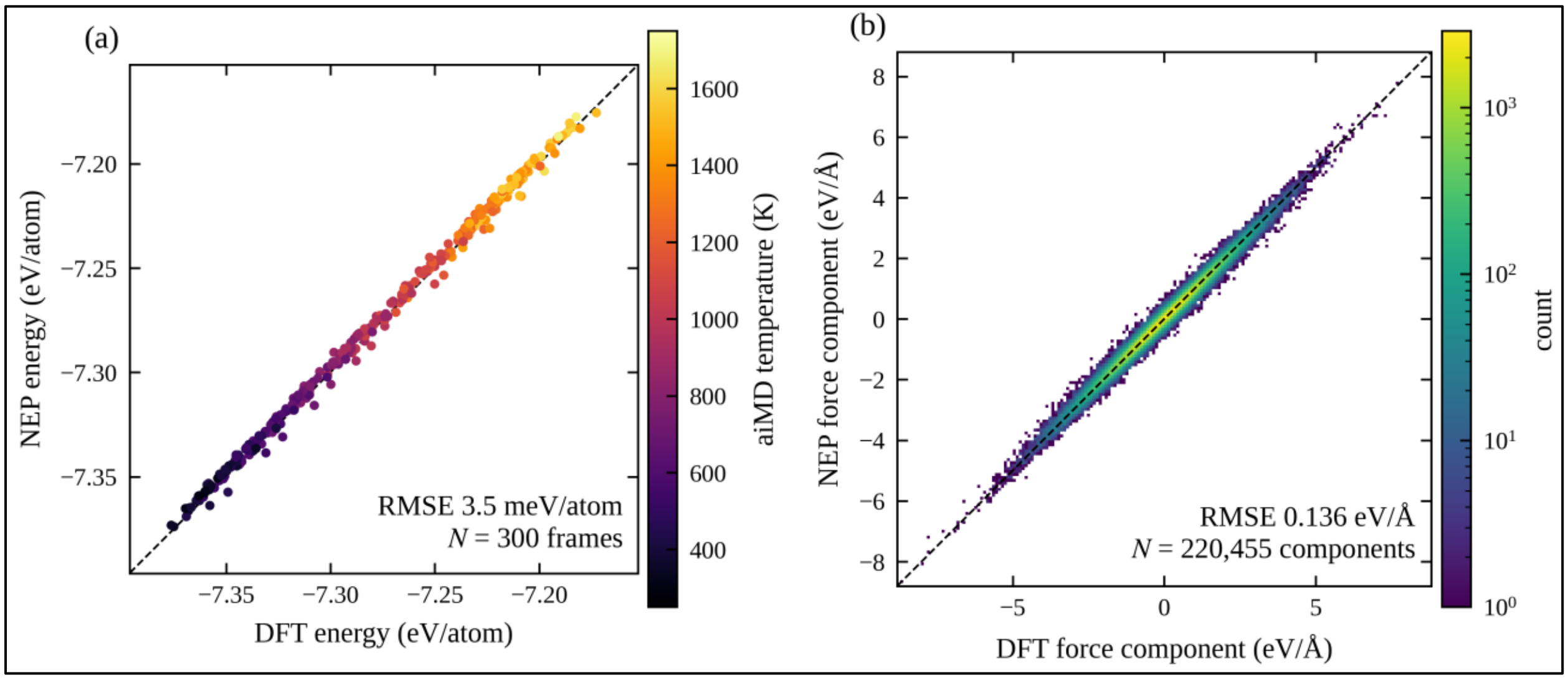

***FIG. S1.*** *NEP versus DFT on high-temperature configurations. (a) Energy parity colored by the aiMD temperature; (b) force-component parity. RMSE: energy 3.5 meV/atom, forces 0.136 eV/Å.*

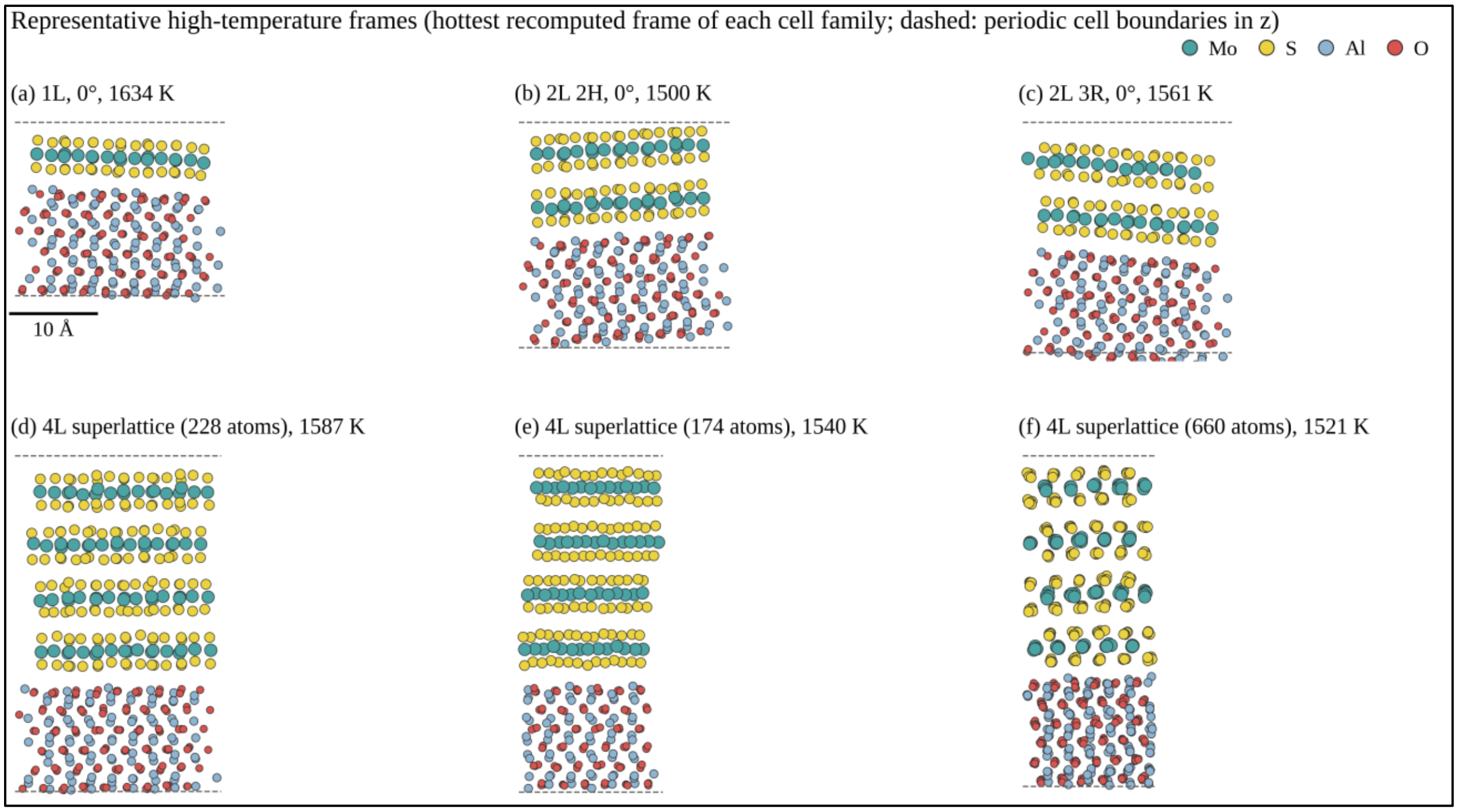

***FIG. S2.*** *Representative frames of the high-temperature set (the hottest recomputed frame of each cell family, side views). All cells are periodic in z (dashed: cell boundaries; about 3 Å to the next slab image, no vacuum): (a) one, (b, c) two (2H, 3R) $MoS_2$ layers on an Al-terminated $Al_2O_3(0001)$ slab, and (d–f) four-layer superlattices of increasing cell size.*

**TABLE S3.** High-temperature validation runs. Each run is a DFT ab initio MD heating ramp of a fully periodic $MoS_2/Al_2O_3(0001)$ cell; about 15 frames per run were taken along the ramp and recomputed as single points with the data-collection settings. '$MoS_2$ on slab': one or two $MoS_2$ layers on the $Al_2O_3(0001)$ slab; '4L superlattice': four $MoS_2$ layers and the slab, periodic along z. $MoS_2$ layers: number and stacking; θ: twist angle of $MoS_2$ relative to the substrate; registry: lateral translation of the $MoS_2$ block as in Note S5.3 (r0–r2); mismatch: the material that takes the coincidence-cell lattice mismatch (0° cells: +0.94 % on $MoS_2$, −0.93 % on $Al_2O_3$, or ±0.47 % shared; 30° cell: +2.45 % on $MoS_2$ and −2.33 % on $Al_2O_3$).

| Family | $MoS_2$ layers | θ | Registry | Mismatch | Atoms | T (K) | Frames |
|---|---|---|---|---|---|---|---|
| $MoS_2$ on slab | 1L | 0° | – | – | 147 | 372–1634 | 15 |
| | 1L | 60° | – | – | 147 | 345–1711 | 15 |
| | 1L | +10.9° | – | – | 111 | 340–1729 | 15 |
| | 1L | −10.9° | – | – | 111 | 352–1631 | 15 |
| | 1L | +19.1° | – | – | 258 | 320–1066 | 15 |
| | 1L | 30° | – | – | 435 | 399–1626 | 15 |
| | 2L 2H | 0° | – | – | 174 | 341–1500 | 15 |
| | 2L 2H | 30° | – | – | 510 | 369–1583 | 15 |
| | 2L 3R | 0° | – | – | 174 | 341–1561 | 15 |
| 4L superlattice | 4L 2H | 0° | r0 | $MoS_2$ | 228 | 368–1587 | 15 |
| | 4L 2H | 0° | r0 | shared | 228 | 363–1613 | 15 |
| | 4L 2H | 0° | r0 | $Al_2O_3$ | 228 | 306–1552 | 15 |
| | 4L 2H | 0° | r1 | shared | 228 | 357–1569 | 15 |
| | 4L 2H | 0° | r2 | shared | 228 | 354–1686 | 15 |
| | 4L 2H | 60° | r0 | shared | 228 | 273–1541 | 15 |
| | 4L 3R | 0° | r0 | shared | 228 | 313–1697 | 15 |
| | 4L 3R | 60° | r0 | shared | 228 | 401–1543 | 15 |
| | 4L 2H | +10.9° | r0 | shared | 174 | 309–1540 | 15 |
| | 4L 2H | −10.9° | r0 | shared | 174 | 293–1565 | 15 |
| | 4L 2H | 30° | r0 | shared | 660 | 354–1521 | 15 |

## Note S4. Bulk Properties

The bulk properties of **Sec. 3.2** were computed on α-$Al_2O_3$ in its 30-atom hexagonal conventional unit cell, 2H-$MoS_2$, 3R-$MoS_2$ and the 1H-$MoS_2$ monolayer (Fig. S3). For the three bulk crystals, the energy was computed at 13 volumes between $V/V_0 = 0.94$ and 1.06, with the cell shape and the ions relaxed at each fixed volume, and fitted to the third-order Birch–Murnaghan equation of state[1],

$$E(V) = E_0 + \frac{9V_0B_0}{16}\left\{\left[(V_0/V)^{2/3}-1\right]^3 B_0' + \left[(V_0/V)^{2/3}-1\right]^2\left[6-4(V_0/V)^{2/3}\right]\right\},$$

where $E_0$ and $V_0$ are the equilibrium energy and volume, $B_0$ is the bulk modulus and $B_0'$ its pressure derivative. We report $V_0$ and $B_0$ only, because the energy–volume curves of the layered crystals are too flat to determine $B_0'$ reliably. To probe the van der Waals interlayer binding directly, the energy of 2H- and 3R-$MoS_2$ was also computed as a function of the interlayer lattice parameter over $c/c_0 = 0.92$–1.15 (11 points) at fixed a with the ions relaxed, and that of the monolayer as a function of its in-plane lattice parameter over $a/a_0 = 0.96$–1.04 (11 points).

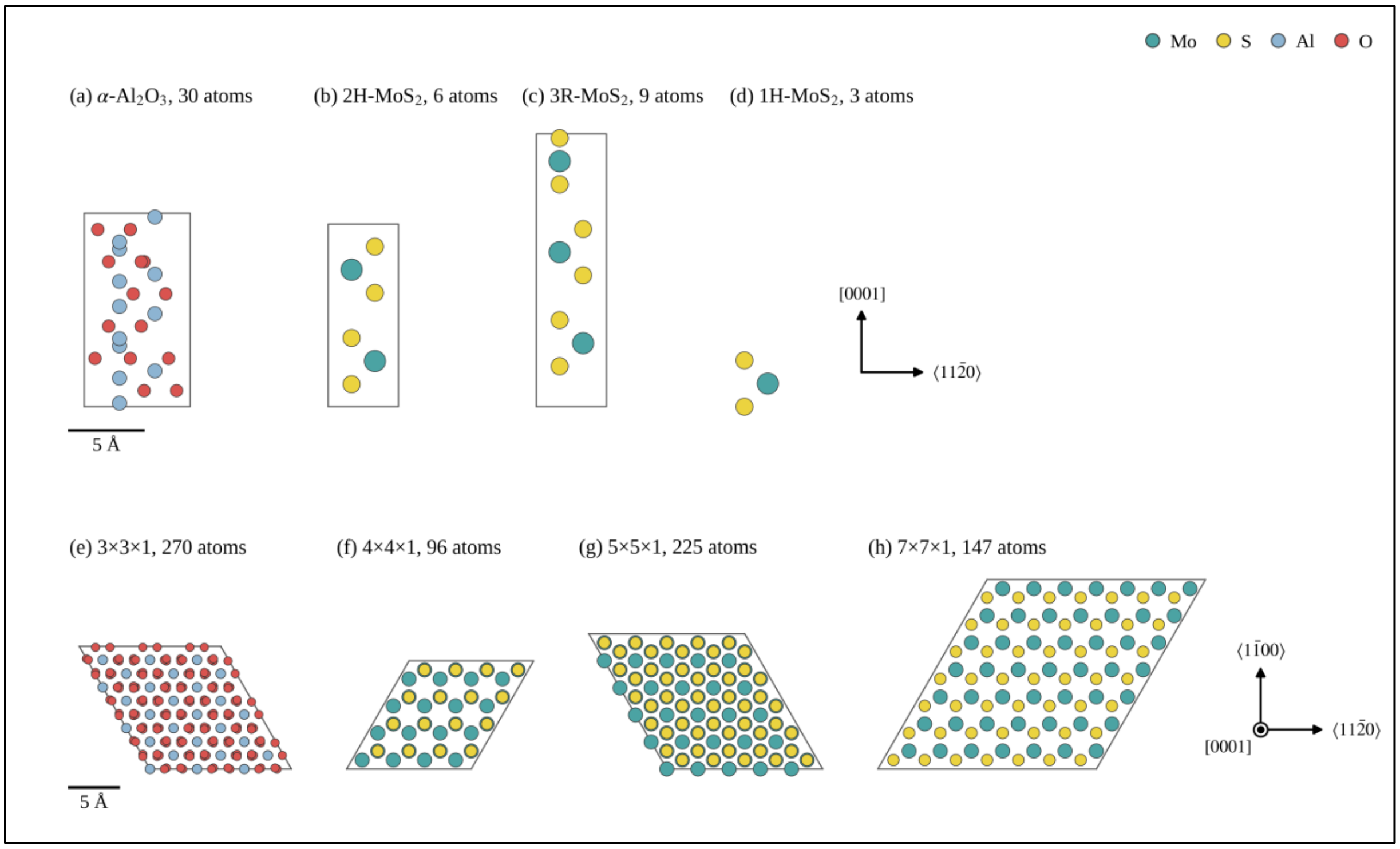

***FIG. S3.*** *DFT reference cells of the bulk properties (Sec. 3.2). (a–d) Relaxed reference cells, side views with the cell outlined: α-$Al_2O_3$ in the hexagonal setting ($Al_{12}O_{18}$, 30 atoms; a = 4.786 Å, c = 13.059 Å), 2H-$MoS_2$ ($Mo_2S_4$, 6 atoms; a = 3.163 Å, c = 12.334 Å), 3R-$MoS_2$ ($Mo_3S_6$, 9 atoms; a = 3.165 Å, c = 18.423 Å) and the 1H-$MoS_2$ monolayer (3 atoms; a = 3.167 Å; about 20 Å of vacuum along z, not drawn). The equation-of-state and elastic-constant calculations strain exactly these cells. (e–h) Phonon supercells: 3×3×1 α-$Al_2O_3$ (270 atoms), 4×4×1 2H (96), 5×5×1 3R (225) and 7×7×1 1H (147); top views with the supercell outlined. Arrows give the common hexagonal frame ($a_1$ along ⟨11-20⟩) of all panels.*

## Note S5. Interfacial Properties

### Note S5.1. Slab Geometries

The surface energies were computed from series of slabs (Fig. S4). Each slab was computed both as cut from the bulk and after relaxation of the ions at fixed cell.

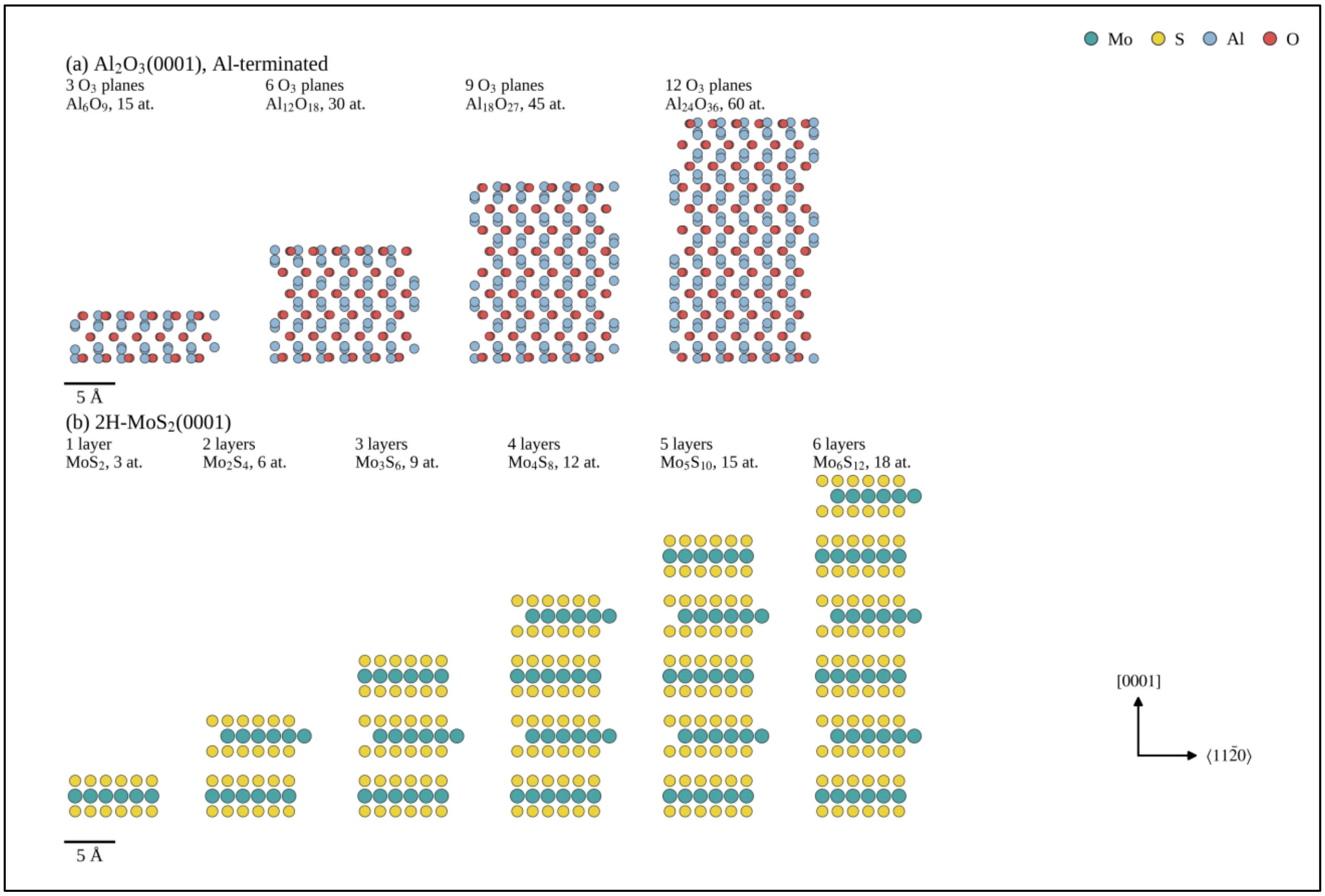

***FIG. S4.*** *Atomic structures of $Al_2O_3$(0001) and 2H-$MoS_2$(0001) slabs with varying thicknesses used to compute surface energy. Relaxed geometries, side views along ⟨1-100⟩ of an orthogonal 3a × √3a window; the vacuum (about 18 Å) is not drawn. A dipole correction is applied along z.*

## Note S5.2. Adhesion Energies

The adhesion energy of the 1H-$MoS_2$ monolayer is computed as

$$E_{\mathrm{ad}} = \frac{E_{\mathrm{het}} - E_{\mathrm{slab}} - E_{\mathrm{layer}}}{N_{\mathrm{MoS_2}}},$$

where $E_{het}$, $E_{slab}$ and $E_{layer}$ are the total energies of the heterostructure, the bare slab and the free $MoS_2$ layer, each relaxed in the same supercell, and $N_{MoS_2}$ is the number of $MoS_2$ formula units. Because the free layer carries the same biaxial strain as the supported one, the strain energy cancels. The 0° cell consists of 3×3 $MoS_2$ on 2×2 $Al_2O_3$ (147 atoms, +0.94 % strain on $MoS_2$), and the 10.9° cell of √7×√7 $MoS_2$ on √3×√3 $Al_2O_3$ (111 atoms, −0.88 %) (Fig. S5). For each cell, 16 lateral registries of the layer were screened with single-point calculations, the two lowest were relaxed, and the lower relaxed energy is used. The rigid adhesion scans of Fig. 4(a) use the same three terms without relaxation: the layer of the relaxed heterostructure is displaced rigidly along z by −0.5 to +10 Å from its relaxed height, and the gap d is measured between the lowest S plane and the topmost O plane.

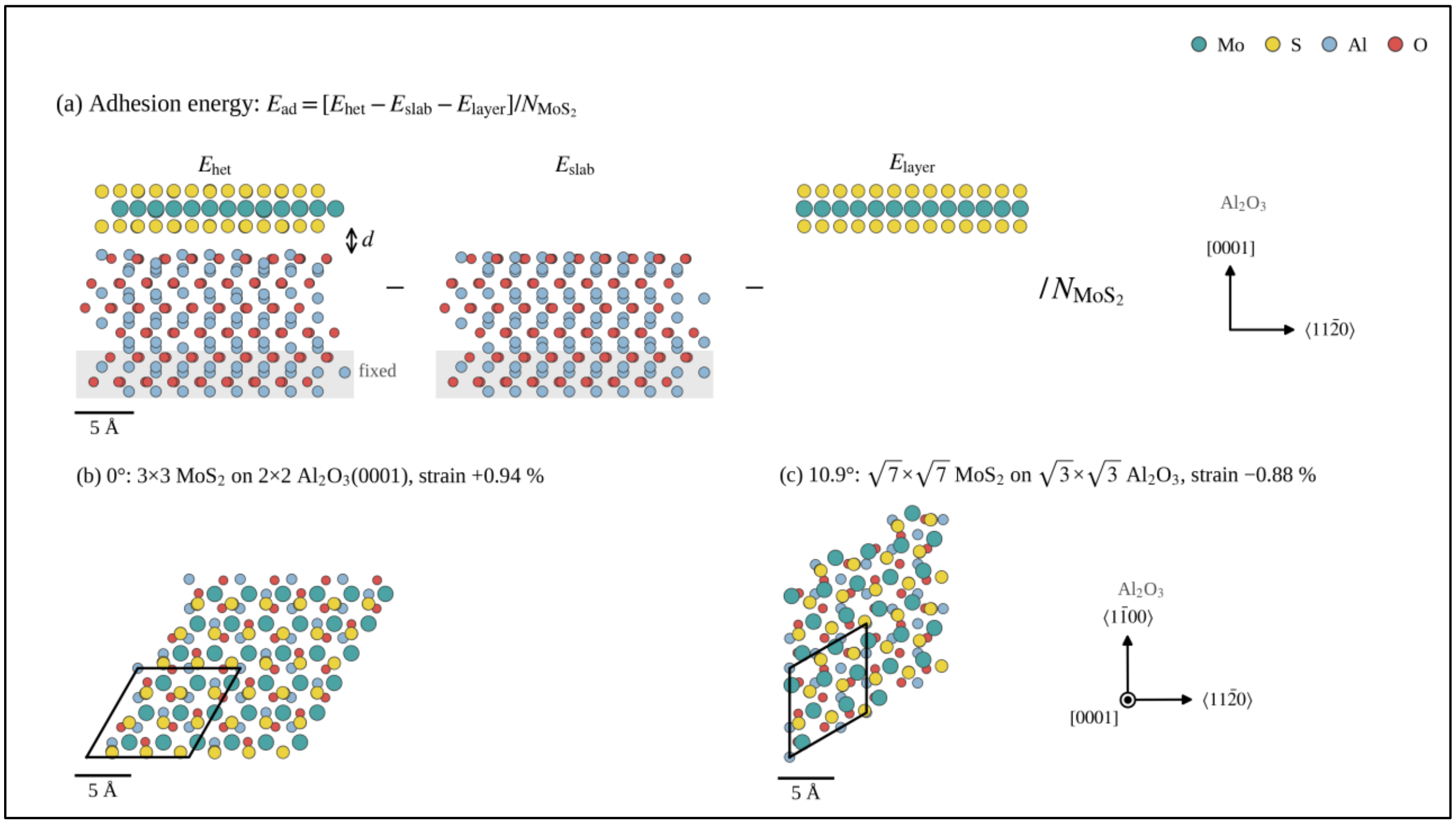

***FIG. S5.*** *DFT simulation cells of the adhesion energy of a 1H-$MoS_2$ monolayer on Al-terminated $Al_2O_3$(0001). (a) $E_{ad} = [E_{het} - E_{slab} - E_{layer}]/N_{MoS_2}$: the heterostructure, the bare slab and the free $MoS_2$ layer are each relaxed in the same supercell; the bottom 36 of the 120 slab atoms (shaded) are fixed, and the lowest-energy registry of the screened in-plane offsets is used. d is the gap between the lowest S plane and the topmost O plane. The rigid scan of Fig. 4(a) uses the same three terms with the layer displaced rigidly along z, without relaxation, all in one 55 Å box. (b, c) Top views (2×2 repeated, supercell outlined) of the commensurate 0° (3×3 $MoS_2$ on 2×2 $Al_2O_3$, +0.94 % strain on $MoS_2$) and 10.9° (√7×√7 $MoS_2$ on √3×√3 $Al_2O_3$, −0.88 %) cells; only the $MoS_2$ layer and the topmost substrate layers are shown. All panels share one length scale. Arrows give the $Al_2O_3$(0001) crystallographic directions.*

## Note S5.3. Interfacial Energy

The interfacial energy was obtained from fully periodic superlattices of the $Al_2O_3(0001)$ slab ($Al_{48}O_{72}$, $n_1 = 24$ formula units) and four 2H-$MoS_2$ layers ($Mo_{36}S_{72}$, $n_2 = 36$), with 3×3 $MoS_2$ on 2×2 $Al_2O_3$ at 0° twist and 228 atoms per cell (Fig. S6),

$$\gamma_{\text{int}} = \frac{E_{\text{SL}} - n_1\, e_{\text{Al}_2\text{O}_3} - n_2\, e_{\text{MoS}_2}}{2A},$$

where $E_{SL}$ is the total energy of the superlattice, $e_{Al_2O_3}$ and $e_{MoS_2}$ the energies per formula unit of the relaxed bulk crystals, A the in-plane cell area, and the factor 2 accounts for the two interfaces per cell. The superlattice was relaxed with the cell free and then recomputed as a single point; it contains no vacuum, so $\gamma_{int}$ includes the residual coherency strain energy. Three lateral registries were considered, in which the $MoS_2$ block is translated by 0, 0.92 and 1.84 Å along $a_1 + a_2$ (r0–r2).

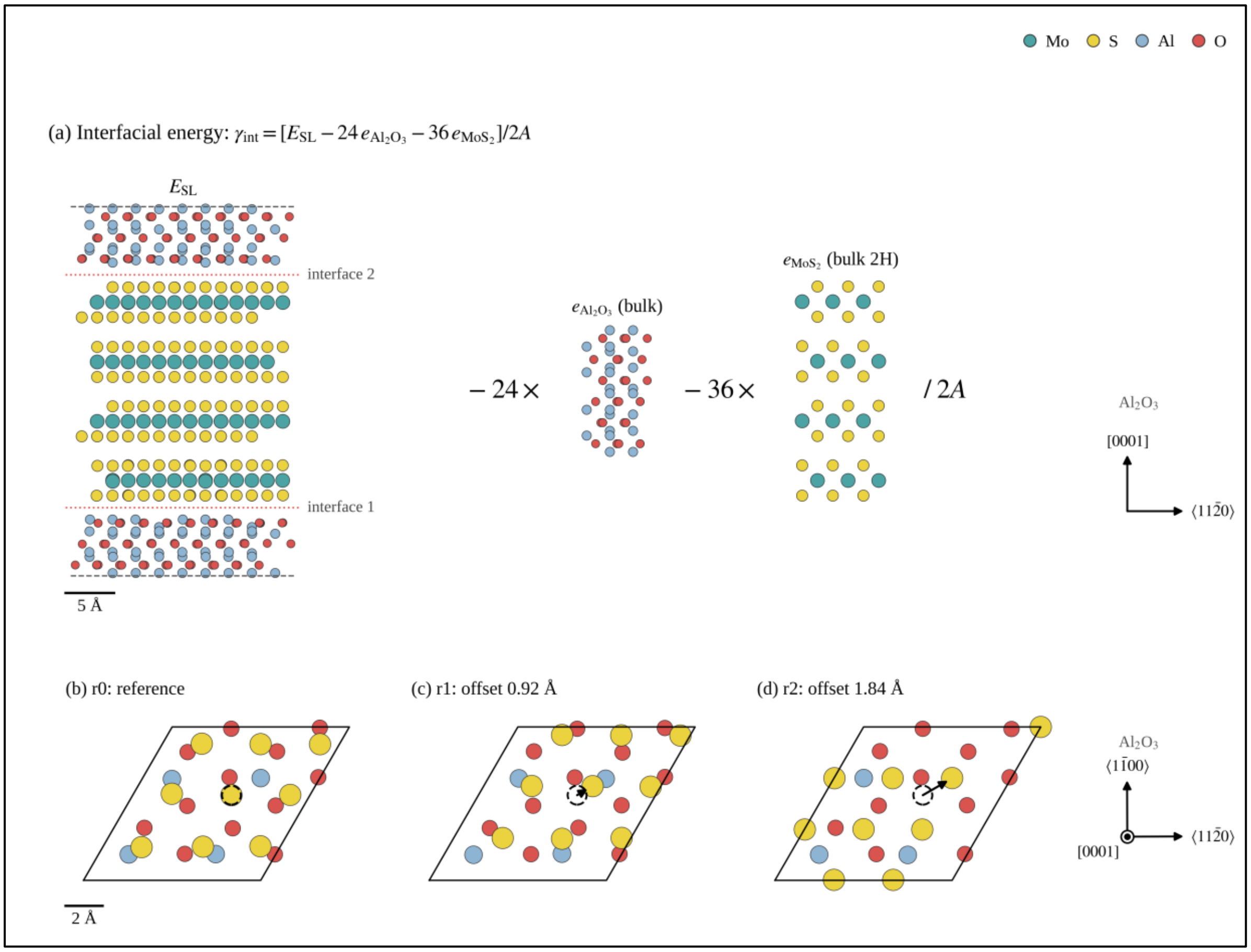

*FIG. S6. DFT simulation cells of the $MoS_2/Al_2O_3(0001)$ interfacial energy. (a) $\gamma_{int} = [E_{SL} - n_1 e_{Al_2O_3} - n_2 e_{MoS_2}]/2A$, where $E_{SL}$ is the energy of a fully periodic superlattice of an Al-terminated $Al_2O_3(0001)$ slab ($Al_{48}O_{72}$, $n_1 = 24$) and four 2H-$MoS_2$ layers ($Mo_{36}S_{72}$, $n_2 = 36$) at 0° twist (228 atoms, in-plane cell 9.59 Å, area A), and $e_{Al_2O_3}$, $e_{MoS_2}$ are the energies per formula unit of relaxed bulk α-$Al_2O_3$ and 2H-$MoS_2$. The superlattice is drawn with the periodic boundary (dashed lines) in the middle of the sapphire slab, so that both $MoS_2/Al_2O_3$ interfaces of the cell (dotted lines) are visible; hence the factor 2A. (b–d) Top views of the interface for the three registries, aligned on the substrate. In r1 and r2 the entire $MoS_2$ stack is displaced rigidly by 0.92 and 1.84 Å along $a_1 + a_2$; the arrow shows the displacement of one tracked S atom from its r0 position (dashed circle). Arrows give the $Al_2O_3(0001)$ crystallographic directions.*

## Note S5.4. Full-Layer Rotations

The twist-angle series of Fig. 5(a) uses commensurate cells at eleven angles θ between the $MoS_2$ and sapphire lattices (Fig. S7). For each angle, the adhesion energy $E_{ad}(\theta)$ is computed as in Note S5.2, and $\Delta E(\theta) = E_{ad}(\theta) - E_{ad}(0°)$. Four cell families are used, listed below; for each cell, 16 lateral registries were screened (4 for the 1002-atom 30° cell) and the two lowest relaxed (only the lowest for 30°). Because the $MoS_2/Al_2O_3$ pair is 120° periodic, −10.9° is equivalent to 109.1°.

| Angles | Cell | Atoms | Strain on $MoS_2$ | Fixed atoms |
|---|---|---|---|---|
| 0°, 60° | 3×3 $MoS_2$ on 2×2 $Al_2O_3$ | 147 | +0.94 % | 36 |
| ±10.9°, 49.1°, 70.9° | √7×√7 $MoS_2$ on √3×√3 $Al_2O_3$ | 111 | −0.88 % | 27 |
| 19.1°, 40.9°, 79.1°, 100.9° | 4×4 $MoS_2$ on √7×√7 $Al_2O_3$ | 258 | +0.15 % | 63 |
| 30° | 8×8 $MoS_2$ on 3√3×3√3 $Al_2O_3$ | 1002 | −1.66 % | 243 |

The 0° and 60° cells share the supercell, strain and reference energies, so their difference contains no strain contribution. At their best registries they differ rigidly by only 0.05 meV/$MoS_2$, consistent with the ≤ 0.1 meV/$MoS_2$ reported previously[2]. The 2.4 meV/$MoS_2$ between them after relaxation is of the same order as the scatter between relaxations of the same cell started from different registries, 0.1–1.3 meV/$MoS_2$ for 0°, 60° and +10.9° and 6.5 meV/$MoS_2$ for −10.9°, so relaxed full-layer differences of a few meV/$MoS_2$ may mark only the resolution of these calculations.

*Comparison with previous DFT studies*

Previous DFT calculations of commensurate $MoS_2$ monolayers on Al-terminated $Al_2O_3(0001)$, with PBE-D2, also find θ = 0° lower in energy than the other orientations considered[2,3]. They find the antiparallel 0° and 60° orientations degenerate within about 0.1 meV/$MoS_2$. In our relaxed cells, θ = 60° lies 2.4 meV/$MoS_2$ above θ = 0°. This is about twice the spread between relaxations from different registries (main text, Sec. 3.3.2), so the two orientations are nearly degenerate here as well. The magnitudes differ: Sakuma et al.[2] obtain 35 meV/$MoS_2$ between 10.9° and 0° compared with 15.8 meV/$MoS_2$ here with PBE-D3.

Computed orientation energies on sapphire depend on the coincidence cell and the strain it imposes on the layer[3,4]. They also depend on the lateral registry[3], the surface termination[5], whether a finite flake or an infinite layer is modeled[6], and the DFT functional and dispersion correction. In a coincidence cell, the elastic energy of the strained layer can be comparable to its van der Waals interaction with the substrate. The energies of strained full layers therefore do not by themselves determine the epitaxial orientation[3,4]. Ohtake et al., for example, find the 0° orientation 0.164 eV/$MoS_2$ below the 90° orientation (the 30° family in our convention); the 90° cell carries a lattice mismatch of 4.18 %, compared with 0.42 % for 0°, and the authors conclude that DFT gives no definitive answer on their relative stability[3]. Here, $E_{ad}(\theta)$ is referenced to a free layer at the same strain, so the elastic energy of the layer approximately cancels and ΔE(θ) contains its interaction only with the substrate. The full-layer rotation series therefore tests the interface energy landscape described by the NEP; it does not necessarily predict the epitaxial orientation.

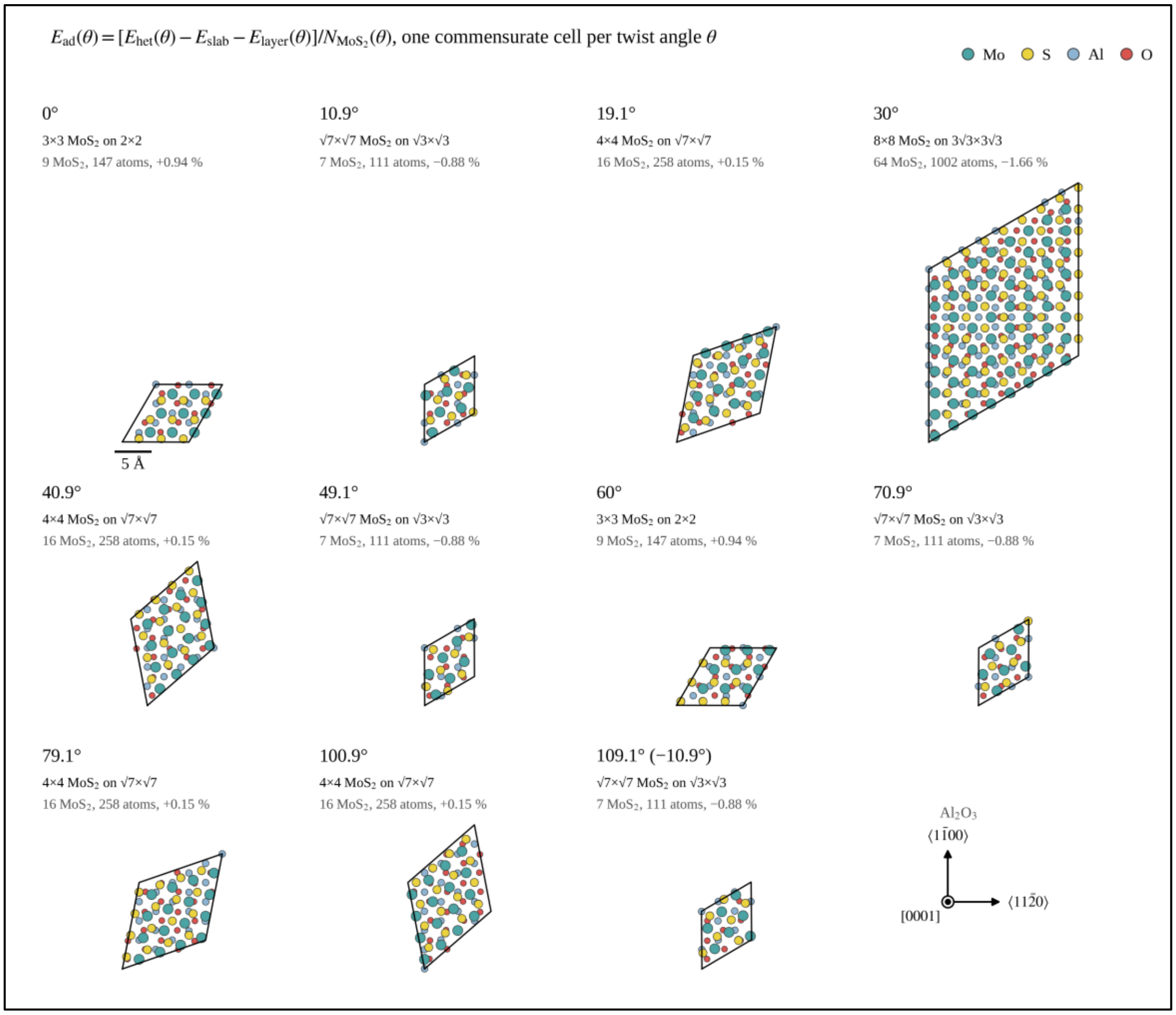

***FIG. S7.*** *Commensurate DFT cells of the full-layer twist-angle series (Fig. 5a; Table S4). For each twist angle θ, $E_{ad}(\theta) = [E_{het}(\theta) - E_{slab} - E_{layer}(\theta)]/N_{MoS_2}(\theta)$, with the heterostructure, the bare Al-terminated $Al_2O_3$(0001) slab and the free $MoS_2$ layer relaxed in the same supercell (Note S5.2). Top views of the relaxed heterostructures ($MoS_2$ layer and topmost substrate layers; one supercell outlined), all at the same length scale. Labels give the $MoS_2$ supercell on the sapphire supercell, the number of $MoS_2$ units, the total number of atoms and the biaxial strain applied to $MoS_2$. The $MoS_2$/sapphire pair is 120° periodic, so −10.9° is drawn as 109.1°; for every angle the lowest-energy relaxed registry is used. All cells are drawn in a common frame with the $Al_2O_3$ ⟨11-20⟩ direction horizontal.*

## Note S5.5. Supported Flake Rotations

The rotation energy of Fig. 5(b) uses a triangular $Mo_{21}S_{66}$ flake (87 atoms) whose zigzag Mo edges are fully covered by S dimers (100 % S; Note S6), on the 6×6 $Al_2O_3(0001)$ slab (1080 atoms; 1167 atoms in total, bottom 324 fixed). The flake was first relaxed in vacuum. For every angle φ, it was then rotated rigidly about its centroid, placed on the relaxed slab and relaxed (Fig. S9). The rotation energy,

$$\Delta E(\varphi) = E(\varphi) - E(\varphi = 0),$$

is the difference between the total energies E(φ) of cells of identical composition, so no reference energies enter. The angle φ follows the convention of the twist angle θ of the full layer (Sec. 3.3.2): at φ = 0°, the flake lattice has the orientation of the full layer at θ = 0°, and its zigzag edges are parallel to $Al_2O_3[11\bar{2}0]$, as they are at φ = 60°. At φ = 30° and 90°, the edges are parallel to $Al_2O_3\langle 1\bar{1}00\rangle$. One lateral registry was used for each of φ = 0, 30, 60 and 90°, taken as the lowest-energy registry of a screening over the five $C_3$-inequivalent registries on a thinner $Al_2O_3(0001)$ slab (½ the thickness of the conventional cell). The relaxed cells are shown in Fig. S8.

*Comparison with previous DFT studies*

Unlike the full layer, the supported flake is most stable in the 30° orientation family (φ = 30° and 90°), in which its zigzag edges run along the sapphire ⟨1-100⟩ direction. Previous DFT calculations of triangular $MoS_2$ flakes on c-plane sapphire also favor this family, all with PBE-D3. Fu et al. find a single minimum at 30° between 0° and 120° on an O-terminated surface[6]. Chen et al. find two equivalent minima at 30° and 90° on single-Al-terminated $\alpha$-$Al_2O_3(0001)$[7]. Zou et al. also find minima at 30° and 90° on an Al-terminated surface, about 0.08 eV/$MoS_2$ below 0° and differing from each other by only 0.01 meV/$MoS_2$[8]. Fu et al. also showed that a periodic layer gives nearly equal energies at 30° and 90°, and that the edge and corner atoms of a finite triangular seed make the two orientations inequivalent[6]. For the flake considered in this study, φ = 30° lies 0.15 eV below φ = 90°. We compare only the preferred family with these studies, not the order of 30° and 90°, which depends on how the triangle is oriented on the substrate. Calculations of a smaller $Mo_{12}S_{24}$ seed instead find 0° and 60° most stable on half-Al-terminated and OH-terminated surfaces, with an additional metastable 30° configuration on the OH-terminated one[5]. The preferred family therefore depends on the surface termination and on the size and edges of the flakes. In the $Mo_{21}S_{66}$ flake, 15 of the 21 Mo atoms lie on its edges or corners, so that its orientation preference is dominated by the edges and corners rather than by the interior, consistent with its differing from that of the full layer. The energy differences between orientations are small, at most 5 meV per flake atom (Table S4).

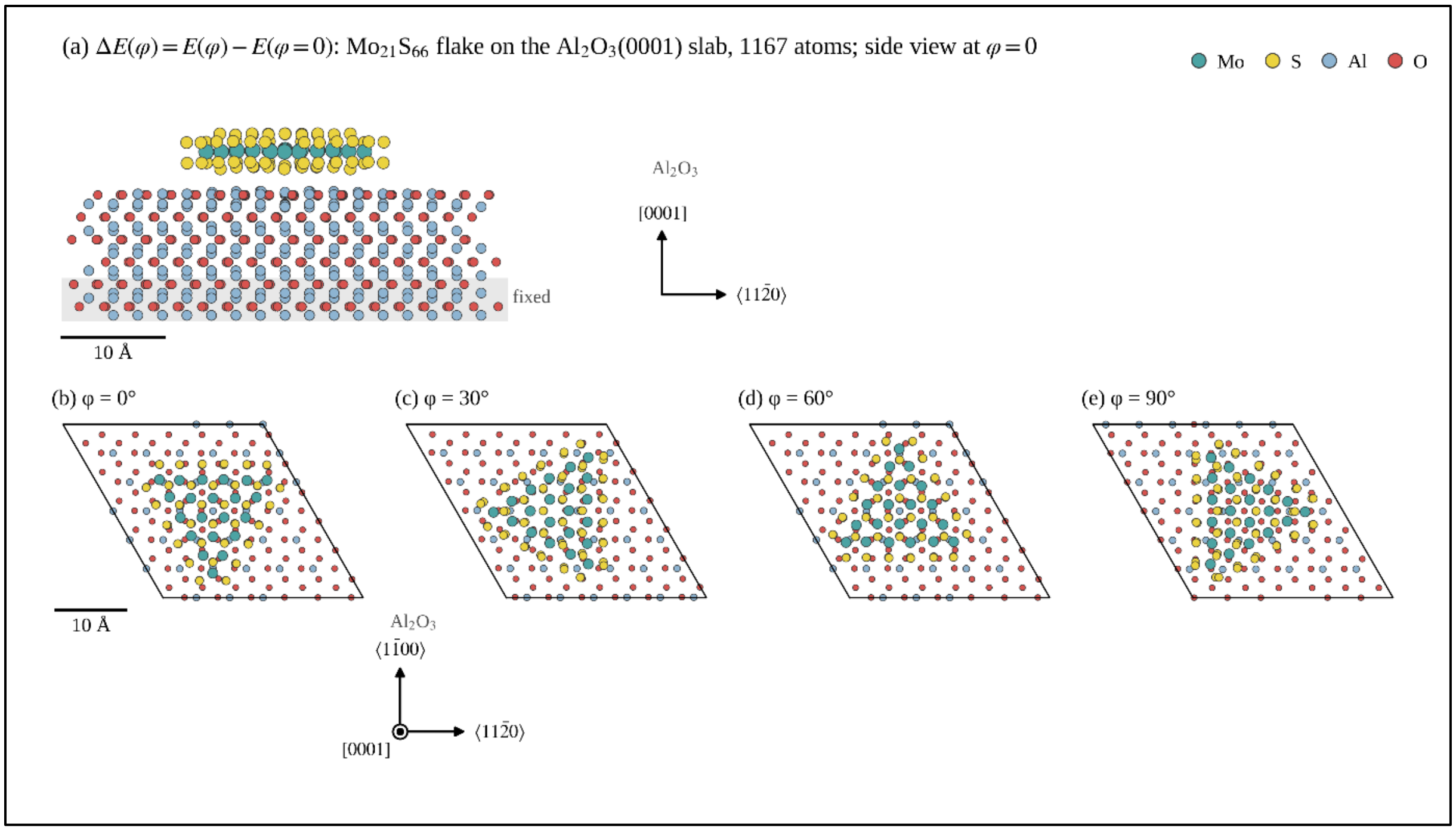

***FIG. S8.*** *DFT cells for the rotation energy of a $Mo_{21}S_{66}$ flake on the 6×6 Al-terminated $Al_2O_3(0001)$ slab, one unit cell thick (1080 atoms; 1167 atoms in total). $\Delta E(\varphi) = E(\varphi) - E(\varphi = 0)$ is the total-energy difference of the same supercell with the flake rotated in plane by φ about its center; the slab and the flake are identical in all cells, so no reference energies enter. (a) Side view of the φ = 0 cell (bottom 324 atoms fixed, shaded). (b-e): top views of the relaxed flake on the topmost substrate layers for each φ (supercell outlined).*

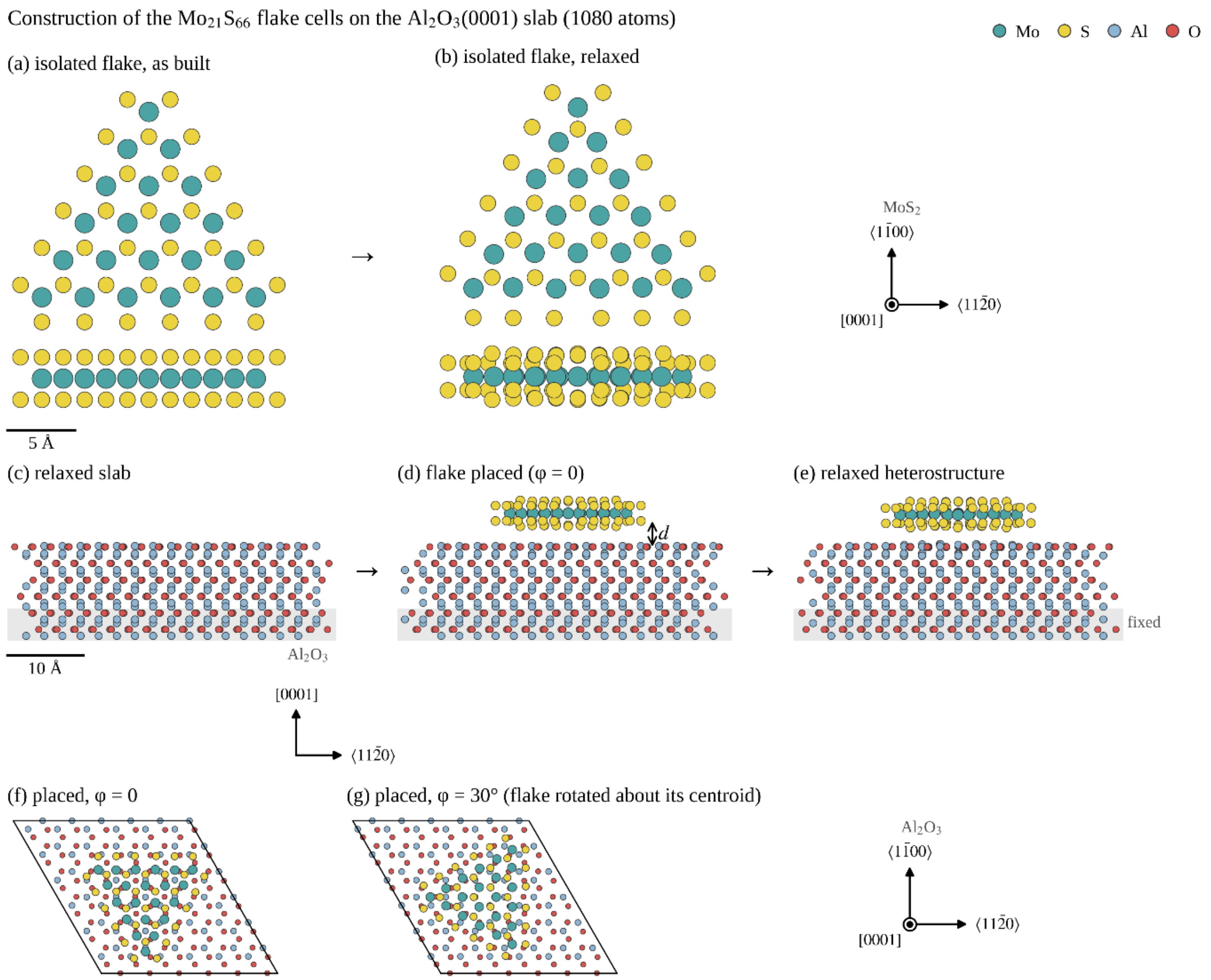

***FIG. S9.*** *Construction of the $Mo_{21}S_{66}$ flake cells on the $Al_2O_3(0001)$ slab (6×6, 1080 atoms; 1167 atoms in total). (a) The isolated flake as built (cut from the 1H monolayer with S-dimer-passivated zigzag edges) and (b) after relaxation in vacuum (RMS / maximum displacement 0.42 / 0.65 Å; the flake bows, with the edge S dimers about 0.7 Å below the center). (c) Relaxed bare slab (fixed atoms shaded). (d) The relaxed flake placed on the relaxed slab at φ = 0 and (e) the relaxed heterostructure (flake atoms move by 0.22 Å RMS, 0.35 Å maximum; the median height d of the bottom-S plane above the top O layer changes from 3.42 to 3.07 Å). (f, g) Top views of the relaxation inputs at φ = 0 and 30°: for every φ the relaxed flake (b) is rotated rigidly about its centroid, placed as in (d), and relaxed as in (d) → (e).*

## Note S5.6. Interlayer Gaps

Figure S10 compares the gap, d, between the $MoS_2$ and the $Al_2O_3(0001)$ surface in the NEP-relaxed and the DFT-relaxed structures of all interfaces of Sec. 3.3 of the main text. The NEP reproduces the gap within 0.035 Å for the full monolayer at all eleven twist angles and for both interfaces of the three superlattices. For the supported $Mo_{21}S_{66}$ flake, it places the interior of the flake 0.02–0.10 Å closer to the substrate than DFT.

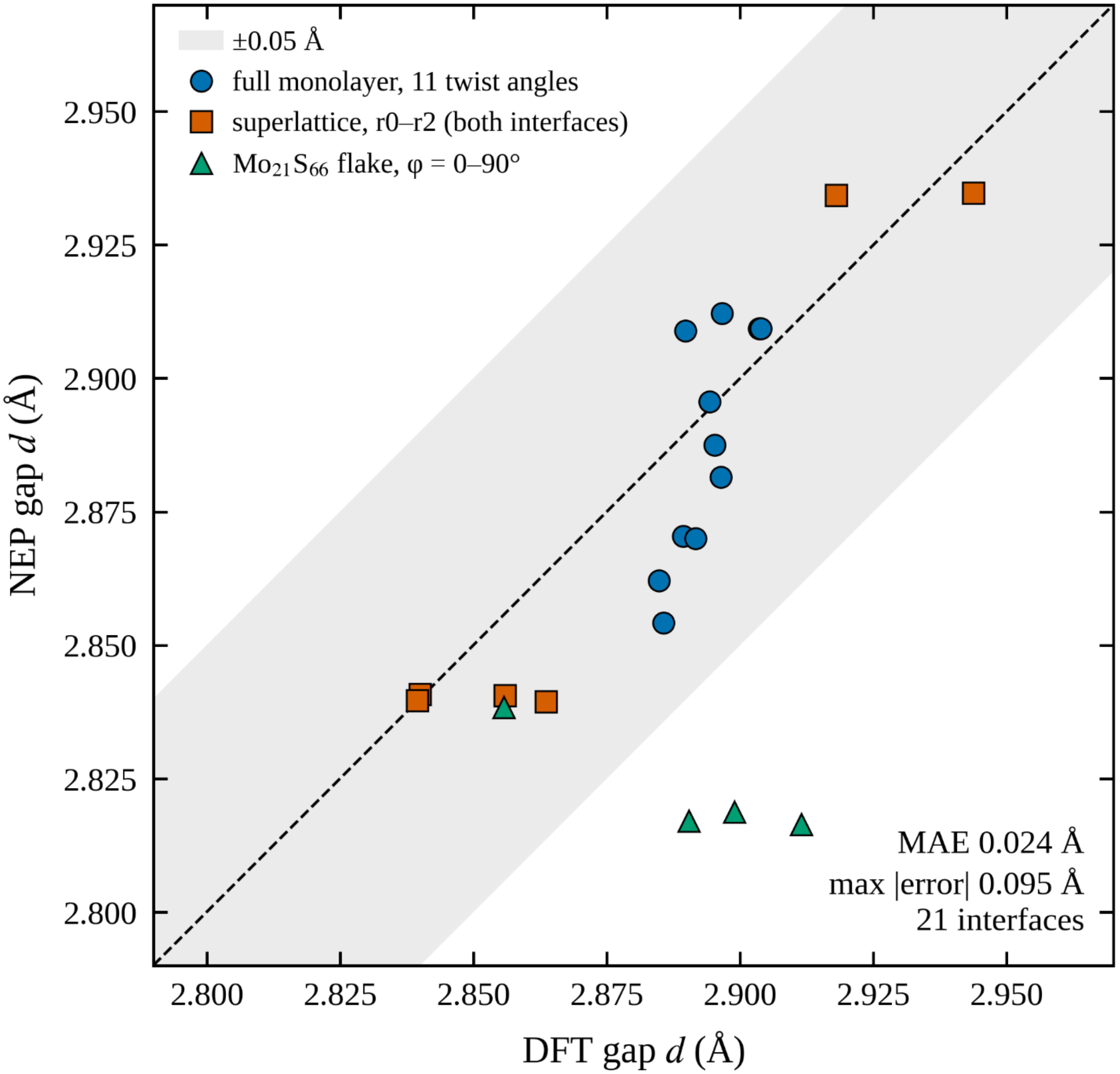

***FIG. S10.*** *Gap, d, between $MoS_2$ and the Al-terminated $Al_2O_3(0001)$ surface, NEP versus DFT, each in its own relaxed structure (the NEP relaxed from the DFT structure): the full monolayer at the eleven twist angles of Fig. 5(a) (circles; lowest-energy registry; Note S5.4), both interfaces of the four-layer 2H-$MoS_2$ superlattices in registries r0–r2 (squares; Note S5.3), and the $Mo_{21}S_{66}$ flake at $\varphi$ = 0, 30, 60 and 90° (triangles; Note S5.5). d is the distance between the mean height of the S plane facing the substrate and the mean height of the topmost O plane; for the flake, only atoms within 6 Å of its centroid are used, away from the bent edges. Dashed line: perfect agreement; shaded band: ±0.05 Å. Over all 21 interfaces, the mean absolute error is 0.024 Å (0.8 %) and the largest error 0.095 Å (3.3 %; flake at $\varphi$ = 60°).*

## Note S5.7. Numerical Values of Interfacial Calculations

**TABLE S4.** Surface, adhesion, interface and rotation energetics. "NEP (DFT geom.)" is a single-point NEP evaluation on the DFT-relaxed structures; "NEP (relaxed)" is the NEP after its own relaxation from the DFT geometry. Rigid scans are unrelaxed by construction. Errors refer to the relaxed NEP where available and are given in % or, for energy differences (†), in the unit of the row; they are computed from unrounded values. $\Delta E(\theta) = E_{ad}(\theta) - E_{ad}(0°)$; flake energies are relative to φ = 0°, with φ defined as θ (Note S5.5). Adhesion, rotation and flake cells use Al-terminated $Al_2O_3(0001)$ slabs one unit cell thick; $\gamma_{int} = [E_{SL} - n_1e_1 - n_2e_2]/2A$ is obtained from periodic four-layer 2H-$MoS_2$/$Al_2O_3$ superlattices (3×3 $MoS_2$ on 2×2 sapphire, two interfaces per cell) in three lateral registries, in which the $MoS_2$ block is translated by (0, 0), (1/9, 1/9) and (2/9, 2/9) of the sapphire 1×1 cell (r0, r1 and r2; 0, 0.92 and 1.84 Å).

| System | Quantity | Unit | DFT | NEP (DFT geom.) | NEP (relaxed) | Error |
|---|---|---|---|---|---|---|
| $Al_2O_3(0001)$, Al-terminated | $\gamma_s$ (relaxed) | J/m² | 1.941 | 1.931 | 1.927 | −0.7 |
| 2H-$MoS_2(0001)$ | $\gamma_s$ (relaxed) | J/m² | 0.242 | 0.239 | 0.239 | −1.3 |
| Rigid scan, $\theta$ = 0° | $E_{ad}$ at minimum | meV/$MoS_2$ | −378.4 | −375.5 | – | +0.8 |
| | equilibrium S–O gap | Å | 2.860 | 2.852 | – | −0.3 |
| | $E_{ad}$ tail, gap ≥ 6 Å | meV/$MoS_2$ | −16.8 | 0.7 | – | +17.5† |
| Rigid scan, $\theta$ = 10.9° | $E_{ad}$ at minimum | meV/$MoS_2$ | −325.6 | −304.3 | – | +6.5 |
| | equilibrium S–O gap | Å | 2.888 | 2.898 | – | +0.3 |
| | $E_{ad}$ tail, gap ≥ 6 Å | meV/$MoS_2$ | −16.4 | 0.6 | – | +17.0† |
| Relaxed full layer | $E_{ad}(0°)$ | meV/$MoS_2$ | −313.1 | −309.8 | −312.6 | +0.2 |
| | $\Delta E(+10.9°)$ | meV/$MoS_2$ | 15.8 | 28.9 | 30.4 | +14.6† |
| | $\Delta E(19.1°)$ | meV/$MoS_2$ | 11.0 | 11.7 | 12.9 | +1.8† |
| | $\Delta E(30°)$ | meV/$MoS_2$ | 13.9 | 37.3 | 37.1 | +23.2† |
| | $\Delta E(40.9°)$ | meV/$MoS_2$ | 11.0 | 11.7 | 13.1 | +2.1† |
| | $\Delta E(49.1°)$ | meV/$MoS_2$ | 16.1 | 29.8 | 21.0 | +4.9† |
| | $\Delta E(60°)$ | meV/$MoS_2$ | 2.4 | 4.9 | 3.2 | +0.8† |
| | $\Delta E(70.9°)$ | meV/$MoS_2$ | 15.5 | 28.4 | 11.2 | −4.2† |
| | $\Delta E(79.1°)$ | meV/$MoS_2$ | 3.6 | 6.5 | 7.0 | +3.4† |
| | $\Delta E(100.9°)$ | meV/$MoS_2$ | 4.2 | 7.5 | 8.2 | +4.0† |
| | $\Delta E(-10.9°) \equiv 109.1°$ | meV/$MoS_2$ | 11.7 | 31.2 | 11.6 | −0.1† |
| $MoS_2$/$Al_2O_3$ interface | $\gamma int, r0$ | J/m² | 1.627 | 1.601 | 1.591 | −2.2 |
| | $\gamma int, r1$ | J/m² | 1.613 | 1.595 | 1.584 | −1.8 |
| | $\gamma int, r2$ | J/m² | 1.627 | 1.600 | 1.591 | −2.2 |
| $Mo_{21}S_{66}$ triangular flake on $Al_2O_3(0001)$ | ΔE(φ = 30°) | eV/flake | −0.42 | 1.17 | −0.40 | +0.02† |
| | ΔE(φ = 60°) | eV/flake | −0.07 | 0.29 | 0.16 | +0.23† |
| | ΔE(φ = 90°) | eV/flake | −0.27 | 0.74 | −0.48 | −0.21† |

**Note S6. Edge Energies**

A Mo edge ends in a row of Mo atoms, and its S coverage counts the S atoms added outside that row: 0 % for a bare Mo edge, 50 % for one S atom and 100 % for one $S_2$ dimer per edge Mo atom, each S bonded to a single Mo. An S edge ends in lattice S columns, either complete (100 %, each S bonded to two Mo) or with every second column removed (50 %).

Isolated ribbons are periodic along the edge, with length L per cell, and separated from their images by vacuum in the other two directions (Fig. S11). Armchair ribbons of widths w = 6, 8 and 10 are stoichiometric and have two equivalent edges, so that

$$\gamma_{\mathrm{AC}} = \frac{E - N\,\mu_{\mathrm{MoS_2}}}{2L},$$

where E is the total energy of the ribbon cell and N the number of formula units. Zigzag ribbons always have one Mo edge and one S edge and therefore give only the sum of the two edge energies,

$$\gamma_{\mathrm{Mo}} + \gamma_{\mathrm{S}} = \frac{E - N_{\mathrm{Mo}}\,\mu_{\mathrm{MoS_2}} - \Delta N_{\mathrm{S}}\,\mu_{\mathrm{S}}}{L}.$$

Ribbons of different widths agree within 0.002 eV/Å, so the ribbon energies are converged with respect to width. Table S5 reports the energy per edge length relative to the pristine monolayer, $E_{edge} = (E_{ribbon} - N_{MoS_2}\,\varepsilon_{ML} - \Delta N_S\,\mu_S^{\alpha})/L_{edges}$, where $\varepsilon_{ML}$ is the energy per formula unit of the monolayer, $L_{edges}$ the total edge length in the cell, and excess or missing S is referenced to α-S. For stoichiometric ribbons, this is exactly the energy of the ribbon relative to the pristine monolayer per edge length.

Individual Mo-edge and S-edge energies were obtained from triangular flakes with three identical edges of n Mo atoms each (Figs. S12–S14). Each of the five terminations was computed for n = 4–8 and relaxed in vacuum, and

$$X_n = E_n - N_{\mathrm{Mo}}\,\mu_{\mathrm{MoS_2}} - \Delta N_{\mathrm{S}}\,\mu_{\mathrm{S}} = \gamma \cdot 3na + 3E_{\mathrm{c}}$$

was fitted linearly against the perimeter 3na at both limits of $\mu_S$. Here $E_n$ is the total energy of the flake, *a* the lattice constant, γ the edge energy (the slope) and $E_c$ the corner energy (the intercept is three corner energies); $N_{Mo}$ and $\Delta N_S$ are counted from each relaxed structure. Each method is evaluated at its own limits (Sec. 2.4): the Mo-rich limit lies at $\Delta\mu_S = -1.20$ eV in DFT and at −1.27 eV with the NEP. The largest fit residuals are listed in Table S6. We note that for the 100 % S edge, the n = 5 flake relaxed to a less symmetric edge.

On $Al_2O_3$(0001), armchair ribbons (w = 6 and 8, 1014–1032 atoms), zigzag ribbons (w = 6, 534–540 atoms) and Mo-edge and S-edge triangles with 100 % S (n = 4–6, 1114–1167 atoms) were computed on the slab (Figs. S15 and S16). In the supported cells, E is replaced by $E - E_{slab}$, with $E_{slab}$ the energy of the bare slab in the same cell, and the monolayer reference by ε, the energy per formula unit of the full monolayer on the same substrate, taken from the 0° adhesion cell of Note S5.2:

$$\gamma_{\mathrm{AC}}^{\mathrm{sub}} = \frac{E - E_{\mathrm{slab}} - N\varepsilon}{2L}, \qquad E_n - E_{\mathrm{slab}} - \Delta N_{\mathrm{S}}\,\mu_{\mathrm{S}} = N_{\mathrm{Mo}}\,\varepsilon + \gamma \cdot 3na + 3E_{\mathrm{c}}.$$

ε is fixed rather than fitted, because a three-parameter fit over three flake sizes is ill-conditioned. Even so, the supported Mo-edge triangle fit leaves a DFT residual of 1.05 eV, and the supported Mo-edge energy used for the grain boundaries (Note S8) is therefore obtained from a zigzag ribbon instead. Tables S5 and S7 list the ribbon and individual edge energies.

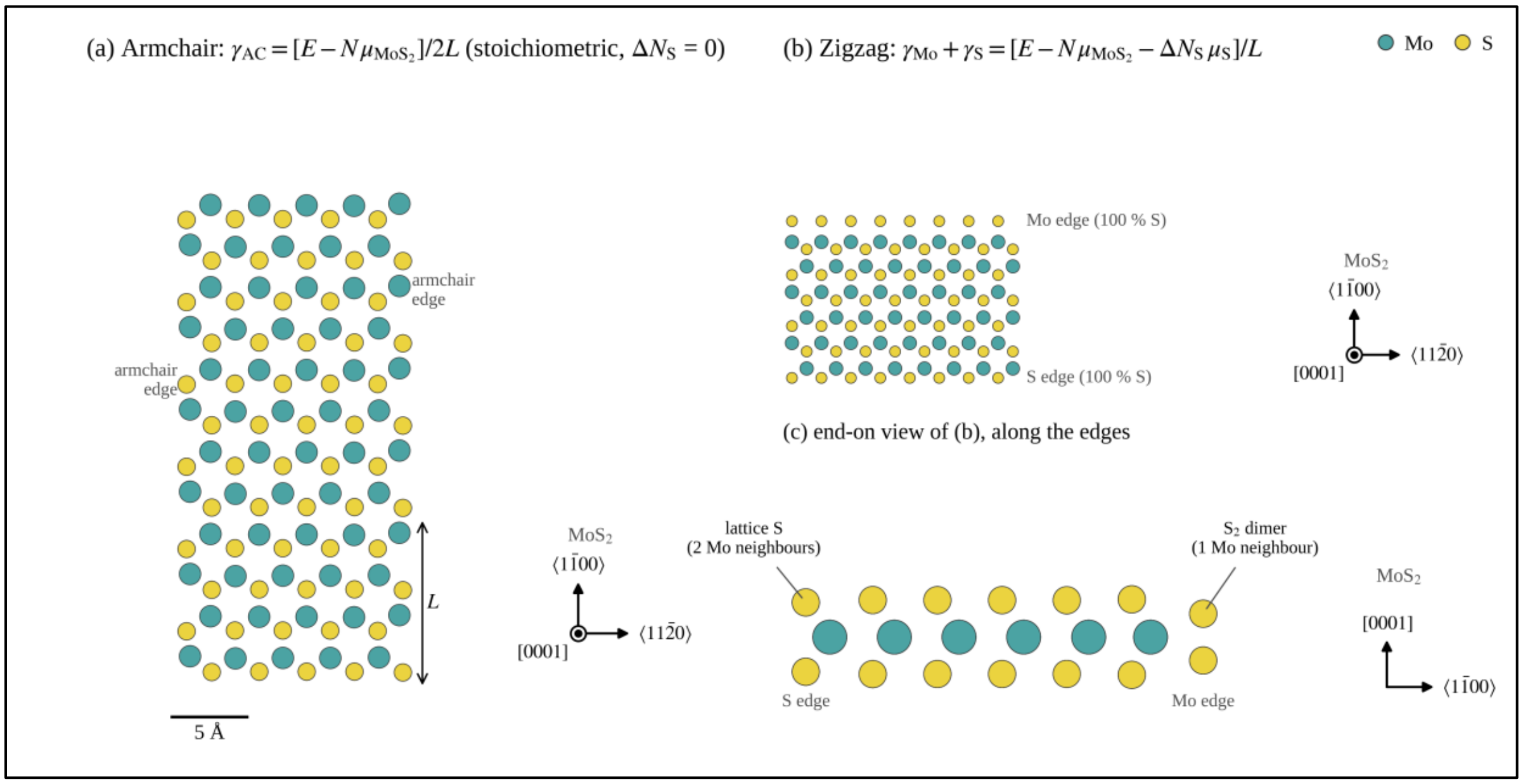

***FIG. S11.*** *DFT ribbon cells of the isolated edges. (a) Armchair ribbon (w = 10, periodic along ⟨1-100⟩ with length L, three periods drawn): stoichiometric with two equivalent armchair edges, so $\gamma_{AC} = [E - N\mu_{MoS_2}]/2L$ is independent of $\mu_S$. (b) Zigzag ribbon (w = 6, Mo edge 100 % S | S edge 100 % S; periodic along ⟨11-20⟩, two periods drawn): a zigzag ribbon always has one Mo edge and one S edge, so it yields only $\gamma_{Mo} + \gamma_S = [E - N_{Mo}\,\mu_{MoS_2} - \Delta N_S\,\mu_S]/L$; individual values come from the triangles. (c) End-on view of (b) along the edges: the Mo edge carries $S_2$ dimers (S–S 1.98 Å, 2.23 Å outside the last Mo row, one Mo neighbor), the S edge ends in lattice S (1.02 Å outside, two Mo neighbors). In top views the two S of a dimer or column overlap, so both edge types end in yellow; the end-on view distinguishes them.*

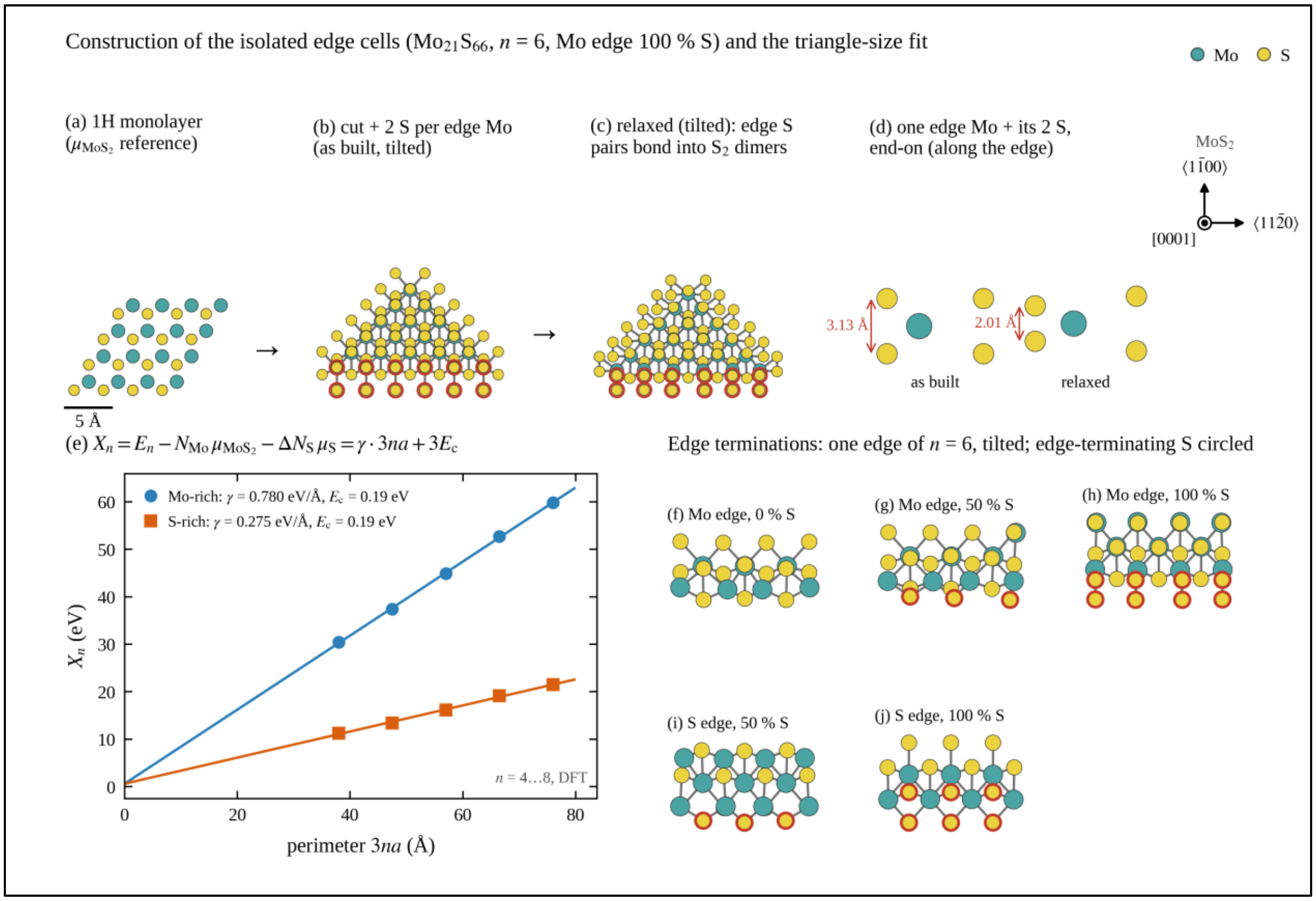

***FIG. S12.*** *Construction of the isolated edge cells and the triangle-size fit (DFT). (a) The relaxed 1H monolayer defines* $\mu_{MoS_2}$*. (b) A triangular flake with n Mo per edge is cut from it and two S are placed on every edge Mo at the lattice S positions (*$Mo_{21}S_{66}$*, n = 6; tilted view, edge S circled), and (c) relaxed in vacuum. (d) End-on view of one edge Mo with its two edge S: the relaxation is almost purely out of plane (S move up to 0.58 Å in z, ≤ 0.31 Å in plane) and pinches each S pair from 3.13 Å (unbonded, lattice heights ±1.57 Å) into an* $S_2$ *dimer (2.01 Å, ±1.00 Å from the Mo plane); because the two S overlap in a top view, the change is invisible from above. (e) For n = 4–8,* $X_n = E_n - N_{Mo}\,\mu_{MoS_2} - \Delta N_S\,\mu_S$ *is linear in the perimeter 3na; the slope is the edge energy γ and the intercept* $3E_c$*, shown at the Mo-rich and S-rich limits. (f–j) Tilted views of one edge for the five terminations (each with its own n = 4–8 series); circled: S outside the last Mo row (Mo edges) or the terminating S row (S edges).*

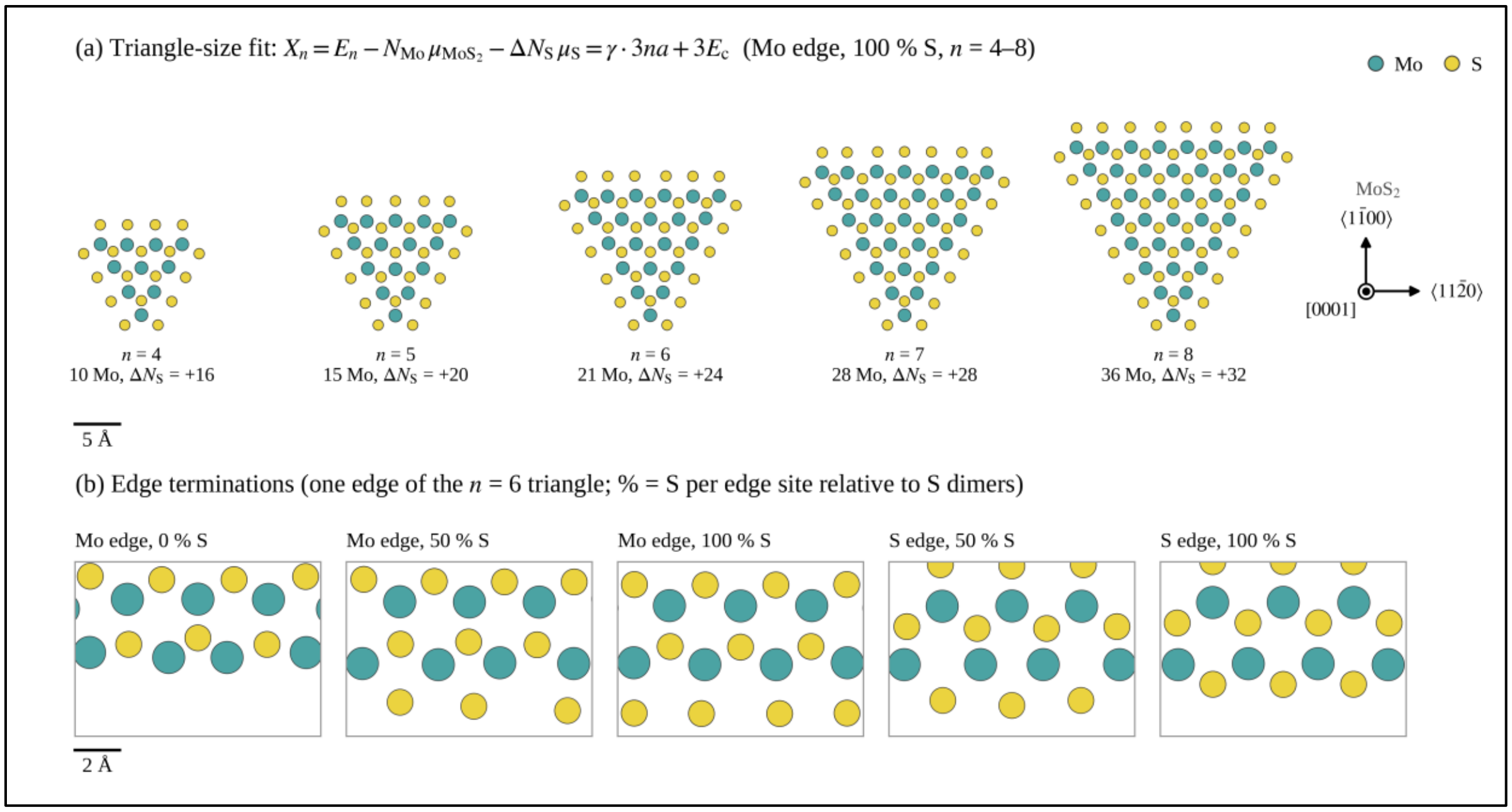

***FIG. S13.*** *DFT cells for the individual edge energies of isolated 1H-$MoS_2$ (Fig. 6). (a) Triangular flakes with three identical edges of n Mo atoms (Mo edge, 100 % S, n = 4–8); $N_{Mo}$ and $\Delta N_S = N_S - 2N_{Mo}$, counted from each relaxed structure, are given below each flake. (b) The five edge terminations, each fitted with its own n = 4–8 series, shown on one edge of the n = 6 triangle: Mo edge with 0 %, 50 % and 100 % S, and S edge with 50 % and 100 % S. Top views; the two S layers overlap in projection. Arrows give the $MoS_2$ crystallographic directions.*

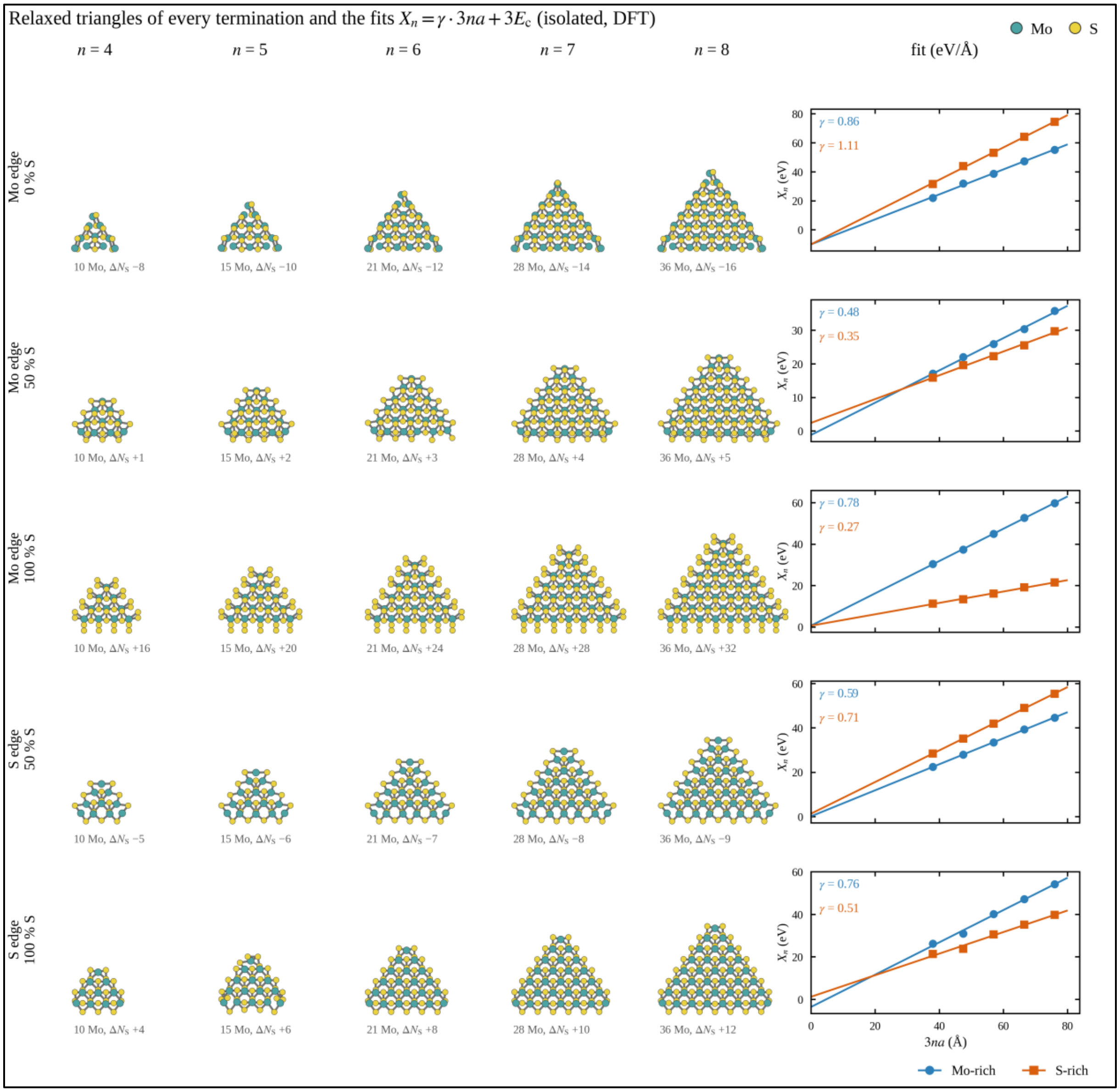

***FIG. S14.*** *All relaxed triangular flakes used for the individual edge energies of isolated 1H-$MoS_2$ (DFT). Rows: the five edge terminations; columns: n = 4–8 Mo per edge (tilted views, common length scale; Mo count and S excess $\Delta N_S$ given under each). Right: for each termination, $X_n = E_n - N_{Mo}\,\mu_{MoS_2} - \Delta N_S\,\mu_S$ versus the perimeter 3na at the Mo-rich (circles) and S-rich (squares) limits, with linear fits $X_n = \gamma \cdot 3na + 3E_c$; the slopes γ are the edge energies of Fig. 6.*

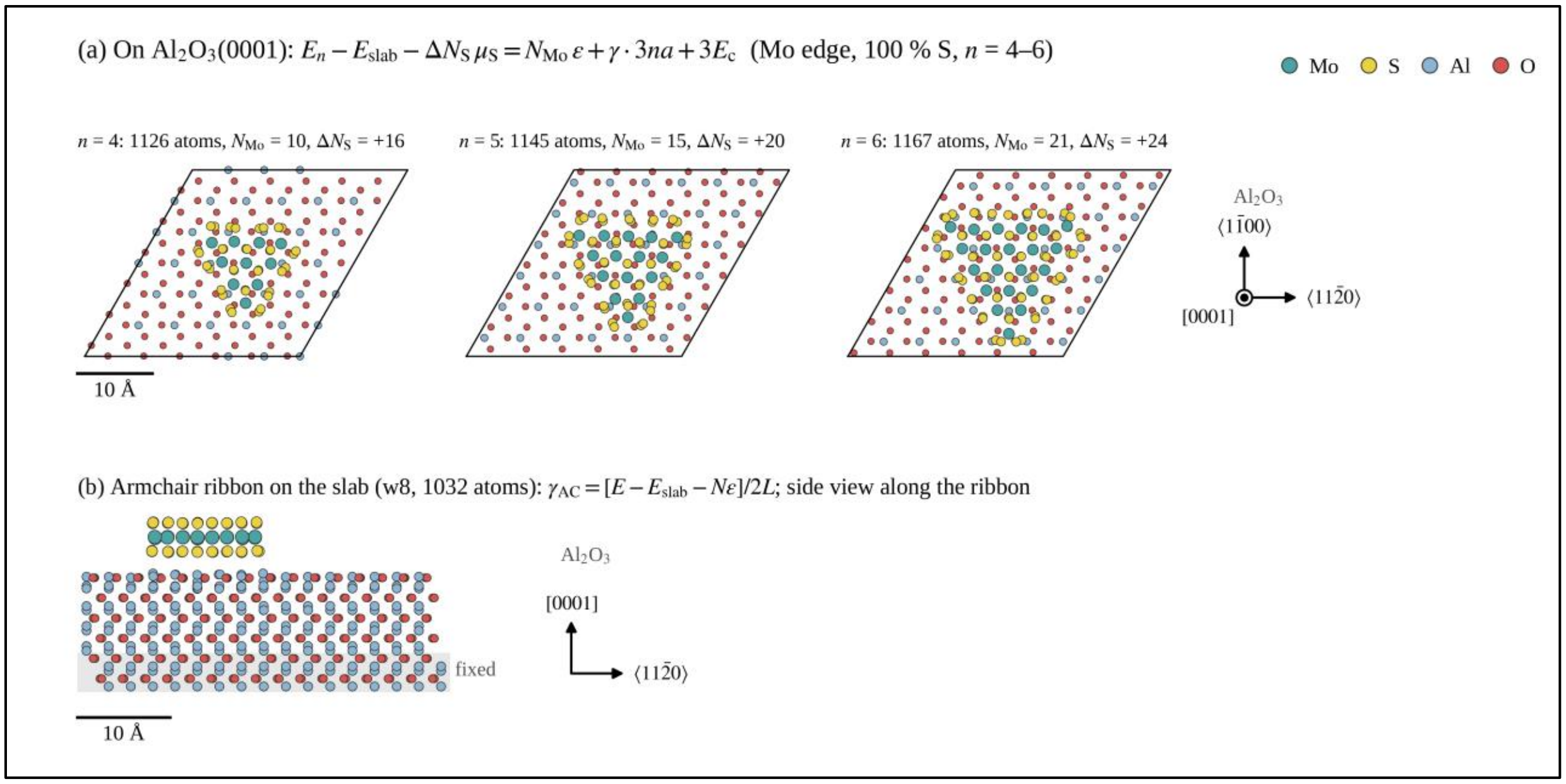

***FIG. S15.*** *DFT cells for the edge energies on $Al_2O_3(0001)$. (a) Mo-edge (100 % S) triangles with n = 4–6 on the 6×6 $Al_2O_3(0001)$ slab (1126–1167 atoms, bottom 324 fixed; $MoS_2$ and topmost substrate layers shown): $E_n - E_{slab} - \Delta N_S\,\mu_S = N_{Mo}\,\varepsilon + \gamma\cdot 3na + 3E_c$, with ε the energy per $MoS_2$ of the full monolayer on the same sapphire (fixed, not fitted). (b) Armchair ribbon (w = 8, 1032 atoms) on the slab, side view along the ribbon: $\gamma_{AC} = [E - E_{slab} - N\varepsilon]/2L$. S-edge triangles and zigzag ribbons on sapphire use the same cells and are not drawn. Arrows give the $Al_2O_3$ crystallographic directions.*

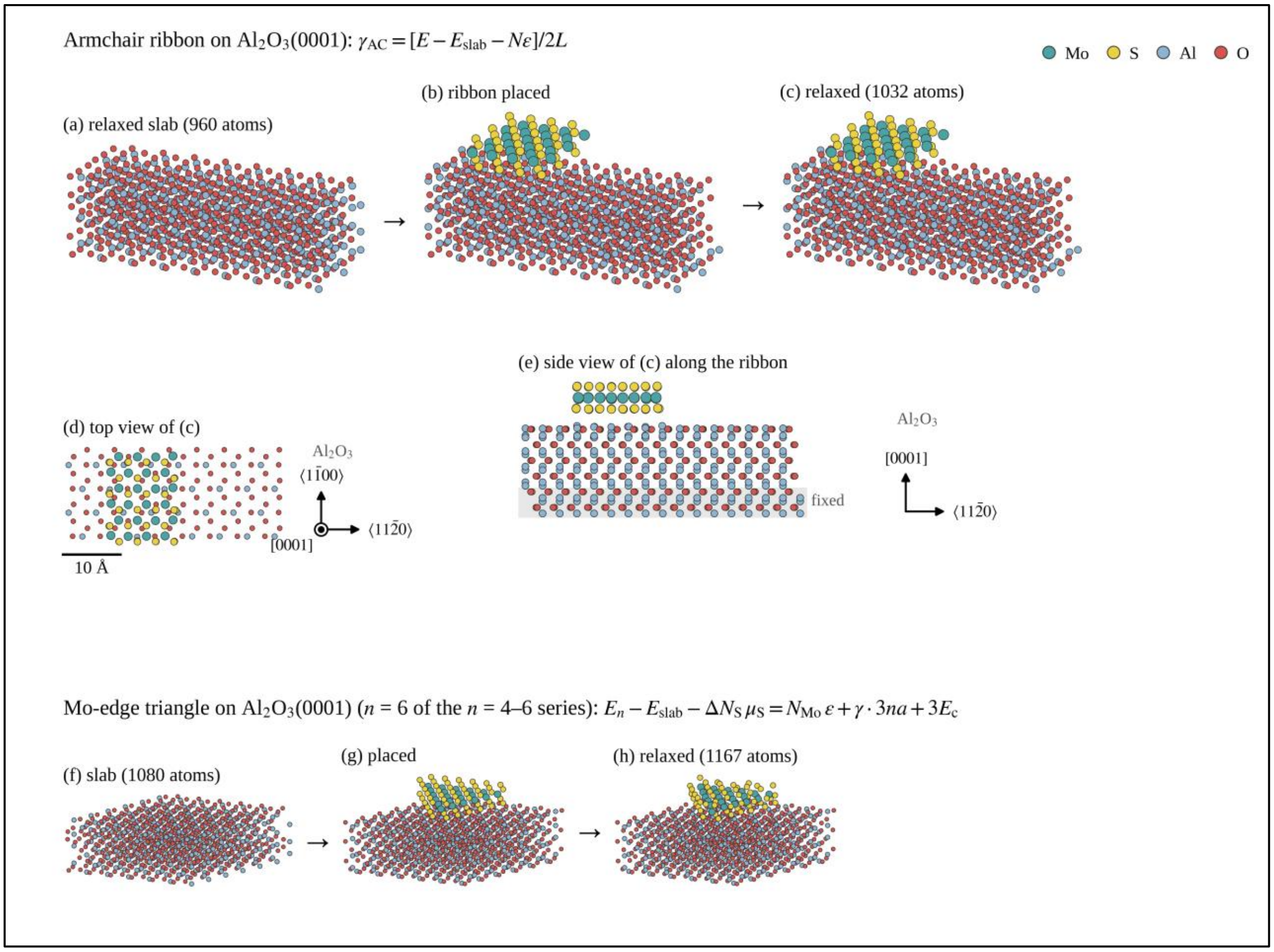

***FIG. S16.*** *Construction of the edge cells on $Al_2O_3(0001)$. Armchair ribbon (top): (a) relaxed $Al_2O_3(0001)$ slab (960 atoms), (b) relaxed isolated w = 8 ribbon placed on it, (c) relaxed heterostructure (1032 atoms; $MoS_2$ atoms move 0.09 Å RMS, 0.26 Å maximum), oblique view, (d) top view and (e) side view along the ribbon (fixed slab atoms shaded). Mo-edge triangle (bottom): (f) 6×6 slab (1080 atoms), (g) $Mo_{21}S_{66}$ placed, (h) relaxed (1167 atoms; 0.61 Å RMS, 2.10 Å maximum).*

**TABLE S5.** Edge energies of 1H-$MoS_2$ nanoribbons, $\Delta E_{edge} = (E_{ribbon} - N \cdot \varepsilon_{ML})/2L$, where $\varepsilon_{ML}$ is the energy per $MoS_2$ of the pristine monolayer (on sapphire: of the supported monolayer, with the slab energy subtracted from $E_{ribbon}$) and $2L$ the total edge length per cell. For stoichiometric ribbons ($\Delta N_S = 0$, S atoms beyond $2N$ per ribbon cell) no chemical potential enters. A zigzag ribbon has one Mo edge and one S edge, so its value is their average; edges are labeled by S coverage. †Off-stoichiometric ribbons: the excess (missing) S is referenced to α-S (S-rich limit), $-\Delta N_S \cdot \mu_S$(α-S) is added to the numerator.

| Ribbon | Edges | $\Delta N_S$ | DFT (eV/Å) | NEP (eV/Å) | Error (%) |
|---|---|---|---|---|---|
| Zigzag ribbon, isolated | Mo edge 0 % / S edge 100 % | 0 | 0.790 | 0.714 | −9.6 |
| | Mo edge 50 % / S edge 50 % | 0 | 0.548 | 0.530 | −3.3 |
| Armchair ribbon, isolated | armchair (both edges) | 0 | 0.790 | 0.774 | −2.0 |
| Zigzag ribbon, on $Al_2O_3$(0001) | Mo edge 0 % / S edge 100 % | 0 | 0.714 | 0.659 | −7.6 |
| Armchair ribbon, on $Al_2O_3$(0001) | armchair (both edges) | 0 | 0.751 | 0.697 | −7.2 |
| Zigzag ribbon, isolated † | Mo edge 0 % / S edge 50 % | −4 | 0.905 | 0.868 | −4.1 |
| | Mo edge 50 % / S edge 100 % | +4 | 0.442 | 0.376 | −14.9 |
| | Mo edge 100 % / S edge 50 % | +4 | 0.504 | 0.498 | −1.1 |
| | Mo edge 100 % / S edge 100 % | +8 | 0.392 | 0.342 | −12.7 |
| Zigzag ribbon, on $Al_2O_3$(0001) † | Mo edge 50 % / S edge 100 % | +3 | 0.391 | 0.272 | −30.6 |
| | Mo edge 100 % / S edge 50 % | +3 | 0.443 | 0.442 | −0.3 |
| | Mo edge 100 % / S edge 100 % | +6 | 0.281 | 0.287 | +2.0 |

**TABLE S6.** Largest residuals of the linear fits of the triangular-flake energies against the perimeter (n = 4–8) used to obtain the individual edge energies of isolated 1H-$MoS_2$ (DFT).

| Termination | Mo 0 % | Mo 50 % | Mo 100 % | S 50 % | S 100 % |
|---|---|---|---|---|---|
| Largest residual (eV) | 1.13 | 0.43 | 0.26 | 0.23 | 1.56 |

**TABLE S7.** Individual edge energies $\gamma$ (eV/Å) of 1H-$MoS_2$ at the Mo-rich and S-rich limits, as plotted in Fig. 6. Mo and S edges are obtained from triangular flakes of increasing size ($n$ = 4–8 Mo per edge isolated, 4–6 on sapphire) by fitting $E_n - N\cdot\varepsilon_{ML} - \Delta N_S\cdot\mu_S = \gamma\cdot 3na + 3E_{corner}$; armchair values come from stoichiometric ribbons and are independent of $\mu_S$. Edges are labeled by their S coverage. Each method is evaluated at its own limits (Mo-rich ΔμS = −1.20 eV in DFT and −1.27 eV in the NEP).

| | | Mo-rich | | | S-rich ($\Delta\mu_S$ = 0) | | |
|---|---|---|---|---|---|---|---|
| Environment | Edge | DFT | NEP | Err. (%) | DFT | NEP | Err. (%) |
| Isolated monolayer | Mo edge, 0 % S | 0.862 | 0.775 | −10.0 | 1.114 | 1.043 | −6.4 |
| | Mo edge, 50 % S | 0.480 | 0.461 | −4.0 | 0.354 | 0.327 | −7.6 |
| | Mo edge, 100 % S | 0.780 | 0.836 | +7.2 | 0.275 | 0.300 | +9.3 |
| | S edge, 50 % S | 0.586 | 0.566 | −3.5 | 0.713 | 0.700 | −1.8 |
| | S edge, 100 % S | 0.760 | 0.651 | −14.4 | 0.507 | 0.383 | −24.5 |
| | armchair | 0.790 | 0.774 | −2.0 | 0.790 | 0.774 | −2.0 |
| On $Al_2O_3$(0001) | Mo edge, 100 % S | 0.624 | 0.684 | +9.7 | 0.119 | 0.149 | +25.2 |
| | S edge, 100 % S | 0.610 | 0.560 | −8.3 | 0.358 | 0.292 | −18.4 |
| | armchair | 0.751 | 0.697 | −7.2 | 0.751 | 0.697 | −7.2 |

## Note S7. Point Defects

All point defects were placed in a 6×6 1H-$MoS_2$ monolayer on a 4×4 $Al_2O_3$(0001) slab (588 atoms in the pristine cell, bottom 144 slab atoms fixed), with the same 3:2 commensuration and strain as the 0° adhesion cell (Fig. S17). The thirteen defects are single S vacancies on the vacuum side and on the interface side, a vacancy of both S atoms of one column ($V_{S_2}$), two separated S vacancies on opposite sides, the Mo vacancy ($V_{Mo}$), the $V_{MoS_3}$ complex on the vacuum side and on the interface side, $V_{MoS_6}$, a Mo atom on an S site on either side ($Mo_S$), a Mo atom on an $S_2$ column ($Mo_{S_2}$), an S atom on a Mo site ($S_{Mo}$) and an $S_2$ dimer on a Mo site ($(S_2)_{Mo}$). Each defect cell was relaxed from the as-cut structure with the same fixed atoms, and the defect energy is the total-energy difference of the relaxed defective and pristine cells, ΔE = E(defect) − E(pristine). The formation energy is

$$E_f = \Delta E + \sum_i n_i \mu_i$$

where $n_i$ is the number of atoms of species i removed (negative if added) and $\mu_i$ its chemical potential, with $\mu_{Mo} + 2\mu_S = \mu_{MoS_2}$ and $\mu_S$ at the Mo-rich or S-rich limit. Table S8 lists the defect energies of Fig. 7 of the main text, and Fig. S18 shows the formation energies.

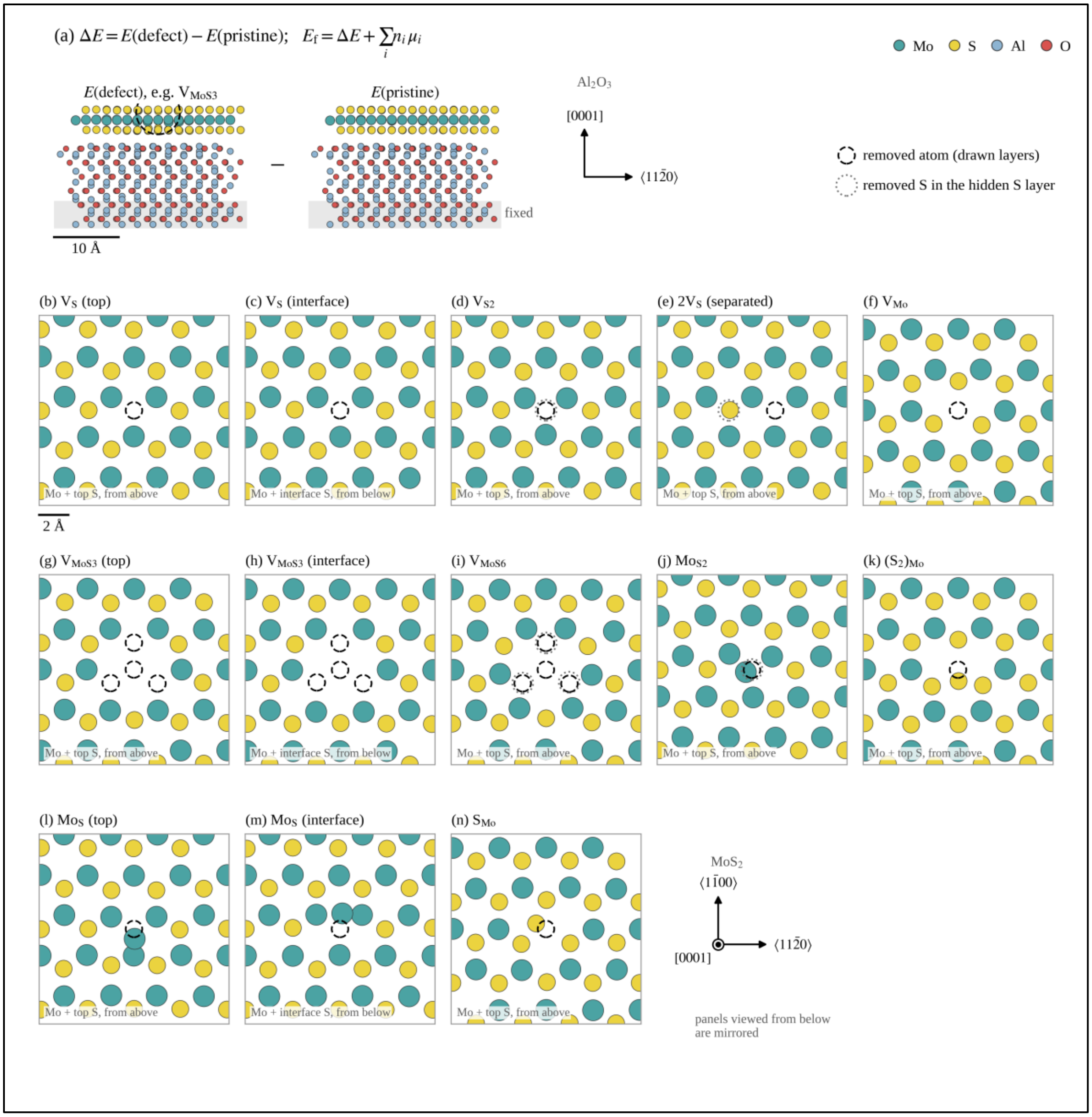

***FIG. S17.*** *DFT cells of the point-defect calculations (Fig. 7). (a) All defects are placed in a 6×6 1H-$MoS_2$ monolayer on a 4×4 Al-terminated $Al_2O_3$(0001) slab (588 atoms pristine; bottom 144 slab atoms fixed, shaded). $\Delta E = E(\text{defect}) - E(\text{pristine})$ and $E_f = \Delta E + \Sigma_i\, n_i\, \mu_i$, with $n_i$ the number of atoms of species i removed (negative if added) and $\mu_{Mo} + 2\mu_S = \mu_{MoS_2}$. (b–n) Relaxed structures of the 13 defects around the defect site. Only the Mo layer and the S layer facing the viewer are drawn, because the two S layers overlap in plan view: vacuum-side (‘top’) defects are viewed from above, interface-side defects from below (mirrored). Dashed circles mark removed atoms of the drawn layers, dotted grey circles removed S atoms of the hidden S layer (pristine positions); antisite atoms are drawn at their relaxed positions.*

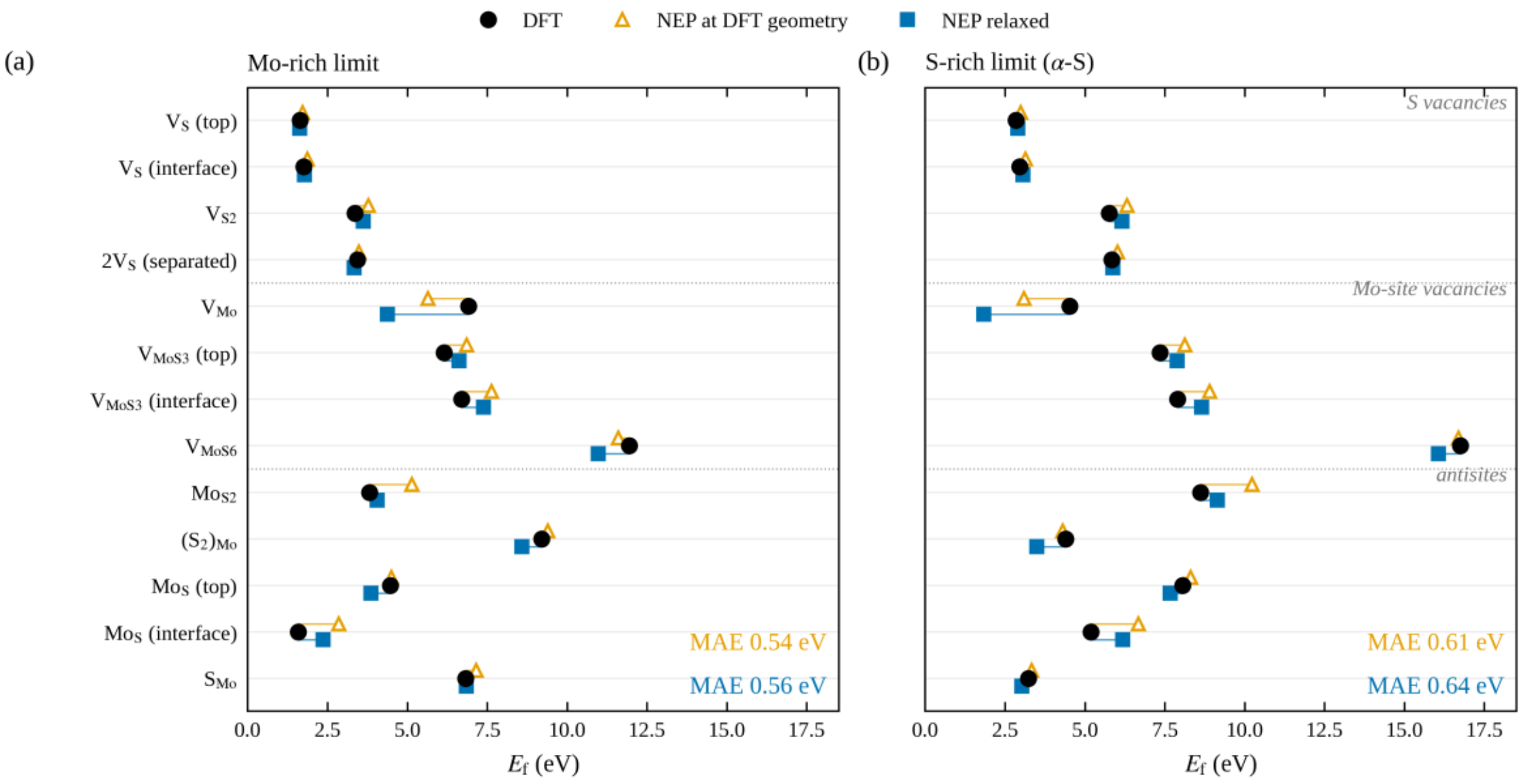

***FIG. S18.*** *Point-defect formation energies of 1H-$MoS_2$ on $Al_2O_3$(0001) at the (a) Mo-rich and (b) S-rich limits: DFT (filled circles), NEP evaluated on the DFT-relaxed structures (open triangles) and after NEP relaxation (filled squares).* $E_f = E(\text{defect}) - E(\text{pristine}) + \Sigma_i n_i\mu_i$, *with* $n_i$ *the number of atoms of species i removed (negative if added) and* $\mu_{Mo} + 2\mu_S = \mu_{MoS_2}$ *(1H monolayer). Each method uses its own reference energies; NEP gives* $\Delta H_f = -2.543$ *eV per formula unit versus −2.399 eV in DFT. Mean absolute errors 0.54–0.64 eV.*

**TABLE S8.** Point-defect energies ΔE = E(defect) − E(pristine) (eV) in a 6×6 1H-$MoS_2$ monolayer on a 4×4 Al-terminated $Al_2O_3$(0001) slab (588 atoms pristine, bottom 144 slab atoms fixed). ΔE is the total-energy difference of the same two cells in both methods (not atom-balanced; NEP energies on the DFT absolute scale via the isolated-atom energies). NEP is evaluated on the DFT-relaxed cells and after NEP relaxation of both cells from the same starting structures; errors are NEP − DFT. 'Top' and 'interface' denote the S layer facing vacuum or sapphire.

| Class | Defect | DFT | NEP (DFT geom.) | Error | NEP (relaxed) | Error |
|---|---|---|---|---|---|---|
| S vacancies | $V_S$ (top) | 7.08 | 7.20 | +0.12 | 7.11 | +0.03 |
| | $V_S$ (interface) | 7.20 | 7.35 | +0.15 | 7.26 | +0.07 |
| | $V_{S2}$ | 14.24 | 14.74 | +0.50 | 14.58 | +0.34 |
| | 2$V_S$ (separated) | 14.32 | 14.44 | +0.12 | 14.29 | −0.02 |
| Mo-site vacancies | $V_{Mo}$ | 18.39 | 17.03 | −1.35 | 15.77 | −2.62 |
| | $V_{MoS3}$ (top) | 33.94 | 34.69 | +0.75 | 34.45 | +0.51 |
| | $V_{MoS3}$ (interface) | 34.49 | 35.47 | +0.98 | 35.21 | +0.72 |
| | $V_{MoS6}$ | 56.06 | 55.88 | −0.19 | 55.25 | −0.82 |
| Antisites | $Mo_{S2}$ | 3.24 | 4.72 | +1.48 | 3.62 | +0.39 |
| | $(S_2)_{Mo}$ | 9.78 | 9.83 | +0.05 | 9.01 | −0.77 |
| | $Mo_S$ (top) | −1.56 | −1.41 | +0.15 | −2.05 | −0.49 |
| | $Mo_S$ (interface) | −4.43 | −3.05 | +1.38 | −3.55 | +0.89 |
| | $S_{Mo}$ | 12.85 | 13.06 | +0.21 | 12.75 | −0.10 |
| | Mean absolute error | | | 0.57 | | 0.60 |

## Note S8. Grain Boundaries

Following Komsa and Krasheninnikov[9], mirror-twin boundary energies are computed per lattice constant along the boundary, from ribbons that are periodic along the boundary (length L) and contain one MTB between two mirror domains and two S-passivated Mo edges (Fig. S19):

$$E_{\mathrm{MTB}} = \left[E - N_{\mathrm{Mo}}\,\mu_{\mathrm{MoS_2}} - \Delta N_{\mathrm{S}}\,\mu_{\mathrm{S}} - 2\gamma_{\mathrm{Mo}}L\right]\frac{a}{L},$$

where E is the total energy of the ribbon cell and $\gamma_{\mathrm{Mo}}$ the Mo-edge energy of the same method and the same outer S coverage from the triangle fit of Note S6. The freestanding 4|4P and 4|4E ribbons contain 112–176 atoms; on sapphire, the ribbons contain 816 atoms, of which 216 of the 720 slab atoms are fixed, and

$$E_{\mathrm{MTB}}^{\mathrm{sub}} = \left[E - E_{\mathrm{slab}} - N_{\mathrm{Mo}}\,\varepsilon - \Delta N_{\mathrm{S}}\,\mu_{\mathrm{S}} - 2\gamma_{\mathrm{Mo}}^{\mathrm{sub}}L\right]\frac{a}{L}.$$

The supported Mo-edge energy is $\gamma_{\mathrm{Mo}}^{\mathrm{sub}} = \Sigma_{zz}^{\mathrm{sub}} - \gamma_{\mathrm{S}}^{\mathrm{sub}}$, where $\Sigma_{zz}^{\mathrm{sub}}$ is the edge-energy sum of the zigzag ribbon on sapphire and $\gamma_{\mathrm{S}}^{\mathrm{sub}}$ the S-edge energy from the supported S-edge triangles. The difference ΔE(4|4E − 4|4P) = [E(4|4E) − E(4|4P)]·a/L compares cells of identical composition and needs neither a chemical potential nor an edge reference.

Symmetric tilt boundaries were computed in fully periodic bicrystals that contain two boundaries of opposite polarity (Fig. S20),

$$\gamma_{\mathrm{GB}} = \frac{E - N_{\mathrm{Mo}}\,\mu_{\mathrm{MoS_2}} - \Delta N_{\mathrm{S}}\,\mu_{\mathrm{S}}}{2L},$$

where E is the total energy of the bicrystal and L the length of each boundary, so that $\gamma_{\mathrm{GB}}$ is the average over the two boundaries. The Σ7 (21.8°), Σ19 (13.2°) and Σ37 (9.4°) bicrystals contain 160–165, 216–221 and 429–431 atoms, and for each the stoichiometric translation and the lowest non-stoichiometric translations (S-deficient or S-rich core) were computed. The Σ7 boundary between ±10.9° grains was also computed freestanding and on the $Al_2O_3(0001)$ slab at the same strain (−1.04 %; 882–884 atoms, bottom 216 fixed); on sapphire, $E - E_{\mathrm{slab}} - N_{\mathrm{Mo}}\,\varepsilon^{\pm 10.9}$ replaces $E - N_{\mathrm{Mo}}\,\mu_{\mathrm{MoS_2}}$, where $\varepsilon^{\pm 10.9}$ is the energy per formula unit of the supported monolayer at the same strain. Each method is evaluated at its own limits (Sec. 2.4): the Mo-rich limit lies at $\Delta\mu_{\mathrm{S}} = -1.20$ eV in DFT and at −1.27 eV with the NEP. Table S9 lists the boundary energies of Fig. 8 of the main text together with the NEP-relaxed values, which are also plotted in Fig. S21.

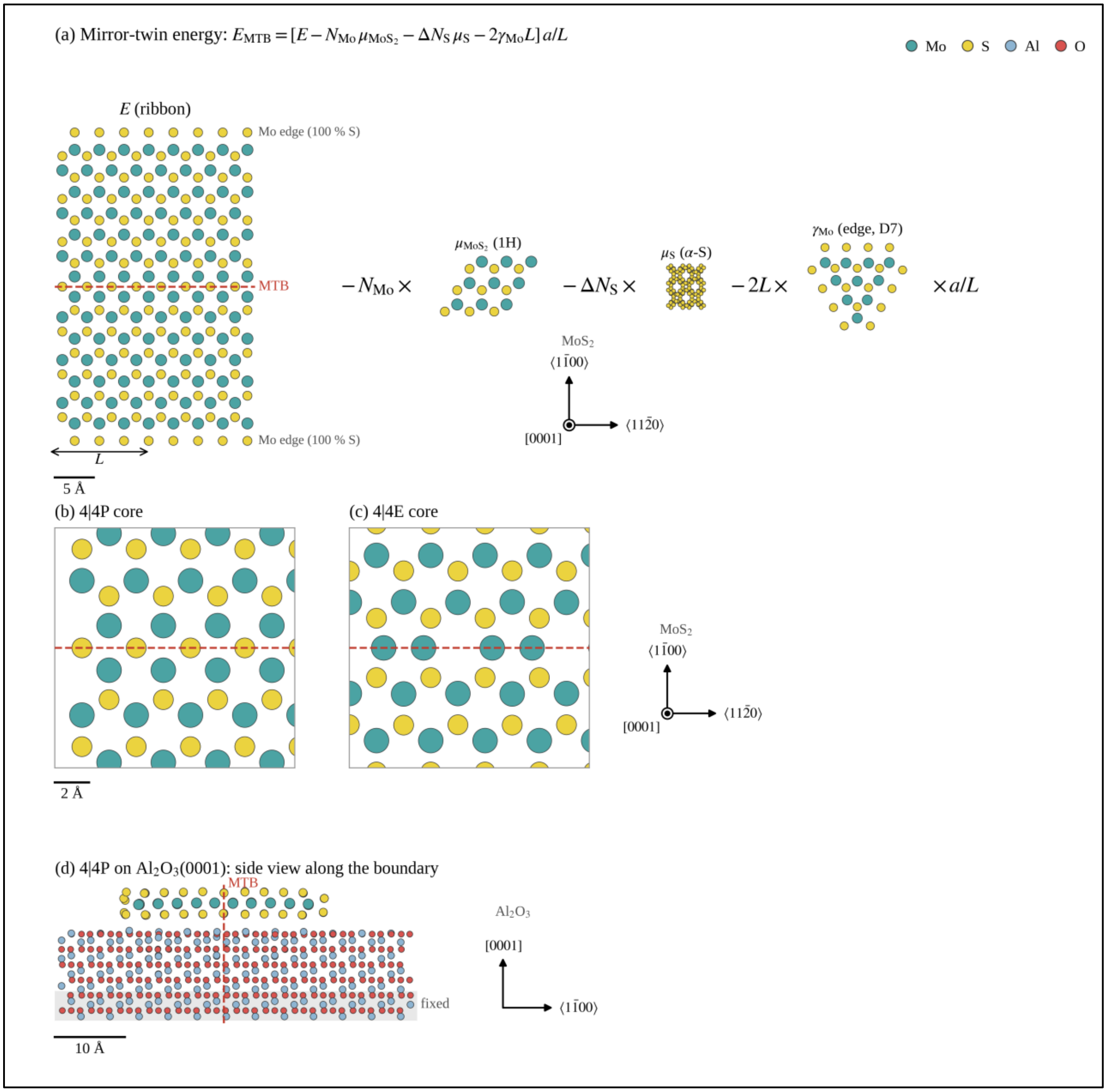

***FIG. S19.*** *DFT cells of the mirror-twin boundary (MTB) energies (Fig. 8, Table S9). (a)* $E_{MTB} = [E - N_{Mo}\,\mu_{MoS_2} - \Delta N_S\,\mu_S - 2\gamma_{Mo}\,L]\cdot a/L$*: E is the energy of a ribbon periodic along the boundary (length L, two periods drawn) containing one MTB (dashed red) between two mirror domains and two S-passivated Mo edges;* $\gamma_{Mo}$ *is the energy of the 100 %-S Mo edge of the same method from the triangle-size fit (Note S6); a is the lattice constant. (b, c) Relaxed cores of the freestanding 4|4P and 4|4E boundaries (top view, two periods; dashed line: boundary). (d) 4|4P ribbon on the* $Al_2O_3$*(0001) slab (816 atoms, bottom 216 fixed, shaded), viewed along the boundary; on sapphire the slab and supported-monolayer energies replace* $N_{Mo}\,\mu_{MoS_2}$ *and the supported Mo-edge energy replaces* $\gamma_{Mo}$*. Arrows give the* $MoS_2$ *(a–c) and* $Al_2O_3$ *(d) crystallographic directions.*

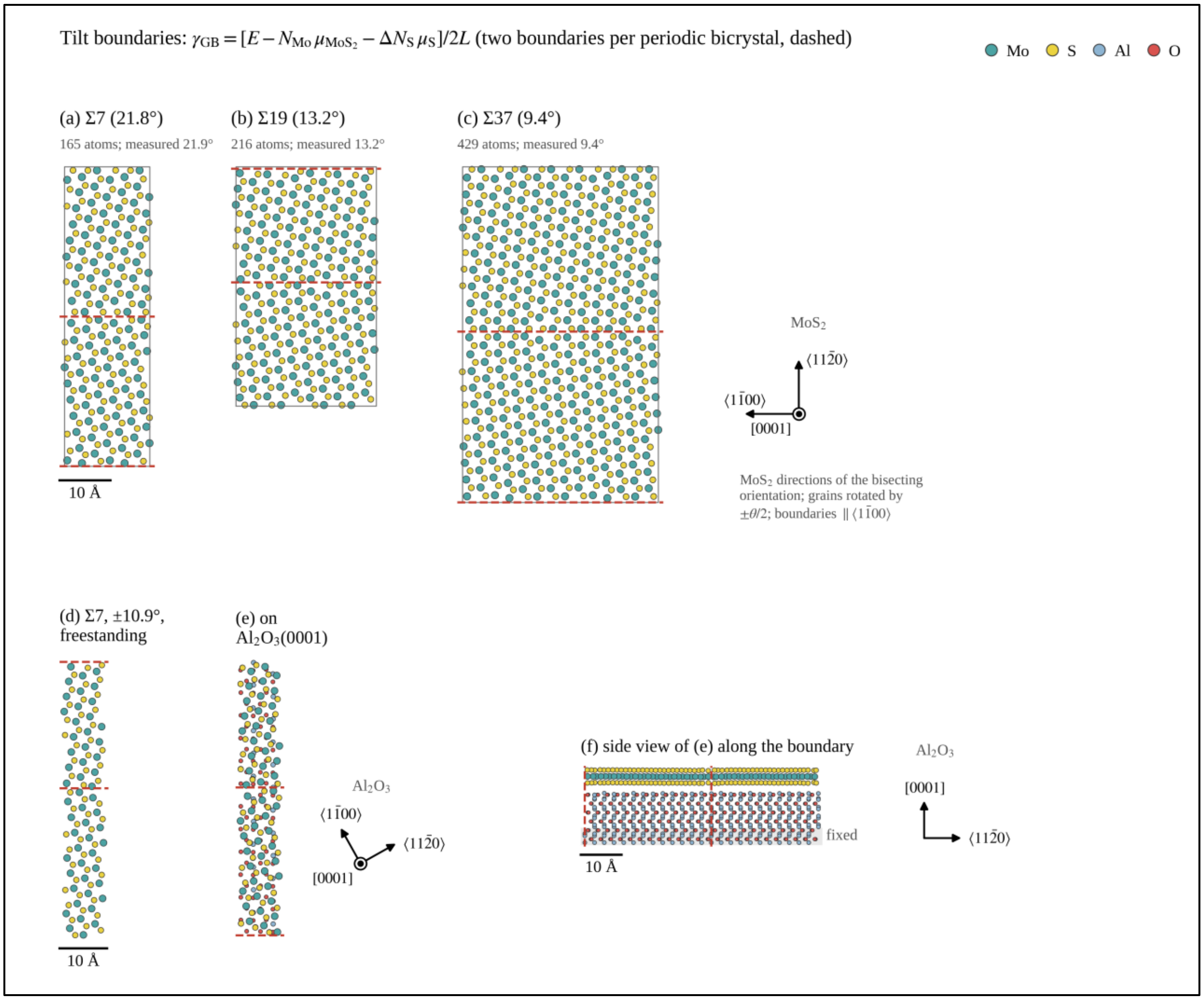

***FIG. S20.*** *DFT cells of the symmetric tilt grain boundaries (Fig. 8, Table S9).* $\gamma_{GB} = [E - N_{Mo}\,\mu_{MoS_2} - \Delta N_S\,\mu_S]/2L$, *where E is the energy of a fully periodic* $MoS_2$ *bicrystal containing two boundaries of length L (dashed red; the two boundaries have opposite polarity, so* $\gamma_{GB}$ *is their average). (a–c) Relaxed stoichiometric Σ7 (21.8°), Σ19 (13.2°) and Σ37 (9.4°) bicrystals (two periods along the boundary; cell outlined); misorientations measured from the Mo–Mo bond directions of the two grains. The grains are rotated by ±θ/2 from a common bisecting orientation, and the boundaries run along its ⟨1-100⟩ direction. (d) Σ7 boundary between ±10.9° grains, freestanding, and (e) the same boundary on the* $Al_2O_3$*(0001) slab (882 atoms, bottom 216 fixed), at equal strain (−1.04 %); (f) side view of (e) along the boundary. On sapphire the slab and supported-monolayer energies replace* $N_{Mo}\,\mu_{MoS_2}$*.*

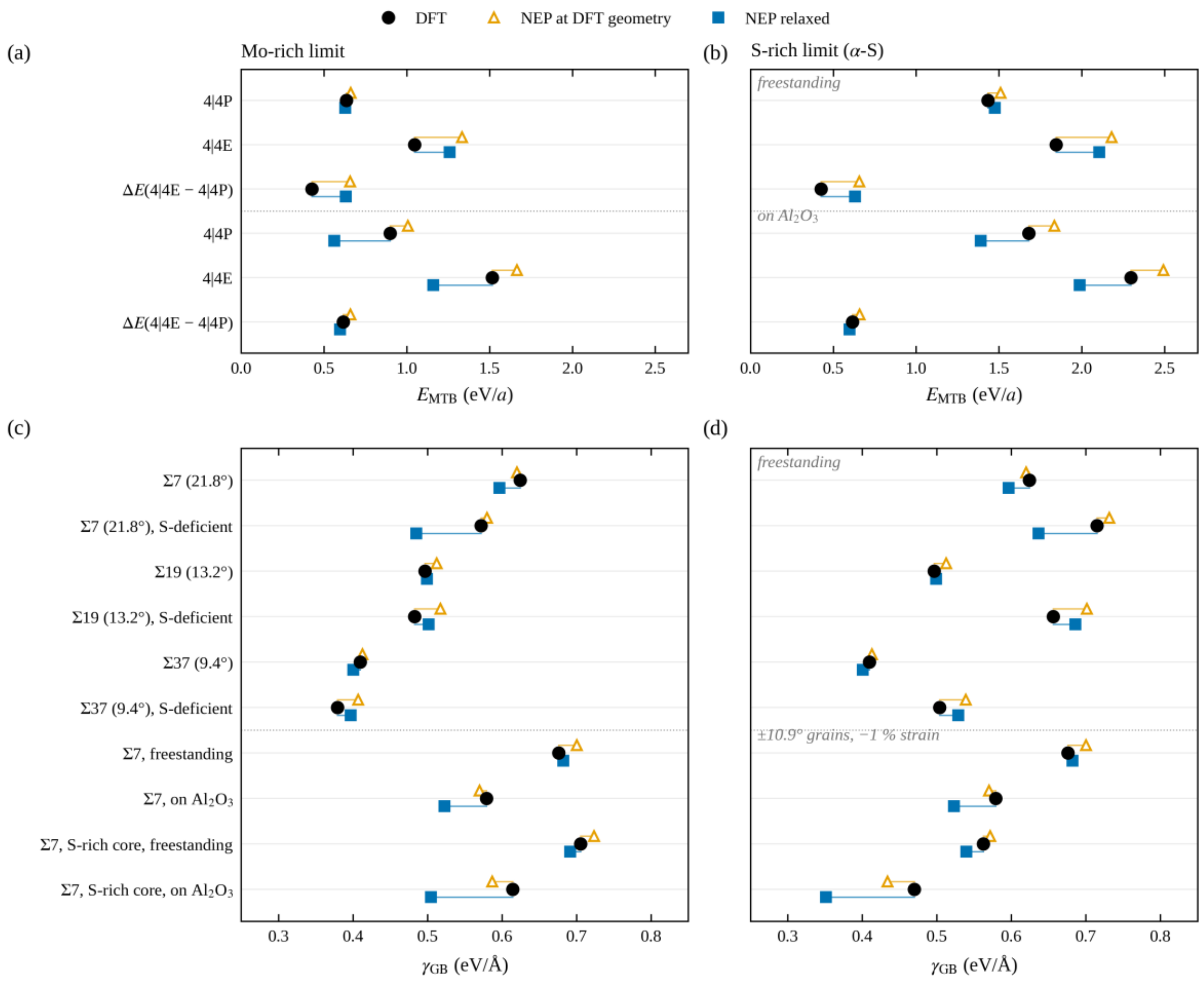

***FIG. S21.*** *Grain-boundary energies of 1H-$MoS_2$ from DFT (filled circles), the NEP evaluated on the DFT-relaxed structures (open triangles) and the NEP after relaxation from the same structures (filled squares). (a, b) Mirror-twin boundary energies per lattice constant a along the boundary (Note S8); ΔE(4|4E − 4|4P) compares cells of identical composition and needs no chemical potential or edge reference. (c, d) Tilt grain-boundary energies from periodic bicrystals with two boundaries; S-deficient and S-rich-core variants are the lowest non-stoichiometric translations. Lower block: Σ7 between ±10.9° grains, freestanding and on $Al_2O_3$(0001) at the same strain (−1.04 %). Left: Mo-rich limit; right: S-rich limit (α-S). On sapphire, NEP relaxation shifts the $MoS_2$ laterally by about 0.5 Å relative to the substrate in these cells, which lowers the NEP-relaxed supported values.*

**TABLE S9.** Grain-boundary energies of 1H-$MoS_2$ at the Mo-rich and S-rich (α-S) limits. $NEP_a$: NEP evaluated on the DFT-relaxed structure; $NEP_b$: NEP relaxed from it. Mirror-twin boundaries (MTB) are given per lattice constant $a$ along the boundary, $E_{MTB} = [E - N_{Mo}\mu_{MoS2} - \Delta N_S\mu_S - 2\gamma_{Mo}L]\cdot a/L$, from ribbons with one MTB and two S-passivated (or bare) Mo edges, $\gamma_{Mo}$ being the Mo-edge energy of the same method; on $Al_2O_3$(0001) the slab and supported-monolayer energies replace $N_{Mo}\mu_{MoS2}$. $\Delta E$(4|4E − 4|4P) compares cells of identical composition and needs no chemical potential or edge reference. Tilt boundaries: $\gamma_{GB} = [E - N_{Mo}\mu_{MoS2} - \Delta N_S\mu_S]/2L$ from periodic bicrystals with two boundaries; stoichiometric rows are independent of $\mu_S$. The ±10.9° rows compare the same Σ7 boundary freestanding and on $Al_2O_3$ at equal strain (−1.04 %). Each method is evaluated at its own limits (Mo-rich $\Delta\mu S$ = −1.20 eV in DFT and −1.27 eV in the NEP).

| Group | Boundary | Mo-rich | | | S-rich (α-S) | | |
|---|---|---|---|---|---|---|---|
| | | DFT | $NEP_a$ | $NEP_b$ | DFT | $NEP_a$ | $NEP_b$ |
| MTB, freestanding (eV/$a$) | 4\|4P | 0.635 | 0.662 | 0.628 | 1.434 | 1.510 | 1.476 |
| | 4\|4E | 1.047 | 1.333 | 1.259 | 1.846 | 2.181 | 2.107 |
| | $\Delta E$(4\|4E − 4\|4P) | 0.426 | 0.657 | 0.631 | 0.426 | 0.657 | 0.631 |
| MTB, bare outer edges (eV/$a$) | 4\|4P | 0.456 | 0.677 | 0.608 | 1.256 | 1.525 | 1.456 |
| | 4\|4E | 0.807 | 1.343 | 1.233 | 1.606 | 2.190 | 2.081 |
| MTB, on $Al_2O_3$(0001) (eV/$a$) | 4\|4P | 0.900 | 1.006 | 0.562 | 1.681 | 1.835 | 1.390 |
| | 4\|4E | 1.515 | 1.666 | 1.160 | 2.297 | 2.494 | 1.988 |
| | $\Delta E$(4\|4E − 4\|4P) | 0.616 | 0.659 | 0.597 | 0.616 | 0.659 | 0.597 |
| Tilt GB, freestanding (eV/Å) | Σ7 (21.8°) | 0.624 | 0.620 | 0.596 | 0.624 | 0.620 | 0.596 |
| | Σ7 (21.8°), S-deficient | 0.572 | 0.580 | 0.485 | 0.715 | 0.732 | 0.637 |
| | Σ19 (13.2°) | 0.496 | 0.512 | 0.499 | 0.496 | 0.512 | 0.499 |
| | Σ19 (13.2°), S-deficient | 0.482 | 0.517 | 0.502 | 0.656 | 0.702 | 0.686 |
| | Σ37 (9.4°) | 0.410 | 0.413 | 0.401 | 0.410 | 0.413 | 0.401 |
| | Σ37 (9.4°), S-deficient | 0.379 | 0.407 | 0.397 | 0.504 | 0.539 | 0.529 |
| Tilt GB, ±10.9° grains (eV/Å) | Σ7, freestanding | 0.676 | 0.700 | 0.682 | 0.676 | 0.700 | 0.682 |
| | Σ7, on $Al_2O_3$ | 0.579 | 0.570 | 0.523 | 0.579 | 0.570 | 0.523 |
| | Σ7, S-rich core, freestanding | 0.705 | 0.724 | 0.691 | 0.562 | 0.572 | 0.540 |
| | Σ7, S-rich core, on $Al_2O_3$ | 0.614 | 0.587 | 0.505 | 0.470 | 0.434 | 0.351 |

**Note S9. Stacking Faults**

Generalized stacking-fault energies were computed for sliding one $MoS_2$ layer over the next, starting from 2H or 3R stacking, in three systems (Fig. S22): a freestanding bilayer (6 atoms, about 25 Å of vacuum), a bulk crystal with one faulted interface (2H: four layers, 12 atoms; 3R: six layers, 18 atoms) and a bilayer on the $Al_2O_3(0001)$ slab (3×3 $MoS_2$ on 2×2 $Al_2O_3$, 174 atoms, 0° registry, bottom 36 atoms fixed). The upper part of the cell is displaced by s = t·a along the zigzag direction ⟨11-20⟩ or by s = t·√3a along the armchair direction ⟨1-100⟩, for t = 0–1 in steps of 1/12. In the bulk cells, the c vector is tilted, c′ = c + s, so that only the interface across the periodic boundary is faulted. The fault energy is

$$\gamma(t) = \frac{E(t) - E(0)}{A},$$

where E(t) is the total energy at displacement t and A the in-plane area of a cell with one faulted interface. Every point was computed both rigidly, at the interlayer spacing of the unfaulted cell, and with the atoms relaxed along z at fixed x and y (force criterion 0.005 eV/Å). The unstable stacking-fault energy $\gamma_{max}$ is the maximum of γ(t) along a path.

All curves are reported for the same sliding direction relative to the faulted interface, with the upper layer moving against the in-plane Mo→S direction of the lower layer. Figure S23 shows all twelve sets of curves, of which Fig. 9 of the main text shows three, and Table S10 lists $\gamma_{max}$ and the energy of the 2H-armchair metastable stacking. The cells of all calculations of Notes S6–S9 are summarized in Table S11.

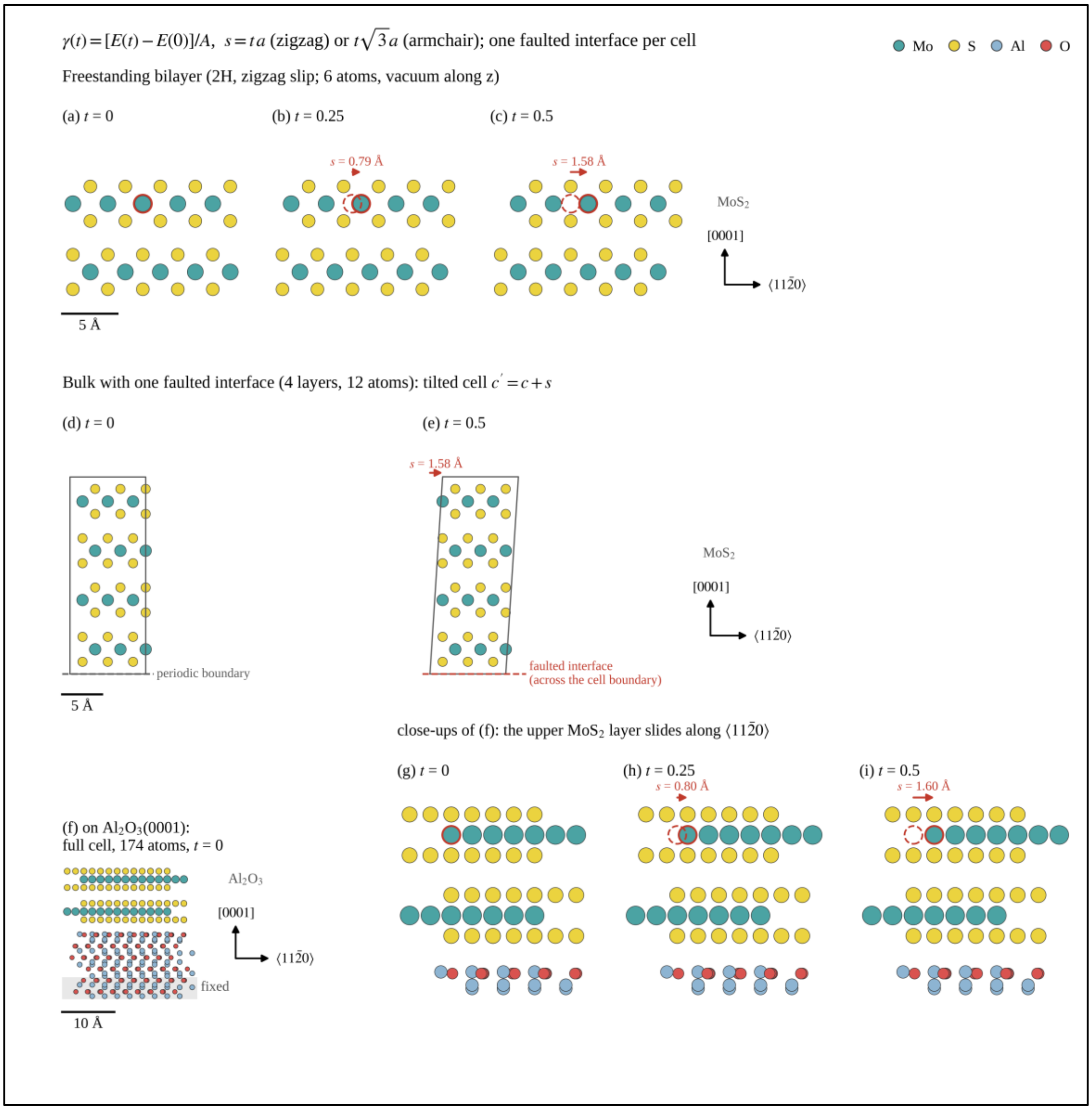

***FIG. S22.*** *DFT cells of the generalized stacking-fault energy (Fig. 9, Table S10). γ(t) = [E(t) − E(0)]/A is the energy per in-plane area A of a cell containing one faulted interface whose upper side is displaced by s = t·a along ⟨11-20⟩ (shown) or s = t·√3a along ⟨1-100⟩, t = 0–1 in 12 steps; each point is computed rigidly and with z relaxed (x, y fixed). (a–c) Freestanding 2H bilayer (6 atoms, vacuum along z) at t = 0, 1/4 and 1/2: the upper layer is displaced (red arrow; dashed and solid circles: one Mo before and after). (d, e) Bulk crystal with one faulted interface: four layers in a cell whose c vector is tilted, c′ = c + s, so that all atoms keep their positions and only the interface across the periodic boundary is faulted. (f) Bilayer on the* $Al_2O_3(0001)$ *slab (174 atoms, 0° registry, bottom 36 fixed), full cell at t = 0; (g–i) close-ups at t = 0, 1/4 and 1/2: only the upper* $MoS_2$ *layer is displaced (0.80 and 1.60 Å); the lower layer and the substrate stay in place. Side views along ⟨1-100⟩.*

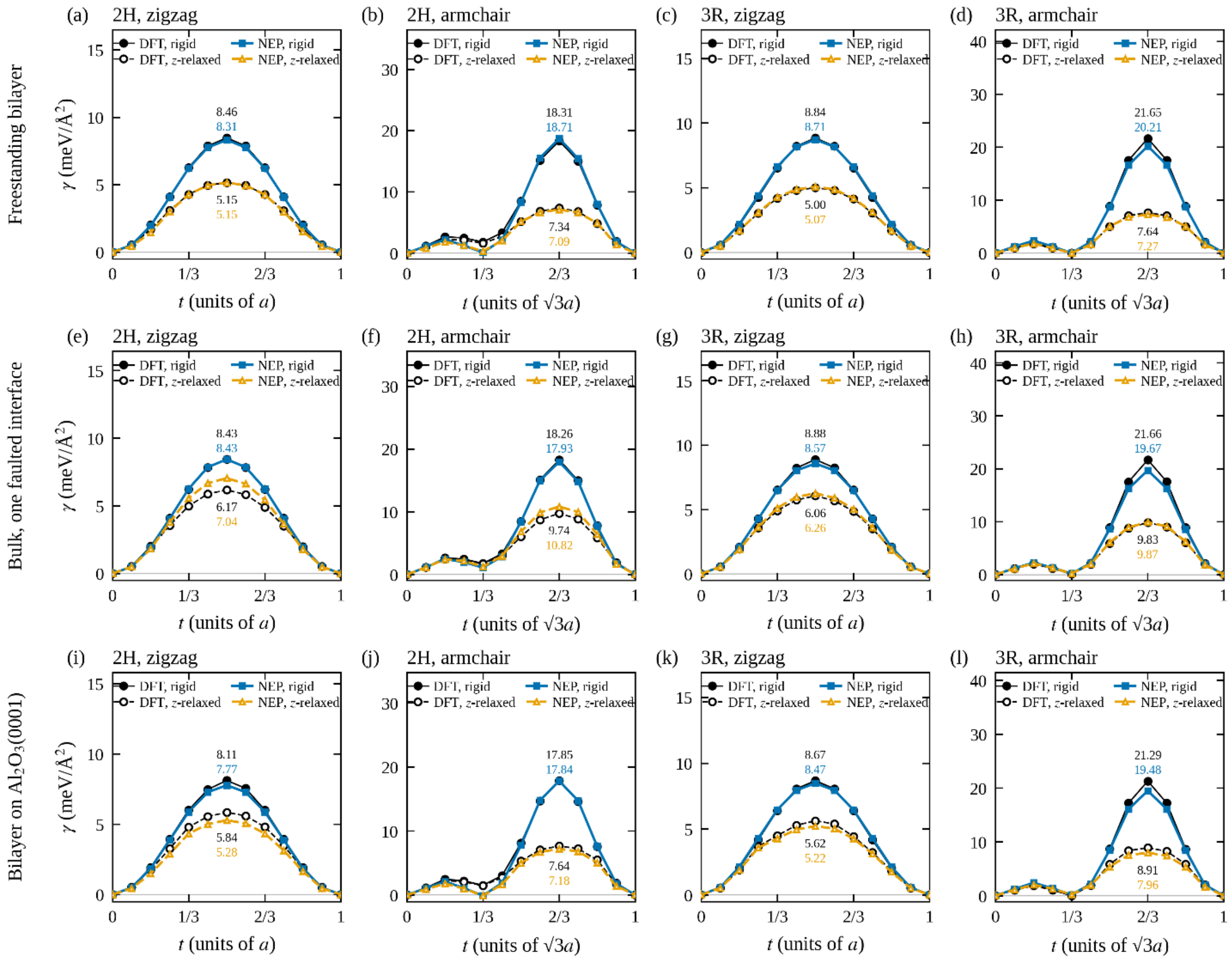

***FIG. S23.*** *Generalized stacking-fault energies of $MoS_2$, DFT versus NEP. Rows: freestanding bilayer, bulk crystal with one faulted interface, bilayer on $Al_2O_3(0001)$ (0° registry). $\gamma(t) = [E(t) - E(0)]/A$ for sliding one layer over the next along the zigzag (t in units of a) and armchair (t in units of √3a) directions, from 2H or 3R stacking. Rigid: all atoms at the spacing of the unfaulted cell; z-relaxed: x, y fixed and z relaxed with DFT. DFT: black circles (filled rigid, open z-relaxed); NEP: blue squares (rigid) and orange triangles (z-relaxed), evaluated on the identical structures. Numbers give the maxima $\gamma_{max}$ of each curve (meV/Å²): DFT in black, NEP in color. All curves are drawn for the upper layer of the faulted pair sliding against the in-plane Mo→S direction of the lower layer; armchair paths computed in the opposite direction are plotted as $\gamma(1 - t)$, which is exact because each path spans one lattice vector.*

**TABLE S10.** Stacking-fault energies (meV/Å$^2$) of $MoS_2$: unstable stacking-fault energy $\gamma_{max}$, the maximum of $\gamma(t) = [E(t) - E(0)]/A$ for sliding one layer over the next along the zigzag or armchair direction from 2H or 3R stacking, and, for 2H armchair, the energy of the metastable intermediate stacking (local minimum of $\gamma$). Rigid: atoms at the equilibrium spacing of the unfaulted cell; $z$-relaxed: $x$, $y$ fixed, $z$ relaxed in DFT. The NEP is evaluated on the identical structures (for $z$-relaxed, the DFT-relaxed ones).

| | | Rigid | | | $z$-relaxed | | |
|---|---|---|---|---|---|---|---|
| System | Stacking, direction | DFT | NEP | Err. (%) | DFT | NEP | Err. (%) |
| Freestanding bilayer | 2H, zigzag | 8.46 | 8.31 | −1.8 | 5.15 | 5.15 | 0.0 |
| | 2H, armchair | 18.31 | 18.71 | +2.2 | 7.34 | 7.09 | −3.5 |
| | 3R, zigzag | 8.84 | 8.71 | −1.4 | 5.00 | 5.07 | +1.4 |
| | 3R, armchair | 21.65 | 20.21 | −6.7 | 7.64 | 7.27 | −4.8 |
| | 2H, armchair, metastable min. | 1.79 | 0.09 | −95.2 | 1.59 | 0.28 | −82.5 |
| Bulk, one faulted interface | 2H, zigzag | 8.43 | 8.43 | 0.0 | 6.17 | 7.04 | +14.1 |
| | 2H, armchair | 18.26 | 17.93 | −1.8 | 9.74 | 10.82 | +11.1 |
| | 3R, zigzag | 8.88 | 8.57 | −3.5 | 6.06 | 6.26 | +3.4 |
| | 3R, armchair | 21.66 | 19.67 | −9.2 | 9.83 | 9.87 | +0.5 |
| | 2H, armchair, metastable min. | 1.75 | 1.10 | −37.5 | 1.72 | 1.23 | −28.5 |
| Bilayer on $Al_2O_3$(0001) | 2H, zigzag | 8.11 | 7.77 | −4.2 | 5.84 | 5.28 | −9.6 |
| | 2H, armchair | 17.85 | 17.84 | −0.1 | 7.64 | 7.18 | −5.9 |
| | 3R, zigzag | 8.67 | 8.47 | −2.3 | 5.62 | 5.22 | −7.1 |
| | 3R, armchair | 21.29 | 19.48 | −8.5 | 8.91 | 7.96 | −10.6 |
| | 2H, armchair, metastable min. | 1.50 | −0.11 | −107.3 | 1.44 | −0.12 | −108.4 |

**TABLE S11.** DFT calculations of the edges, point defects, grain boundaries and stacking faults (Notes S6–S9), all with the data-collection settings of Table S1. Atoms: range over the set; in-plane cell: |a| (× |b| if different); vacuum: empty gap along z of the final structure; relaxation: degrees of freedom relaxed ('static': single points, including the screening calculations); fixed: atoms fixed in all directions; N: calculations used.

| Property | Cell set | Atoms | In-plane cell (Å) | Vacuum (Å) | Relaxation | Fixed | N |
|---|---|---|---|---|---|---|---|
| Edges | Isolated ribbons (zigzag, armchair) | 36–120 | 12.67–29.25 × 10.97–40.60 | 14.9–15 | ions | 0 | 16 |
| | Isolated triangles n = 4–8 | 22–140 | 24.50–40.33 × 23.23–36.94 | 14.2–15 | ions | 0 | 25 |
| | Ribbons on sapphire | 534–1032 | 9.57–38.29 × 16.58–33.16 | 15.5 | ions | 144–288 | 6 |
| | Triangles n = 4–6 on sapphire | 1114–1167 | 28.72 | 15.4–15.7 | ions | 324 | 6 |
| | Sapphire slab references (edges) | 480–1080 | 9.57–38.29 × 16.58–33.16 | 20.9 | static | 144–324 | 3 |
| Chem. potentials | Chemical-potential references | 2–128 | 3.13–10.68 × 3.13–13.05 | – | static, ions + cell | 0 | 7 |
| Point defects | Defect-cell references (slab, strained 1H) | 108–480 | 19.15 | 20–20.7 | ions | 0–144 | 2 |
| | Pristine 1H/Al2O3 defect cell | 588 | 19.15 | 25.1 | ions | 144 | 1 |
| | 13 point defects on sapphire | 581–589 | 19.15 | 24–25.2 | ions, static | 144 | 15 |
| Grain boundaries | Freestanding MTB ribbons (4\|4P, 4\|4E) | 112–176 | 12.67 × 39.07–53.69 | 14.9 | ions | 0 | 8 |
| | Freestanding tilt bicrystals Σ7/19/37 | 160–431 | 8.38–19.26 × 47.01–66.72 | 19.8–20 | ions | 0 | 8 |
| | MTB ribbons on sapphire | 816 | 9.57 × 49.74 | 15.4 | ions | 216 | 2 |
| | Σ7 bicrystal on sapphire | 882–884 | 8.29 × 57.44 | 15.2–15.4 | ions | 216 | 2 |
| | Sapphire slab references (GB) | 720 | 8.29–9.57 × 49.74–57.44 | 20.9 | static | 216 | 2 |
| Stacking faults | Freestanding bilayers (2H, 3R) | 6 | 3.16 | 24.4–25 | static, ions (z only) | 0 | 300 |
| | Bulk with one fault, tilted c (2H, 3R) | 12–18 | 3.16 | – | static, ions (z only) | 0 | 104 |
| | Bilayer on sapphire (2H, 3R) | 174 | 9.57 | 15.1–15.5 | static, ions (z only) | 36 | 104 |

## Supplemental Material References